\documentclass[a4paper,11pt]{article}
\pdfoutput=1 % if your are submitting a pdflatex (i.e. if you have
\usepackage{jheppub} % for details on the use of the package, please
\usepackage[T1]{fontenc} % if needed
\usepackage{ulem}

\usepackage{graphicx}
\usepackage{epsfig,amssymb,amsmath}
\usepackage{slashed, color}

\newcommand{\beq}{\begin{equation}}
\newcommand{\eeq}{\end{equation}}
\newcommand{\bea}{\begin{eqnarray}}
\newcommand{\eea}{\end{eqnarray}}
\newcommand{\bear}{\begin{array}}
\newcommand {\eear}{\end{array}}
\newcommand{\bef}{\begin{figure}}
\newcommand {\eef}{\end{figure}}
\newcommand{\bec}{\begin{center}}
\newcommand {\eec}{\end{center}}

\newcommand{\dis}[1]{\begin{equation}\begin{split}#1\end{split}\end{equation}}

\newcommand{\lsim}{
\mathrel{\hbox{\rlap{\hbox{\lower4pt\hbox{$\sim$}}}\hbox{$<$}}}}
\newcommand{\gsim}{
\mathrel{\hbox{\rlap{\hbox{\lower4pt\hbox{$\sim$}}}\hbox{$>$}}}}

\newcommand{\Mpl}{M_{\rm Pl}}

\title{\boldmath General Continuum Clockwork }

\preprint{CTPU-17-37, APCTP Pre2017-012}
\author[a]{Kiwoon Choi,}
\author[b]{Sang Hui Im,}
\author[a,c]{Chang Sub Shin}

\affiliation[a]{ Center for Theoretical Physics of the Universe, Institute for Basic Science (IBS), \\Daejeon 34051, Korea}
\affiliation[b]{Bethe Center for Theoretical Physics and Physikalisches Institut der Universit\"at Bonn,\\Nussallee 12, 53115 Bonn, Germany}
\affiliation[c]{Asia Pacific Center for Theoretical Physics, Pohang 37673, Korea}

\emailAdd{kchoi@ibs.re.kr}
\emailAdd{shim@th.physik.uni-bonn.de}
\emailAdd{csshin@ibs.re.kr}

\abstract{The continuum clockwork is an extra-dimensional set-up to realize certain features of the clockwork mechanism  generating  exponentially suppressed or hierarchical couplings  of light particles.
We study the continuum clockwork in a general scheme  in which  large volume,  warped geometry, and localization of zero modes in extra dimension are described  by  independent parameters.
For this, we propose a generalized 5-dimensional linear dilaton model which can realize such set-up as a solution of the model, and examine the KK spectrum and the couplings of zero modes and massive KK modes to boundary-localized operators for
 the bulk graviton, Abelian gauge bosons and periodic scalar fields.  We discuss how those KK spectra and couplings vary as a function of the volume, warping and localization parameters, and highlight the behavior in the parameter region corresponding to the clockwork limit. 
  We discuss also the field range of 4-dimensional axions originating from either 5-dimensional periodic scalar field or the 5-th component of an Abelian gauge field, and comment on the limitations of continuum clockwork compared  to the discrete clockwork.   }

\begin{document} 
\maketitle
\flushbottom

%%%%%%%%%%%%%%%%%%%%%%%%%%%%%%%%%%%%%%%%%%%%%%%%%%%%%%%%%%%%%%%%
\section{Introduction
\label{sec:introduction}}
%%%%%%%%%%%%%%%%%%%%%%%%%%%%%%%%%%%%%%%%%%%%%%%%%%%%%%%%%%%%%%%%

The clockwork (CW) is a mechanism to generate exponentially  suppressed or hierarchical couplings of light particles with $N$ massive states having comparable masses near the  threshold scale of the mechanism, where the suppression or hierarchy factor is given by ${\cal O}(q^{-N})$ for a CW parameter  $q>1$ 
 \cite{Choi:2014rja, Choi:2015fiu, Kaplan:2015fuy}.  Among the many possible implementations of the mechanism \cite{Saraswat:2016eaz, Giudice:2016yja, Higaki:2015jag, Fonseca:2016eoo, Kehagias:2016kzt, Farina:2016tgd, Ahmed:2016viu, Hambye:2016qkf, vonGersdorff:2017iym, Teresi:2017yrp, Coy:2017yex, Ben-Dayan:2017rvr, Hong:2017tel, Park:2017yrn, Lee:2017fin, Agrawal:2017eqm, Kim:2017mtc,Agrawal:2017cmd, Ibarra:2017tju,Patel:2017pct}, the CW axions \cite{Choi:2015fiu, Kaplan:2015fuy} and $U(1)$ gauge bosons \cite{Saraswat:2016eaz, Giudice:2016yja} are  particularly interesting as the key features of the mechanism can be understood in terms of a specific pattern of symmetry breaking of the underlying $N+1$  (global or local) $U(1)$ symmetries $[U(1)]^{N+1}=\prod_{i=0}^{N} U(1)_i$, which is (explicitly or spontaneously) broken down to a $U(1)_{\rm CW}$ subgroup.  
Furthermore, the key
model parameters such as  the CW parameter $q$ and the involved axion-instanton couplings and $U(1)$ gauge charges, are required to be integer-valued  (in appropriate units) by the compact $[U(1)]^{N+1}$, so the model has a built-in criterion for natural size of these model parameters.
% which would make the resulting hierarchical structure a robust prediction of the model. 
The existence of light axion or $U(1)$ gauge boson having hierarchical couplings 
can be explained by 
the unbroken $U(1)_{\rm CW}$ generated by the charge operator  \bea
Q_{\rm CW}\propto \sum_{i=0}^N \frac{Q_i}{q^i},
\eea which has a localized distribution along the linear quiver 
of the $U(1)$ charges $Q_i$. As it is a subgroup of the compact $[U(1)]^{N+1}$, the unbroken $U(1)_{\rm CW}$ is yet compact, but has an exponentially large range enhanced by $q^N$ relative to the range of $U(1)_i$, which is achieved by a series of discrete monodromy  between the nearest neighbor $U(1)$ symmetries.  This allows an exponentially enhanced axion field range or an exponential hierarchy among  the quantized charges of the unbroken $U(1)_{\rm CW}$ gauge symmetry,
which might  have an interesting implication for the fundamental issues such as the weak gravity conjecture \cite{ArkaniHamed:2006dz, Saraswat:2016eaz, Ibanez:2017vfl}.
% on the axion-instanton couplings and the $U(1)$ gauge couplings of charge %particle, as well as the possibility of
%trans-Planckian axion field range.  

Recently it has been pointed out that the CW mechanism can be implemented in an extra-dimensional setup by taking a limit $N\rightarrow \infty$, while identifying the label $i$ as the coordinate of an extra spacial dimension \cite{Giudice:2016yja}.  Then,  depending upon which aspects of the discrete clockwork (DCW) one likes to reproduce, there can be 
two different approaches to the continuum clockwork (CCW) \cite{Giudice:2016yja,Craig:2017cda,Giudice:2017fmj}\footnote{Refs. \cite{Giudice:2016yja}  and \cite{Craig:2017cda} are using different criterion about what one would call the CW mechanism, causing a certain amount of confusion. In this paper, we will use a broad notion for CW, accommodating the both criteria adopted  in \cite{Giudice:2016yja}  and \cite{Craig:2017cda}.}:  one approach  based on a specific form of 5D geometry involving an exponentially large volume together with  an exponential warp factor \cite{Giudice:2016yja}, 
which can be obtained as a solution of the 5D linear dilaton model \cite{Antoniadis:2011qw}, 
and another approach  based on exponentially localized  zero mode profile over the 5th dimension in a field basis defined in terms of 5D $U(1)$ or discrete gauge symmetry, which is generated by appropriately tuned bulk and boundary mass parameters \cite{Craig:2017cda}. As was briefly discussed in \cite{Giudice:2016yja,Craig:2017cda, Giudice:2017suc} and will be discussed in more detail in this paper, each approach has its own limitations, and as a result  can reproduce only certain partial features of the original DCW axions and $U(1)$ gauge bosons. 
 For instance, assuming a 5D discrete shift symmetry which would assure that the involved axion is a periodic scalar, and also taking into account the $U(1)$ gauge symmetry associated with the involved 5D vector field, in CCW
either the localized CW symmetry to protect zero mode is not respected by gravitational interactions, or an exponential hierarchy between the zero mode couplings at different boundaries can not be generated. 
As other limitations relative to DCW, CCW could yield  neither an exponentially enhanced trans-Planckian field range of the zero mode axion, nor an exponential hierarchy among the 4D gauge charges, while keeping the quantized nature of gauge charges of an unbroken 4D $U(1)$ gauge symmetry.

Yet the CCW is interesting by itself as it offers a setup to realize some features of the DCW mechanism, while incorporating  the well known extra-dimensional solution of the hierarchy problem \cite{Randall:1999ee, Antoniadis:2011qw}.
% e.g. the Randall-Sundrum solution based on warped geometry or the linear dilaton solution.  
Motivated by this,  in this paper we study the CCW in a general setup  involving  large volume,  warped geometry, and localized zero mode profile, which are described by  independent parameters.
 Our setup makes it possible to examine how the patterns of low energy couplings (to boundary operators)
and Kaluza-Klein (KK) mass spectra behave when the model parameters are varying from the CW limit to other limits such as the large extra dimension limit \cite{ArkaniHamed:1998rs} and the Randall-Sundrum limit \cite{Randall:1999ee},  from which we can identify the distinctive features of the CCW.  We also clarify
the limitations of each approach to the CCW to sharpen the differences from the DCW, particularly in connection with the possibility of trans-Planckian axion field range, and also the origin of  unbroken $U(1)_{\rm CW}$ which would protect the zero mode axion or $U(1)$ gauge boson.

The outline of this paper is as follows. In the next section, we first review the DCW with a specific model for clockwork axions and $U(1)$ gauge bosons \cite{Choi:2015fiu, Kaplan:2015fuy, Saraswat:2016eaz, Giudice:2016yja},  and discuss the possible continuum limits following \cite{Giudice:2016yja} and \cite{Craig:2017cda}.
We then introduce a general continuum clockwork which may realize the CCW
in more general ground  involving large volume, warped geometry, and localized zero mode profile simultaneously.
In Sec.~\ref{sec:5Dmodel}, we discuss a generalized linear dilaton model yielding  the background metric and dilaton profile for general CCW  as a solution of the model. We study also specific 5D models for CCW axions and $U(1)$ gauge bosons in Sec.~\ref{sec:5Dmodel}, and finally conclude in Sec.~\ref{sec:conclusion}.

\section{Clockwork and its continuum limits} \label{sec:DCW}

\subsection{Discrete clockwork axions and Abelian gauge bosons}

In this section, we review the key features of the CW mechanism using an example of discrete clockwork axions \cite{Choi:2015fiu,Kaplan:2015fuy} and $U(1)$ gauge bosons \cite{Saraswat:2016eaz,Giudice:2016yja}, and discuss the possible continuum limits following 
\cite{Giudice:2016yja} and \cite{Craig:2017cda}

%As noticed before \cite{},  this example provides the simplest implementation of the clockwork mechanism, possibly with %interesting implications for the fundamental issues such as the weak gravity conjecture and the possibility of trans-Planckian %field excursion.

Generic CW model involves two pieces: (i) a clockwork sector living on a quiver with $N+1$ sites equipped  with  asymmetric nearest-neighbor interactions, and (ii) an external sector which couples to the clockwork sector through a specific site in the quiver, e.g. the first and/or the last site. For clockwork axions and $U(1)$ gauge bosons, the CW sector lagrangian is given by  
\bea
\label{dcw}
{\cal L}_{\rm CW}&=&-\frac{1}{2}\sum_{i=0}^{N} f^2 \partial_\mu U_i^*\partial^\mu U_i+\frac{1}{2}\sum_{j=0}^{N-1} \frac{1}{2}\Lambda^4\left(U_{j+1}^qU_j^*+{\rm h.c}\right) \nonumber \\
&-&\sum_{i=0}^{N}
\frac{1}{4g^2}F^{i\mu\nu}F^i_{\mu\nu}
-\frac{1}{2}\sum_{j=0}^{N-1}v^2\left|(\partial_\mu -i\tilde q A^{j+1}_\mu + iA_\mu^j)\Sigma_{j+1}\right|^2,
\eea
where $U_i$ and $\Sigma_i$ are $U(1)$-valued fields:
\bea
\label{original_axion}
U_i= e^{i\phi_i/f}, \quad \Sigma_i=e^{i\omega_i}.\eea
As usual, here we assume for simplicity that the model parameters, e.g. $f, \Lambda, q,$ etc.,  have site-independent values. 
As for the external sector, we consider the Yang-Mills (YM) gauge fields $G^I_{\mu}$ and the $U(1)$-charged fermions $\psi_I$ $(I=0,N)$, which couple to the CW sector through the 0-th and $N$-th sites as follows:\footnote{One can consider more general external sector on arbitrary sites with $ 0< I < N$, whose feature of interactions with the CW sector can be readily inferred from the subsequent discussions for the $0$-th and $N$-th sites.}
\bea
\label{dcw_e}
{\cal L}_{\rm ext}=\sum_{I=0,N}\left(-\frac{1}{4g_I^2}G^I_{\mu\nu}G^{I\mu\nu}+\frac{\kappa_{I}}{32\pi^2}\frac{\phi_I}{f} G^I_{\mu\nu}\tilde G^{I\mu\nu}
+i\bar\psi_I\bar\sigma^\mu\partial_\mu\psi_I
+ Q_IA^I_\mu\bar\psi_I\bar\sigma_\mu\psi_I \right),
\eea 
where $G^I_{\mu\nu}$ and $\tilde G^I_{\mu\nu}$ are the YM gauge field strength and its dual, respectively, and we ignore the $U(1)$ gauge anomalies which are not relevant for our discussion\footnote{One can easily avoid
the $U(1)$ gauge anomalies by introducing additional fermions with opposite gauge charge, without affecting any of our subsequent discussions.}.

A key component of the  model is the clockwork gear composed of asymmetric nearest-neighbor interactions in the CW sector,  e.g. the axion potentials and St\"uckelberg gauge boson mass terms in (\ref{dcw}),  which is responsible for the following (explicit or spontaneous) symmetry breaking: \bea
\label{cwsb}
\left[U(1)\right]_{\rm global}^{N+1}\times \left[U(1)\right]_{\rm local}^{N+1}\,\, \rightarrow \,\, 
\left[U(1)_{\rm CW}\right]_{\rm global}\times \left[U(1)_{\rm CW}\right]_{\rm local},\eea
where $\left[U(1)\right]^{N+1}$ denotes the $N+1$ {\it compact} (global or local) $U(1)$ symmetries of the model: 
\bea
\label{dcw_s}
\left[U(1)\right]_{\rm global}^{N+1}:  && U_i \rightarrow  e^{i\alpha_i}U_i,\nonumber \\
\left[U(1)\right]_{\rm local}^{N+1}: && A^i_\mu \rightarrow  A^i_\mu +\partial_\mu\beta_i(x),\quad \Sigma_i\rightarrow e^{i(\tilde q\beta_i-\beta_{i-1})}\Sigma_i, \quad \psi_I\rightarrow e^{iQ_{iI}\beta_i}\psi_I,\eea
and the unbroken $U(1)_{\rm CW}$ symmetries are given by
\bea
\label{dcw_us}
\left[U(1)_{\rm CW}\right]_{\rm global}:  && U_i\rightarrow e^{i\alpha/q^{i}}U_i,\nonumber \\
 \left[U(1)_{\rm CW}\right]_{\rm local}: &&  
A^i_\mu \rightarrow  A^i_\mu + \frac{1}{\tilde q^{i}}\partial_\mu\beta(x),
\quad \Sigma_i\rightarrow \Sigma_i, \quad \psi_I\rightarrow e^{iQ_{iI}\beta/\tilde q^{i}}\psi_I.
\eea
Note that $\alpha_i$ and $\beta_i$ are periodic variables with the periodicity $2\pi$, while the periodicity of the unbroken $U(1)_{\rm CW}$ symmetries, i.e. the range of $\alpha$ and $\beta$, are enlarged to $2\pi q^N$ and $2\pi \tilde q^N$, respectively.
The axion potential responsible for the explicit symmetry breaking: $[U(1)]^{N+1}_{\rm global}\rightarrow [U(1)_{\rm CW}]_{\rm global}$  might be generated by non-perturbative dynamics as discussed in \cite{Choi:2015fiu}, or introduced simply by hand \cite{Kaplan:2015fuy}, while  the St\"uckelberg  mass terms for the spontaneous symmetry breaking: $[U(1)]^{N+1}_{\rm local}\rightarrow [U(1)_{\rm CW}]_{\rm local}$ 
can be achieved by introducing the complex scalar fields $\sigma_j$ ($j=1, ..., N$) which carry the $U(1)_{j-1}\times U(1)_j$ gauge charge $Q_{\sigma_j}=(-1, \tilde q)$, while having the vacuum expectation values $\langle \sigma_j\rangle = ve^{i\omega_j}$. 
In order for both  the effective lagrangian  (\ref{dcw}) and the associated symmetries (\ref{dcw_s}) well defined over the full range of the field variables and the symmetry group, the CW parameters $q$ and $\tilde q$, the $U(1)$ gauge charges $Q_{0}$ and $Q_N$, and the axion-instanton couplings $\kappa_0$ and $\kappa_N$ should be all integer-valued.
%\footnote{Here $\psi_I$ are canonically normalized Weyl fermions, and we choose the conventional normalization for the YM gauge %fields $G_{\mu}$, so that the instanton numbers $\frac{1}{32\pi^2}\int d^4 G\tilde G$ are integer-valued.}.
% Note that the axions $\phi_i\equiv \phi_i+2\pi f$ are periodic fields by construction. 

The symmetry breaking (\ref{cwsb}) due to the CW gear  assures that there exist a massless axion and $U(1)$ gauge boson associated with the unbroken $U(1)_{\rm CW}$, in addition to the $N$ massive  gear modes associated with the broken symmetries.
 %which are encoded in the original field variables as $\phi_i = O_{i0}\hat \phi_0+...$ and 
 %$A_\mu^i=\tilde O_{i0}\hat A^0_\mu+...$,where $O_{i0}={\cal N}_0/q^i$ and $\tilde O_{i0}=\tilde{\cal N}_0/\tilde q^i$ 
 %for ${\cal N}_0=\sum_i 1/q^i$  and $\tilde{\cal N}_0=\sum_i 1/\tilde q^i$, and the ellipses denote the massive CW gear states. 
For the particular case that all model parameters are site-independent, one can diagonalize the full $(N+1)\times (N+1)$  mass matrices to find the mass eigenstates.  This results in\footnote{Even when the model parameters are site-dependent, it is still straightforward to find the explicit form of the unbroken $U(1)_{\rm CW}$ symmetries and the associated  massless components, which are given by $Q_{\rm CW} = \sum_i O_{i0} Q_i $ with $O_{i0}\propto 1/ \prod_{k=1}^i q_k$ for site-dependent CW parameters $q_k$ ($k=1, \dots, N$). In such generic situation, one can not get an explicit form of the mass eigenvalues and the mixing matrices for massive modes. However, as long as parameters at different sites are comparable to each other, the qualitative features of the spectrum and mixings are expected to be same as the site-independent case.} 
\bea
\phi_i = \sum_{\ell =0}^N O_{i\ell} \hat\phi_\ell, \quad A^i_\mu = \sum_{\ell=0}^N \tilde O_{i\ell} \hat A^\ell_\mu,\eea
where $\hat\phi_0,\hat A^0_\mu$ denote the canonically  normalized massless modes, while $\hat\phi_n,\hat A^n_\mu$ ($n=1,...,N$) stand for the massive gear modes, and the corresponding mixing matrices are given by \cite{Giudice:2016yja, Kaplan:2015fuy}
\bea \label{mixing}
 && O_{i0}=\frac{{\cal N}_0}{q^i}, \quad O_{in} ={\cal N}_n \left(q\sin\frac{i n\pi}{N+1}-\sin\frac{(i+1)n\pi}{N+1}\right) \quad \left(n=1,...,N\right)\nonumber \\
&& \tilde O_{i0}= gO_{i0}(q\rightarrow \tilde q), \quad \tilde O_{in} = gO_{in}(q\rightarrow \tilde q),
 \eea
where
\bea \label{lambda}
{\cal N}_0=\sqrt{\frac{q^2-1}{q^2-q^{-2N}}}, \quad {\cal N}_n =\sqrt{\frac{2}{(N+1)\lambda_n}}, \quad
\lambda_n = q^2+1-2q\cos\frac{n\pi}{N+1}.\eea	
One finds also the mass eigenvalues
\bea \label{masses}
m_{\hat \phi_n}^2=\lambda_n M_{\rm CW}^2, \quad m_{\hat A_n}^2 =\tilde \lambda_n \tilde M_{\rm CW}^2  \quad (n=1,...,N) \eea
where
\bea 
  M_{\rm CW}=\frac{\Lambda^2}{f},\quad \tilde M_{\rm CW}=g v, \quad \tilde\lambda_n =\lambda_n(q\rightarrow \tilde q).\eea
%correspond to the threshold scale of the CW axions and $U(1)$ gauge bosons. 

Notice that the mixing matrix element $O_{k0}$ in (\ref{mixing}), which is the component of the massless mode in the $k$-th site of the clockwork gear,  is suppressed by the factor $q^{-k} (= e^{-k \ln q})$, 
while all the other components of the mixing matrices and $\lambda_n$ are essentially of order unity.
This exponentially localized distribution of $O_{i0}$ along the quiver of the original field variables $\phi_i$ and $A^i_\mu$ is directly responsible for generating small or hierarchical couplings of light particles in the limit when  $N$ is moderately large, e.g  $N={\cal O}(10)$. More explicitly,
one can write down the couplings to the external sector in (\ref{dcw_e}) in terms of the mass eigenstates:
\bea
 &&  -\frac{1}{2}(\partial_\mu\hat\phi_0)^2 -\frac{1}{2}\sum_n \left(\left(\partial_\mu\hat\phi_n\right)^2 + m^2_{\hat\phi_n}\hat\phi_n^2+ ...\right) \nonumber \\
 && 
-\frac{1}{4}\left(\hat F^0_{\mu\nu}\right)^2 +\sum_n \left(-\frac{1}{4}\left(\hat F^n_{\mu\nu}\right)^2 -m_{\hat A_n}^2 \left(\hat A^n_\mu\right)^2\right)
 \nonumber \\
 && +\frac{1}{32\pi^2}\frac{\hat \phi_0}{F_0}\left(\kappa_0 G^0_{\mu\nu}\tilde G^{0\mu\nu} +\frac{\kappa_N}{q^N} G^N_{\mu\nu}\tilde G^{N\mu\nu}\right)\nonumber \\
 &&-\sum_n
\frac{1}{32\pi^2}\frac{\hat \phi_n}{F_n}\left(\kappa_0 G^0_{\mu\nu}\tilde G^{0\mu\nu} +(-1)^{n}q\kappa_N G^N_{\mu\nu}\tilde G^{N\mu\nu}\right)
\nonumber \\
&& +\hat g_0\hat A^0_\mu\left( Q_0\bar\psi_0\bar\sigma_\mu\psi_0 +\frac{Q_N}{q^N}\bar\psi_N\bar\sigma_\mu\psi_N \right) \nonumber \\
&&    -\sum_n\hat  g_n \hat A^n_\mu\left(Q_0\bar\psi_0\bar\sigma_\mu\psi_0 +(-1)^n q Q_N\bar\psi_N\bar\sigma_\mu\psi_N  \right), 
\eea	
where we choose the gauge $\Sigma_i=1$, and
\bea
&& \frac{1}{F_0} = \frac{{\cal N}_0}{f}, \quad \,\, \frac{1}{F_n} = \frac{{\cal N}_n\sin \frac{n\pi}{N+1}}{f},  \quad\,\, \hat g_0 = g \tilde{\cal N}_0, \,\, \quad \hat g_n = g \tilde {\cal N}_n \sin \frac{n\pi}{N+1}.\eea
Here one can see that the massless modes $\hat{\phi}_0$ and $\hat{A}_\mu^0$ couple to the external sector in the $N$-th site with an exponentially small coupling suppressed by $q^{-N}$, while all other couplings are typically of  ${\cal O}(1/ f)$ for axions and ${\cal O}(g)$ for gauge bosons, although the couplings of massive modes  can be suppressed by the factor $1/\sqrt{N}$ for large $N$.\footnote{Even for more general external sector located at an arbitrary site $I$, it can be easily shown that  
only the massless modes have suppressed couplings to the $I$-th site by the factor $q^{-I}$.}

The above lagrangian written in terms of the mass eigenstates shows that  the couplings of massless modes to the operators at the $N$-th site are exponentially suppressed by $1/q^N$  compared to the couplings to the similar operators at the $0$-th site,
if $\kappa_0$ and $\kappa_N$ ($Q_0$ and $Q_N$) are comparable to each other, which would be a natural choice for integer-valued $\kappa_{0,N}$ and $Q_{0,N}$.
%the compact $[U(1)]_{\rm global}^{N+1}$ $\left([U(1)]_{\rm local}^{N+1}\right)$ symmetries.
There is another exponential hierarchy of ${\cal O}(1/q^N)$
between the massless mode couplings and
the massive mode couplings to the operators at the $N$-th site. Notice that this second hierarchy is independent of the relative size of $\kappa_0 \,(Q_0)$ compared to $\kappa_N \,(Q_N)$. 
We will see that this distinction is important when discussing continuum limit of the DCW.
Here the suppressed coupling of the zero mode axion to the YM gauge fields at the $N$-th site is a consequence of
the enlarged periodicity of $[U(1)_{\rm CW}]_{\rm global}$ in (\ref{dcw_us}), which results in
the enlarged field range of the zero mode axion
\dis{
\Delta \hat\phi_0\equiv 2\pi F_{\rm eff} = 2\pi q^N F_0,
}
which is exponentially bigger than the original axion field range $2\pi f$ in (\ref{original_axion}). 
Another notable feature of the DCW is  the spectrum of massive modes.
The parameter $\lambda_n$ in (\ref{lambda}) for the mass eigenvalues (\ref{masses}) satisfies
\dis{
(q-1)^2 < \lambda_n < (q+1)^2,
}
while $\lambda_n$ is increasing in $n$. Therefore, in the limit $N\gg 1$ the mass eigenvalues $m_n$ in (\ref{masses}) become approximately degenerate around the clockwork threshold scale $M_{\rm CW}$, and their mass gaps behave as
$\delta m \sim M_{\rm CW}/N$, which might have interesting phenomenological implications \cite{Giudice:2017fmj}. 

Before moving to the discussion of continuum clockwork, we summarize the key features of the discrete CW axions and $U(1)$ gauge bosons. 
\begin{enumerate}
\item The key model parameters such as the CW parameter $q$, the axion-instanton couplings $\kappa_I$ and the $U(1)$ gauge charges $Q_I$ have integer values as required  by the underlying compact $[U(1)]^{N+1}$ symmetry. Therefore the model is equipped with an unambiguous criterion for natural size of the relevant UV parameters, providing a basis for the subsequent discussion of the hierarchical structure of the low energy effective couplings of zero mode axion  and zero mode $U(1)$ gauge boson.
\item Zero mode axion and $U(1)$ gauge boson are protected by the unbroken compact $U(1)_{\rm CW}$ generated by a charge operator $Q_{\rm CW}\propto \sum_i Q_i/q^i$ which has a localized distribution along the linear quiver of the charge operators $Q_i$ of $[U(1)]^{N+1}=\prod_{i=0}^N U(1)_i$. The localized feature of the unbroken symmetry
leads to a multiplicative monodromy structure, enhancing the range of $U(1)_{\rm CW}$
by $\prod_{j=1}^N q = q^N$.  This results in also an exponential hierarchy between the couplings of zero modes at the $0$-th and the $N$-th sites, as well as an exponentially enhanced field range of the zero mode axion. 
\item Exponential hierarchy ($\propto q^{-N}$) between the zero mode couplings and the massive mode couplings to the external sector at the  $N$-th site.
\item Approximately  degenerate $N$ massive modes around the threshold scale, with $m_n\sim M_{\rm CW}$ and the small mass splittings
$\delta m\sim m_{n+1}-m_n \sim m_n/N$ in the limit $N\gg 1$. 
\end{enumerate}

\subsection{Continuum limit}

Starting from a discrete clockwork model with $N+1$ sites, its continuum version can be obtained by taking the limit $N\rightarrow \infty$, while identifying the site index $i$ as the coordinate of the 5-th spacial dimension with a finite length $\pi R = N\Delta r$, where $\Delta r$ is the lattice spacing.   In order to keep  the hierarchy factors $q^N, \tilde q^N$ finite, while having a nonzero mass gap between the zero modes and the massive gear states,  one should take also   $q, \tilde q \rightarrow 1$ with the CW mass scales $M_{\rm CW}= \Lambda^2/f$ and $\tilde M_{\rm CW}=gv$ approaching to the lattice cutoff scale $1/\Delta r$. More explicitly, the continuum limit takes
\bea
N\rightarrow \infty, \quad   q, \tilde q \rightarrow 1, \quad  M_{\rm CW}, \tilde M_{\rm CW} \rightarrow \frac{1}{\Delta r} =\infty,\eea
with the following parameter combinations keeping a finite nonzero value:
\bea
\pi R = N\Delta r ,\quad \mu=M_{\rm CW}\ln q, \quad \tilde \mu =\tilde M_{\rm CW}\ln \tilde q,\eea
%where 
%$M_0$ and $M_1$ are the CW threshold scales for axions and $U(1)$ gauge bosons, respectively:
%\bea
%M_0=\frac{\Lambda^2}{f},\quad M_1=gv.\eea
which results in 
\bea
\label{CW_correspondence}
q^N\rightarrow e^{\mu\pi R},\quad \tilde q^N\rightarrow e^{\tilde\mu\pi R}.\eea

To get the continuum limit of the lagrangian densities (\ref{dcw}) and (\ref{dcw_e}), one can make the following substitutions: 
\bea\sum_i \,\rightarrow\, \frac{1}{\Delta r}\int_0^{\pi R} dy, \quad\phi_{i+1}-\phi_i\,\rightarrow\,  \Delta r\, \partial_y \phi, \eea
together with the field and parameter redefinitions:\bea
 && M_{\rm CW}=\frac{\Lambda^2}{f} \,  \rightarrow\, \frac{1}{\Delta r}, \quad \tilde{M}_{\rm CW}=gv \,\rightarrow\, \frac{1}{\Delta r}, \quad q-1\,\rightarrow\, \mu\, \Delta r, \quad \tilde q-1 \,\rightarrow\, \tilde\mu \,\Delta r,  \nonumber \\
&& \hskip 1cm \phi_i(x)\,\rightarrow\, \Phi(x,y) \,\Delta r^{1/2},  \quad A^i_\mu(x)
\,\rightarrow\, A_\mu(x,y), \quad \omega_i(x) \,\rightarrow\, \Omega (x,y)\,  \Delta r, \nonumber \\
&& \hskip 3cm f \,\rightarrow\, {f_5^{3/2}}\, {\Delta r}^{1/2}  ,\quad  g^2 \,\rightarrow\, \frac{g_5^2}{\Delta r},\eea
and finally take the limit $\Delta r \rightarrow 0$.
One then finds 
\bea
\label{ccw}
&&\int dy \left[-\frac{1}{2} \left(\partial_\mu \Phi\right)^2 -\frac{1}{2}\left(\partial_y \Phi +\mu\Phi\right)^2  -\frac{1}{4g_5^2}\left( (F_{\mu\nu})^2 +(\partial_\mu \Omega-\partial_y A_\mu -\tilde\mu A_\mu)^2\right)\right.\nonumber \\
&& \quad \qquad +\,\frac{1}{32\pi^2}\frac{\Phi}{f_5^{3/2}}\left(\kappa_0\,\delta(y)\,G^0_{\mu\nu}\tilde G^{0\mu\nu} +\kappa_\pi \,\delta(y-\pi R)\, G^\pi_{\mu\nu}\tilde G^{\pi\mu\nu}\right)\nonumber \\
&&\left.\quad \qquad  +\,A_\mu\left(Q_0 \,\delta(y)\,  \bar\psi_0\bar\sigma^\mu \psi_0 +Q_\pi \,\delta(y-\pi R)\, \bar\psi_\pi\bar\sigma^\mu \psi_\pi\right)\right].
\eea
where we changed the index for the last site from $N$ to $\pi$, e.g. $\kappa_N \rightarrow \kappa_\pi$.
% assumed that the external sector exists only on the boundaries (the 1st and last sites in DCW model). 
%As for the coup0linDepending upon which issue we are interested in, we can set some of the boundary couplings $\kappa_{0,\pi}$ and %$Q_{0,\pi}$ to be zero.

Taking the continuum limit of $[U(1)]^{N+1}$ in (\ref{dcw_s}) and $U(1)_{\rm CW}$ in (\ref{dcw_us}), we also find 
the associated symmetry transformations are given by 
\dis{
\label{ccw_s}
& [U(1)_{\rm 5D}]_{\rm global}:\,\,\delta \Phi = f_5^{3/2}\alpha(y), \\
&[U(1)_{\rm 5D}]_{\rm local}:\,\,\delta A_\mu=\partial_\mu \beta(x,y), \,\, \delta \Omega = \partial_y \beta + \tilde\mu \beta, \,\, \delta\psi_{0,\pi}=iQ_{0,\pi}\beta(x, y=0, \pi R)\,\psi_{0,\pi}\,,
}
which are explicitly or spontaneously broken down to
\dis{
\label{ccw_us}
 &[U(1)_{\rm CW}]_{\rm global}:   \delta \Phi = f_5^{3/2}  e^{-\mu y}\, \alpha_0,  \\
 & [U(1)_{\rm CW}]_{\rm local}:
 \delta A_\mu =e^{-\tilde\mu y}\partial_\mu\beta_0(x) ,\,\,  \delta \Omega=0,\,\,
   \delta\psi_0=iQ_0\beta_0(x)\, \psi_0,\,\, \delta\psi_\pi=ie^{-\tilde\mu \pi R}\,Q_\pi\beta_0(x)\, \psi_\pi\,,
   }
where $\alpha_0$ is a constant, and $\beta_0(x)$ is a function of the 4D spacetime coordinate $x^\mu$.

Here we see an important difference between the original DCW and its continuum limit. In the DCW, one starts with compact $[U(1)]^{N+1}$ symmetries which are  broken down to $U(1)_{\rm CW}$, and both the full  $U(1)$ symmetries  and the unbroken $U(1)_{\rm CW}$ are perfectly compatible with the 4D diffeomorphism and Lorentz symmetries. 
On the other hand, in the continuum clockwork (CCW) limit, the correspondence  (\ref{CW_correspondence}) implies that the CW parameters $q,\tilde q$ can not be integer-valued anymore, except for the trivial case of $\mu=\tilde\mu =0$.
This raises a question if the $U(1)$ symmetries (\ref{ccw_s}) and (\ref{ccw_us}) in the CCW 
%the corresponding $U(1)_{\rm CW}$ symmetries are defined as a localised %symmetry along the spatial coordinate $y$, which raises a question whether 
can be identified as a sensible compact symmetry compatible with the 5D spacetime symmetries. 
To answer this question, one should embed the continuum lagrangian (\ref{ccw}) into a theory manifestly invariant under the 5D 
diffeomorphism and Lorentz symmetry.
 As we will see, there are two different  embeddings of the continuum lagrangian (\ref{ccw}) to a 5D
diffeomorphism and Lorentz invariant theory, and for each embedding one can retain only different partial features of the DCW.
Note that some issues in clockwork models, e.g. the periodicity (or field range) of axions and the natural size (or quantized nature) of the axion-instanton  couplings $\kappa_0, \kappa_\pi$ and the $U(1)$ gauge charges $Q_0, Q_\pi$, crucially depend on
how the compact nature of the associated $U(1)$ symmetries are introduced in the underlying theory.

We remark that the above subtle feature of CCW partly originates from the fact that we replace the discrete field index $i$ in the DCW model with the spatial coordinate $y$ of extra spacial dimension
in the continuum limit. While the index $i$ is a frozen label in the DCW, the spatial coordinate $y$ describes a dynamical extra dimension
which imposes non-trivial restrictions on the clockwork lagrangian via the 5D spacetime symmetries.
In the following, we will discuss two different prescriptions of CCW originally proposed in \cite{Giudice:2016yja,Craig:2017cda}, and subsequently introduce general continuum
clockwork  which can incorporate these two prescriptions within a common framework.

\subsubsection{CCW-I : CW from localized zero mode profile}
% induced by bulk and boundary masses}

To embed the continuum lagrangain (\ref{ccw}) in a 5D diffeomorphism and Lorentz invariant theory,  we
first notice following \cite{Craig:2017cda} that the lagrangian can be rewritten as follows by %%doing appropriate integration by parts and choosing the gauge $\Omega =0$.
\dis{
\label{ccw-i}
& \int dy \left[-\frac{1}{2}\left(\left(\partial_\mu\Phi\right)^2 + \left(\partial_y\Phi\right)^2 +\left(\mu^2-\mu [ \delta(y)-\delta(y-\pi R) ] \right)\Phi^2\right)\right.
  \\
  &~~\qquad -\frac{1}{4g_5^2}\left( \left(F_{\mu\nu}\right)^2 +\left(\partial_y A_\mu\right)^2 +\left(\tilde\mu^2-\tilde\mu [ \delta(y)-\delta(y-\pi R) ] \right) 
  A_\mu^2
  \right) \\
& ~~~~~\qquad +\frac{\Phi}{f_5^{3/2}}\left(\kappa_0\,\delta(y)\,G_0\tilde G_0 +\kappa_\pi\,\delta(y-\pi R)\,G_\pi\tilde G_\pi\right)
\\
&~~~~~\qquad  + A_\mu\left(Q_0 \,\delta(y)\,  \bar\psi_0\bar\sigma^\mu \psi_0 +Q_\pi \,\delta(y-\pi R)\, \bar\psi_\pi\bar\sigma^\mu \psi_\pi\right)\Big],
}
where the continuum clockwork parameters $\mu, \tilde \mu$ appear as boundary and bulk masses.
The above lagrangian can be easily made to be invariant under the 5D spacetime symmetries by introducing appropriate metric dependence:
\dis{ \label{ccw-1}
\int d^5x \sqrt{-G} &\left[-\frac{1}{2} G^{MN}\partial_M \Phi\partial_N \Phi
-\frac{1}{2}\left(\mu^2-\mu\left[\frac{\delta(y)}{\sqrt{G_{55}}}-\frac{\delta(y-\pi R))}{\sqrt{G_{55}}}\right]\right)\Phi^2
 \right. \\
  & \hskip -2cm -\frac{1}{4g_5^2}G^{MN}G^{PQ}F_{MP}F_{NQ} -\frac{\tilde\mu^2}{4g_5^2} G^{MN}A_M A_N-\frac{\tilde\mu}{4g_5^2} \left(\frac{\delta(y)}{\sqrt{G_{55}}}-\frac{\delta(y-\pi R))}{\sqrt{G_{55}}}\right)G^{\mu\nu}A_\mu A_\nu
 \\
 &\hskip -1cm +\frac{\Phi}{f_5^{3/2}}\left(\kappa_0\, \frac{\delta(y)}{\sqrt{G_{55}}} \, G_0\tilde G_0 +\kappa_\pi \, \frac{\delta(y-\pi R)}{\sqrt{G_{55}}} \, G_\pi\tilde G_\pi\right)
\\
&\hskip -1cm + A_\mu\left(Q_0\, \frac{\delta(y)}{\sqrt{G_{55}}} \, \bar\psi_0\bar\sigma^\mu \psi_0 +Q_\pi \, \frac{\delta(y-\pi R)}{\sqrt{G_{55}}} \, \bar\psi_\pi\bar\sigma^\mu \psi_\pi\right)\Big],
}
%where we have introduced additional term $-(\tilde{\mu}^2/4g_5^2) A_5^2$ to %the original lagrangian (\ref{ccw-i})  in order to have 
%the 5D gauge multiplet $A_M$, 
%but {\color{red} the additional field $A_5$ can be integrated out to be zero %with the constraint $\partial^\mu F_{\mu 5} = 0$}.
where the continuum lagrangain (\ref{ccw-i}) can be obtained when the spacetime metric $G_{MN}$ is replaced by the flat background:
\bea
\label{flat_metric}
ds^2 =\langle G_{MN} \rangle dx^M dx^N =  \eta_{MN}dx^Mdx^N.
\eea

However, once one includes the metric dependence,  certain feature of the model is lost.
For the bulk and boundary masses which are tuned to be equal as in (\ref{ccw-1}),  the CW symmetry (\ref{ccw_us}) is respected when the metric $G_{MN}$ is simply replaced by its 
background value $\eta_{MN}$, however {\it not}  by the interactions of the metric fluctuation $h_{MN}=G_{MN} -\eta_{MN}$.
%It can be also shown that the symmetry cannot be respected by arbitrary %background. 
For instance, upon ignoring the off-diagonal part of $G_{MN}$,  the 5D action of $\Phi$ can be rewritten as
\dis{ \label{ccw-1_axion}
\int d^5x \sqrt{-G} \left[-\frac{1}{2} G^{\mu \nu}\partial_\mu \Phi\partial_\nu \Phi -\frac{1}{2} \left( \frac{1}{\sqrt{G_{55}}} \partial_y \Phi + \mu \Phi \right)^2
+\frac{1}{2} \frac{1}{\sqrt{-G}} \partial_y \left( \frac{\sqrt{-G}}{\sqrt{G_{55}}} \right) \mu \Phi^2 \right], 
}
where the last term shows that the CW symmetry (\ref{ccw_us}) is not respected by the $y$-dependent fluctuation of $G_{\mu\nu}$.
%\bea
%G_{55} &=& 1, \label{cws_1} \\
% \frac{\partial_y \sqrt{-G}}{\sqrt{-G}} &=& \textrm{constant}. \label{cws_2}
% \eea
%The condition (\ref{cws_2}) ensures the third term in (\ref{ccw-1_axion}) to %be a bulk mass term which can be cancelled by introducing another bulk mass.
%Therefore, the symmetry can be respected only by particular background %geometries satisfying the above conditions such as the flat background or RS %background.\footnote{Later we will show that the conditions can be alleviated %by introducing a dilaton field when discussing general CCW so that more %general background is compatible with the CW symmetry.} 
This means that generically the continuum CW symmetry (\ref{ccw_us}) cannot be identified as a good symmetry compatible with 5D spacetime symmetries. 
Rather, it should be regarded as an approximate accidental symmetry which holds  for a particular metric background, but is broken by higher dimensional interactions of the metric fluctuation. 
As a consequence, the particular relations among the bulk and boundary masses, which  are crucial for the existence of massless modes protected by the CW symmetry (\ref{ccw_us}), are potentially unstable against radiative corrections involving the couplings of the metric fluctuation \footnote{It might be still possible that those relations are protected by other symmetry such as supersymmetry \cite{Gherghetta:2000qt}.}.
Nevertheless, if we accept it, while ignoring the fine tuning issue on the bulk and boundary masses, 
the 5D model (\ref{ccw-1}) can reproduce many, although not all, features  of the DCW model as we will show in detail in Sec. \ref{sec:5Dmodel}.

As noticed before, one needs an appropriate periodicity condition for $\Phi$  to address the field range of zero mode axion  and the natural size (or quantized nature) of the boundary axion-instanton couplings $\kappa_0$, $\kappa_\pi$. For this, one can simply impose the condition 
\bea \label{5d_period}
\Phi\equiv \Phi+ 2\pi f_5^{3/2},\eea
 while making the following replacement in the 5D action (\ref{ccw-1}):
\dis{\label{replacement}
 \mu \Phi \,\rightarrow \,  \mu f_5^{3/2} \,\sin\left(\frac{\Phi}{f_5^{3/2}}\right).
} 
%so that it reproduces the original $\mu \Phi$ by leading order expansion %around $\Phi =0$ while having the symmetry $\Phi\rightarrow \Phi+ 2\pi f_5^{3/%2}$.
Then the infinitesimal CW symmetry (\ref{ccw_us}) is modified as follows
\dis{ \label{non-linear_sym}
 [U(1)_{\rm CW}]_{\rm global}\,: ~  \delta \Phi = f_5^{3/2} \alpha_0\, e^{-\mu \int dy \cos(\Phi/f_5^{3/2})},
 }
%While this symmetry is not respected by the interactions of the metric %fluctuation 4D kinetic term $G^{\mu \nu} \partial_\mu \Phi \partial_\nu \Phi$. 
%However,
and the zero mode of the fluctuation $\delta \Phi$ around $\Phi=0$ is protected
to be light by this symmetry as far as we are not concerned with the symmetry breaking by the gravitational couplings of metric fluctuation.

The 5D diffeomorphism invariant theory (\ref{ccw-1}) with
the replacement  (\ref{replacement}) reproduces the continuum lagrangian (\ref{ccw-i}) for a small field fluctuation $\delta\Phi$ around $\Phi =0$ when the 5D metric is fixed as the flat background (\ref{flat_metric}). 
 It is then expected that certain perturbative features of this 5D theory, e.g. the spectrum and couplings of small field fluctuations,  are similar to those of the original DCW model. 
For instance, the bulk and boundary mass terms of $\delta\Phi(x,y)$ and $\delta A_\mu(x,y)$ enforce their zero modes to have $y$-dependent profile exponentially localized at one boundary, e.g. at $y=0$, which results in an exponential hierarchy between the zero mode couplings at $y=0$ and their counterpart couplings  at $y=\pi R$,   as well as a similar hierarchy between the zero mode coupling and the massive mode coupling to the same operator at $y=\pi R$.
Although the 5D theory  (\ref{ccw-1}) reproduces certain perturbative features of the original DCW model,
 large field behavior of the theory such as the field range of the zero mode axion  can be quite different from the DCW model. 
We will discuss these issues in Sec.~\ref{sec:5Dmodel} in terms of the general CCW to be defined later. 
For a moment, we note the key weak point of the 5D diffeomorphism invariant CCW-I  theory (\ref{ccw-1}) which realizes certain features of the DCW model through localized zero mode  profile: {\it  the model has a problem with the CW symmetry (\ref{ccw_us}) which 
is not respected by gravitational interactions}.

In fact, our notion of localized zero mode is not basis-independent, and the localized feature of the zero mode of $\delta\Phi$ disappears for instance if one makes the $y$-dependent field redefinition: $\Phi(x,y)\rightarrow e^{-\mu y} \Phi(x,y)$.
  To avoid an ambiguity arising from such field redefinition,
one needs to specify the field basis used to address the localization of zero mode in the 5th dimension. 
%\footnote{
%If we consider the probability measure $dP(y)$ discussed in \cite{Giudice:2016yja}, 
%the probability density $dP/dy$ is invariant under the naive field redefinition. However, the correct probability density %should be $dP/dl_5$, where $dl_5 = dy \sqrt{ G_{55}}$ is the measure of the extra dimensional length. This yields different %densities for the continuum clockworks related by field redefinition.}. 
Throughout this paper, we will consider the localization in a specific field basis defined in terms of 
the discrete shift symmetry which would ensure the periodicity of $\Phi$ and the compact 5D $U(1)$ gauge symmetry for $A_M$,
i.e. the localization of the zero-mode wavefunctions
 $\delta\Phi(x,y)/\delta \phi_0(x)$ and $\delta A_\mu(x,y)/\delta A_{0\mu}(x)$, where the 5D fields $\Phi(x,y)$ and $A_M(x,y)$ transform under those gauge symmetries as  
 \bea\label{5d_gauge_symmetry}
 \Phi\rightarrow \Phi +2\pi f_5^{3/2}, \quad A_M \rightarrow A_M+\partial_M\Lambda,
 \eea
 where $f_5$ is an $y$-independent constant and the $U(1)$ gauge transformation function $\Lambda(x,y)$ obeys the $y$-independent periodicity condition as $\Lambda\equiv \Lambda+2\pi$. Note that in this standard field basis for 5D gauge field, the two boundary gauge charges $Q_0$ and $Q_\pi$ are quantized in a {\it common} unit as $\Lambda(y=0)$ and $\Lambda(y=\pi R)$ have the common periodicity. The same is true for the axion-instanton couplings $\kappa_0$ and $\kappa_\pi$ as $\Phi(y=0)$ and $\Phi(y=\pi R)$ has the common periodicity. One can then start with  $Q_0\sim Q_\pi={\cal O}(1)$ and $\kappa_0\sim \kappa_\pi={\cal O}(1)$, which are perfectly natural in view of the underlying symmetries,  and examine the exponential hierarchies between the effective couplings of  zero modes at different boundaries.

%{\color{red}  
%It is also noted that different embeddings compatible with five dimensional symmetries remove the degeneracy  between the %continuum clockworks related by {\it naive} field redefinition 
%$(\hat\Phi = e^{ A(y)} \Phi)$, and provide different physical interpretations.}

\subsubsection{CCW-II :  CW from geometry}

As we noticed in the previous subsection, in an approach to implement continuum
CW via localized zero mode profile, 
 the CW symmetry  (\ref{ccw_us})  is not respected by the couplings of metric fluctuation. %which are  a good symmetry of the corresponding 5D diffeomorphism and 
%Lorentz invariant theory (\ref{ccw-1}). 
This problem arises  from non-vanishing bulk and boundary mass parameters which are identified as the origin of the CCW parameters $\mu, \tilde{\mu}$.
In fact, one can easily avoid this problem 
by embedding (\ref{ccw}) into a 5D theory without any bulk or boundary mass term \cite{Giudice:2016yja}.
The continuum lagrangian (\ref{ccw}) can be written in the following 
form by redefining the fields and boundary couplings:
\dis{ \label{ccw-ii}
& \int dy \left[-\frac{1}{2} e^{-2\mu y}\left(\left(\partial_\mu\hat\Phi\right)^2  +\left(\partial_y \hat\Phi\right)^2\right)  -\frac{1}{4g_5^2}e^{-2\tilde\mu y}\left( \left(\hat F_{\mu\nu}\right)^2 +\left(\partial_\mu \hat A_5-\partial_y \hat A_\mu\right)^2\right)\right. \\
&~~~~ \qquad +\frac{\hat\Phi}{f_5^{3/2}} \left( \hat\kappa_0\,\delta(y)\,G_0\tilde G_0 +\hat\kappa_\pi \,\delta(y-\pi R)\,G_\pi\tilde G_\pi\right)\\
&~~~~\left.\qquad  + \hat A_\mu \,  \left( \hat Q_0\, \delta(y) \, \bar\psi_0\bar\sigma^\mu \psi_0 + \hat Q_\pi\,\delta(y-\pi R)\, \bar\psi_\pi\bar\sigma^\mu \psi_\pi\right)\right], 
} 
where the redefined fields and couplings are related to the original fields and couplings as follows.
\dis{ \label{f_redef}
\hat \Phi=e^{\mu y}\,\Phi,\quad  \hat A_\mu=e^{\tilde\mu y}A_\mu,\quad  \hat A_5=e^{\tilde\mu y}\,\Omega
}    
\dis{ \label{c_redef}
\hat \kappa_0 = \kappa_0, \quad \hat \kappa_\pi = \kappa_\pi\, e^{-\mu \pi R}, \quad \hat Q_0 = Q_0, \quad \hat Q_\pi = Q_\pi\, e^{-\mu \pi R}
}
In this prescription,  the CCW parameters
$\mu, \tilde{\mu}$ appear in the wave function coefficients, while the bulk and boundary mass parameters are all vanishing.
For the case with $\mu=\tilde\mu$,  the above form of lagrangian can be embedded into a 5D diffeomorphism and Lorentz invariant linear dilaton model as discussed in \cite{Giudice:2016yja, Giudice:2017suc}:
\dis{ \label{ccw-2}
\int d^5x \sqrt{-G} &\left[-\frac{1}{2} G^{MN}\partial_M\hat\Phi\partial_N\hat\Phi
 -\frac{1}{4g_5^2}\,e^{2S/3}\,G^{MN}G^{PQ}\hat F_{MP}\hat F_{NQ}\right. \\
 &\quad +\frac{\hat\Phi}{f_5^{3/2}}  \left(\hat\kappa_0\, \frac{\delta(y)}{\sqrt{G_{55}}} \, G_0\tilde G_0 +\hat\kappa_\pi \, \frac{\delta(y-\pi R)}{\sqrt{G_{55}}} \, G_\pi\tilde G_\pi\right)
\\
&\quad  \left.+ \hat A_\mu \left(\hat Q_0\, \frac{\delta(y)}{\sqrt{G_{55}}} \, \bar\psi_0\bar\sigma^\mu \psi_0 +\hat Q_\pi \, \frac{\delta(y-\pi R)}{\sqrt{G_{55}}} \, \bar\psi_\pi\bar\sigma^\mu \psi_\pi\right)\right],
}
where the continuum lagrangian (\ref{ccw-ii}) is reproduced
when the metric and dilaton field are replaced 
by the following solution of the linear dilaton model\footnote{We can have $\mu \neq \tilde \mu$ if the bulk dilaton couplings are different from the couplings 
of the original linear dilaton model. This can be also realized in the general CCW as we will discuss later.}:
\dis{ \label{curved_metric}
ds^2= \langle G_{MN} \rangle dx^Mdx^N = e^{-\frac{4}{3} \mu y}\left( dx^2+ dy^2 \right),\quad \langle e^S \rangle = e^{-2 \mu y}.
}

The CW symmetry (\ref{ccw_us}) can be also expressed in terms of the redefined fields $\hat \Phi$ and $\hat A_M$ as 
\dis{
 &[U(1)_{\rm CW}]_{\rm global}:   \delta \hat \Phi = f_5^{3/2}  \alpha_0,  \\
 & [U(1)_{\rm CW}]_{\rm local}:
 \delta \hat A_\mu =\partial_\mu\beta_0(x) ,\,\,  \delta \hat A_5=0,\,\,
   \delta\psi_0=i\hat Q_0\beta_0(x)\, \psi_0,\,\, \delta\psi_\pi=i\hat Q_\pi\beta_0(x)\, \psi_\pi\,.
 }
Now $[U(1)_{\rm CW}]_{\rm global}$ can be identified as a global shift symmetry for the redefined  scalar field $\hat \Phi$, $[U(1)_{\rm CW}]_{\rm local}$ as the 4D subgroup of a 5D  $U(1)$  gauge symmetry for the redefined gauge field $\hat A_M$, and both CW symmetries  are obviously  
compatible with the 5D spacetime symmetries, e.g. respected by the couplings of the spacetime metric and dilaton field. 
Moreover, under the assumption that the redefined boundary couplings $\hat \kappa_{0,\pi}$ and $\hat Q_{0,\pi}$ are integer-valued,
one can impose the following periodicity condition without any difficulty:
\bea 
\label{5d_periodicity}
\hat \Phi\equiv \hat \Phi+ 2\pi f_5^{3/2}, \quad \hat\Lambda(x,y) \equiv \hat\Lambda+2\pi,\eea  
where $\hat\Lambda$ is the transformation function of the underlying 5D 
$U(1)$ gauge symmetry, under which \bea
\hat A_M\rightarrow \hat A_M+\partial_M\hat\Lambda.\eea  
Thus it seems that this construction apparently solves all problems of the CCW-I.

However this construction also has its own limitation. Notice that the redefined couplings $\hat \kappa_0,\hat \kappa_\pi$ and $\hat Q_0,\hat Q_\pi$ reveal an exponential hierarchy in (\ref{c_redef})
if the original couplings $\kappa_0, \kappa_\pi$ and $Q_0, Q_\pi$ were of order unity in order to reproduce the continuum version (\ref{ccw}) of the original DCW model. 
On the other hand, in the 5D theory (\ref{ccw-2}) of $\hat\Phi$ and $\hat A_M$ obeying the periodicity condition (\ref{5d_periodicity}), the redefined couplings $\hat \kappa_{0,\pi}$  and $\hat Q_{0,\pi}$, {\it not} the original couplings $\kappa_{0,\pi}$  and $Q_{0,\pi}$, are required to be integer-valued, so have  natural size of order unity.
% by the 5D discrete gauge symmetry of the axion field $\hat \Phi$ and the
%5D {\it compact} $U(1)$ gauge symmetry of the abelian gauge field $\hat A_M$.
Furthermore, the zero modes of $\hat\Phi$ and $\hat A_M$ in (\ref{ccw-2}) have flat profile over the 5th dimension because of the absence of bulk and boundary masses\footnote{
Note that we are considering the localized zero mode profile in the field basis defined by the 5D gauge symmetries (\ref{5d_gauge_symmetry}), and  the corresponding field basis for CCW-II is provided by the redefined fields $\hat\Phi$ and $\hat A_M$, not by the original fields $\Phi$ and $A_M$.}.
As a consequence, the {\it CCW-II  cannot reproduce the exponential hierarchy among the zero mode couplings at different boundaries}, which is one of the primary features of the original DCW model. 
One may wonder whether the situation can be changed by introducing dilaton-dependent couplings ($\propto e^{-\gamma S}$) to the boundary operators in order to realize the desired exponential hierarchy among the zero mode couplings at different boundaries. 
However, again once one imposes the periodicity condition (\ref{5d_periodicity}), such dilaton-dependent couplings are forbidden by the discrete shift symmetry of $\hat \Phi$ and the 5D gauge symmetry of $\hat A_M$.  
Nevertheless, as was stressed in \cite{Giudice:2017suc},
many of the other features of the DCW can be successfully reproduced in the CCW-II realized through the background geometry of the linear dilaton model. 

The main reason for the above limitation of the CCW-II is attributed to the absence of localization of  zero mode of 5D bosonic fields without bulk and boundary masses. 
There is still an example that localized zero mode profile occurs by warped geometry alone: the 5th component of a 5D gauge field $C_M$ with odd orbifold-parity
whose zero mode can be interpreted as a 4D axion field \cite{Choi:2003wr, Flacke:2006ad}.
In this case, there is a corresponding localized CW symmetry protecting the zero mode axion as we will discuss in Sec.~\ref{sec:5Dmodel}. 
However,  as we will see,  the 5D gauge symmetry of $C_M$ forbids certain boundary couplings of $C_5$, so still prohibits the full realization 
of the characteristic features of the DCW such as the exponentially enlarged axion field range and/or the exponential hierarchy among the zero mode axion couplings to instantons at different boundaries. 

To summarize, both CCW-I and CCW-II have their own  limitations and realize only certain partial features of the original DCW model. For appropriately chosen 5D metric backgrounds, i.e. the flat background (\ref{flat_metric}) for CCW-I and the linear dilaton  background (\ref{curved_metric}) for CCW-II, two models are related to each other by 
the field and parameter redefinitions (\ref{f_redef}) and (\ref{c_redef}). This suggests that two schemes share certain features, for instance the KK spectrum which are independent of the boundary couplings in our approximation. Yet, CCW-I and CCW-II are totally different models. In CCW-I, the CW symmetry which is supposed to protect the zero mode is not respected by the gravitational interactions of the metric fluctuation, which would raise a question about the naturalness of the scheme. In CCW-II, zero modes have flat profile over the 5th dimension in the field basis for which the 5D $U(1)$ and discrete gauge symmetries take the standard form as \ref{5d_gauge_symmetry}, so there is no hierarchy between the zero mode couplings at different boundaries. Related to this point, in CCW-I the {\it $y$-independent} periodicities of the 5D axion field and the 5D $U(1)$ gauge symmetry transformation (\ref{5d_gauge_symmetry}) are imposed on $\Phi$ and $A_M$, and as a consequence the corresponding boundary couplings $\kappa_{0,\pi}$ and $Q_{0,\pi}$ are quantized in the common unit, suggesting that they all have integer values of order unity. On the other hand, in CCW-II, the same applies for the redefined fields $\hat \Phi$ and $\hat A_M$ and boundary couplings $\hat\kappa_{0,\pi}$ and $\hat Q_{0,\pi}$, {\it not} for the original fields and couplings. This is the reason why the two models predict  different hierarchical patterns for the effective couplings of zero modes at different boundaries. In the following, we will introduce {\it general CCW} in which  CCW-I and CCW-II are incorporated into common framework which can realize the continuum CW
in a most general way.

\subsubsection{General CCW}

Inspired by the previous discussions, let us define the general continuum clockwork (general CCW) as an extra-dimensional setup yielding light mode by means of a localized symmetry. In this perspective, we can start with the following most general quadratic lagrangian which might be obtained after replacing the background geometry and dilaton with their vacuum expectation values which are supposed to have an exponential $y$-dependence:
\bea
&&-\frac{1}{2} \int_{0}^{\pi R} dy\left[
e^{-2\mu_1 y}(\partial_\mu\Phi)^2 + e^{-2\mu_2 y}(\partial_y \Phi)^2 + e^{-2\mu_3 y}m_B^2\Phi^2\right. \nonumber \\
&& \left. \qquad \qquad   + e^{-2\mu_4 y}\left(m_0\delta(y)+m_\pi\delta(y-\pi R)\right)\Phi^2\right].\eea
Requiring to be invariant under the following {\it localized infinitesimal} shift symmetry  \bea
\label{localized_sy}
 \delta \Phi = c_0 e^{-\mu y},\eea
where $c_0$ is an infinitesimal constant, we find the model parameters should obey the 
following relations  
\bea
\mu_2=\mu_3=\mu_4, \quad m_B^2=\mu(\mu+2\mu_2), \quad -m_0=m_\pi=\mu,\eea
for which the lagrangian can be written in a form which is manifestly invariant under (\ref{localized_sy}):
\bea
\label{gccw_axion}
-\frac{1}{2}\int dy \left[ e^{-2\mu_1 y}(\partial_\mu\Phi)^2 + e^{-2\mu_2 y}(\partial_y \Phi +\mu\Phi)^2\right].
\eea
Similarly one can consider a general CCW theory for $U(1)$ gauge bosons invariant under a localized infinitesimal $U(1)$ gauge symmetry:
\bea
\label{localized_sy_1}
\delta A_\mu = e^{-\tilde\mu y}\partial_\mu\beta(x),\eea
which result in
\bea
\label{ggcw_photon}
\int dy \left[-\frac{1}{4g_5^2}\left\{e^{-\tilde\mu_1 y}(F_{\mu\nu})^2 + e^{-2\tilde \mu_2 y}(\partial_\mu\omega -\partial_y A_\mu-\tilde\mu A_\mu)^2\right\} + ...  \right]
\eea 
Yet we have a freedom to redefine the fields as
\bea
\label{redef}
\Phi\rightarrow e^{\mu_0 y}\Phi,\quad (A_\mu, \omega)\rightarrow e^{\tilde\mu_0y}(A_\mu, \omega),\eea
while keeping the lagrangian and localized symmetry to take the same form, but with the  redefined fields and parameters. 
This makes the physical interpretation (or the physical origin) of each of the CCW parameters, e.g. $\mu_1, \mu_2, \mu$ for the CCW axion, obscure as they are not invariant under the field redefinition (\ref{redef}).  

On the other hand, we are interested in the possibility that the CCW lagrangians (\ref{gccw_axion}) and
(\ref{ggcw_photon}) originate from a 5D theory in which $\Phi$ can be identified as a periodic 5D scalar field associated with a discrete gauge shift symmetry of the form:
\bea \label{disc_sym}
\Phi\rightarrow \Phi+ 2\pi f_5^{3/2},\eea
 and the 5D vector field $A_M$ is introduced  as the gauge field of compact ($Z_2$ even or odd) $U(1)$ gauge symmetry, under which
 \bea \label{comp_sym}
 A_M \rightarrow A_M+\partial_M\Lambda \quad \left(\Lambda(x,y)\equiv \Lambda(x,y)+2\pi\right).\eea
 Once $\Phi$ and $A_M$  are identified as such field variables in the underlying UV theory, we do not have anymore a freedom to make the field redefinition (\ref{redef}), and can attempt to embed (\ref{gccw_axion}) and
(\ref{ggcw_photon}) into a 5D diffeomorphism and Lorentz invariant theory
defined in the field basis fixed by the above form of 5D discrete shift and $U(1)$ gauge symmetries. As we will see, in such prescription, one can make an unambiguous physical distinction between $\mu_1,\mu_2$ and $\mu$. 
%It turns out that  
%$\mu_1$ and $\mu_2$ describe the effects of background geometry, while $\mu$ is associated with the localization of zero mode induced by the bulk and boundary masses, which satisfy certain specific relations to be invariant under the localized CW symmetries (\ref{localized_sy}) and (\ref{localized_sy_1}).  
Note that the 5D discrete shift and compact $U(1)$ gauge symmetries, which appear to be inevitable in any sensible UV completion of CCW, ensure the quantization of the $U(1)$ gauge charges 
and the axion-instanton couplings in the underlying UV theory, and therefore provide also a criterion for  natural size of relevant UV parameters, which would be essential for a proper interpretation of the exponential hierarchies generated by the CW mechanism.  

As will be explained in the next section in detail,
the most natural way to implement the general CCW within the framework of 5D diffeomorphism and Lorentz invariant theory is the {\it generalized linear dilaton model}. For instance,  the general CCW axion lagrangian (\ref{gccw_axion}) at quadratic order
can be obtained from the following 5D action
\dis{
-\int d^5x \sqrt{-G} \left[ \frac{1}{2} G^{MN}\partial_M\Phi\partial_N\Phi \right. &+ \frac{1}{2} e^{-2cS}   \,f_5^3\, V_B\left(\frac{\Phi}{f_5^{3/2}} \right)  \\
&\left.- e^{-cS}  \, f_5^3  \left(\frac{\delta(y)}{\sqrt{G_{55}}}-\frac{\delta(y-\pi R)}{\sqrt{G_{55}}}
\right) V_b\left(\frac{\Phi}{f_5^{3/2}} \right) \right], 
}
%\dis{
%-\int d^5x \sqrt{-G} \left[ \frac{1}{2} G^{MN}\partial_M\Phi\partial_N\Phi \right. &+ \frac{1}{2} e^{-2cS} m_B^2 \,f_5^3\, V_B\left(\frac{\Phi}{f_5^{3/2}} \right)  \\
%&\left.- e^{-cS} m_b\, f_5^3  \left(\frac{\delta(y)}{\sqrt{G_{55}}}-\frac{\delta(y-\pi R)}{\sqrt{G_{55}}}
%\right) V_b\left(\frac{\Phi}{f_5^{3/2}} \right) \right], 
%}
where $V_B(\theta)$ and $V_b(\theta)$ are even functions of the periodic angle $\theta \equiv \theta + 2\pi$, which are  manifestly invariant under the discrete gauge symmetry (\ref{disc_sym}),  while the background geometry and dilaton turn out to have the following expectation values:
\bea
\label{metric_expec} 
ds^2=\langle G_{MN}\rangle dx^Mdx^N = e^{2k_1 y} \eta_{\mu\nu}dx^\mu dx^\nu + e^{2k_2 y} dy^2, \quad \langle e^{cS}\rangle =e^{k_2 y}.\eea
Then the CCW parameters $\mu_1, \mu_2$ are determined by the geometric parameters $k_1$ and $k_2$ describing the exponential warp factor and the exponentially  large volume, respectively,
\bea
\mu_1=-\left(k_1+\frac{k_2}{2}\right), \quad \mu_2 = -\left(2k_1 -\frac{k_2}{2}\right), \eea
while the CCW parameter $\mu$ for localized symmetry originates from the bulk and boundary masses for  $V_B(\theta)\simeq V_{B0}+ m_B^2 \theta^2$, 
$V_b(\theta) \simeq V_{b0}+ m_b \theta^2$ in the vicinity of $\Phi =0$, which are tuned to satisfy the following relations to have the localized symmetry (\ref{localized_sy}) 
\footnote{For the arbitrary value of $\Phi$, 
the condition becomes $V_B = \frac{1}{4} (V'_b)^2 + (4k_1 - k_2) V_b + {\rm Const.}$.
The localized symmetry must be generalized as (\ref{non-linear_sym}) for arbitrary value of $\Phi$ to have the discrete gauge symmetry (\ref{disc_sym}). }:
\bea
\mu=m_b, \quad m_B^2=m_b\left(m_b+4k_1- k_2\right).\eea
Therefore, the generalized linear dilaton model implements the general CCW with three independent parameters $k_1, k_2$ and $\mu$. We can also see that CCW-I corresponds to $\mu \neq 0$ and $k_1 = k_2=0$, while 
CCW-II corresponds to $\mu = 0$ and $k_1 = k_2 \neq 0$. Thus the general CCW incorporates the two schemes into a single framework and can exhibit continuous deformation among all possible implementations of the CCW.

As we will show in the next section, the three independent parameters are responsible for the following three characteristic features of the CW mechanism: i) the bulk and boundary mass parameter $\mu$ determines the hierarchy between the CW zero mode couplings at different boundaries,
ii) a certain combination of   $\mu$ and the warp factor parameter $k_1$ (for a given $k_2$) determines the hierarchy between the  zero mode coupling and the massive KK mode couplings at the same boundary, and iii) the difference between the warp factor and the volume factor, i.e.  $k_1-k_2$,  controls the mass gap between the massive gear modes (KK modes).

As noticed in the previous discussion, for non-zero $\mu, \tilde \mu$
%an important difference between CCW-I and CCW-II was shown to be due to whether the bulk and boundary mass parameter $\mu$ vanishes or not. Here the same problem occurs 
 the localized symmetries (\ref{localized_sy}) and (\ref{localized_sy_1}) to protect zero modes are not respected by the couplings of  metric and dilaton fluctuations, while an exponential hierarchy between the zero mode couplings at different boundaries is  possible only with non-zero $\mu, \tilde \mu$. 
On the other hand, if $\mu, \tilde \mu$ vanish, the zero modes are protected  by an unbroken symmetry respected by the metric and dilaton fluctuations, while they lose the localized feature for generating an exponential hierarchy among the couplings at different boundaries.   

Before closing this section, let us briefly summarize the main features of the continuum CW. 
\begin{enumerate}
\item General CCW is defined as a generic 5D theory producing light 4D modes protected by a symmetry localized in extra dimension. It can be considered as a generalization of the naive continuum limit of the DCW models, which can accommodate both the CCW from geometry and the CCW associated with localized zero mode profile  over the extra dimension, which is achieved by fine-tuned bulk and boundary masses.   However, there are always certain limitations of CCW, so even general CCW can not produce all of the interesting features of DCW.  For instance, general CCW could yield  neither an exponentially enhanced trans-Planckian field range of the zero mode axion, nor an exponential hierarchy among the 4D gauge charges, while keeping the quantized nature of gauge charges of an unbroken 4D $U(1)$ gauge symmetry. 
\item Sensible UV theory for axions and Abelian gauge bosons should be based on 
a discrete gauge shift symmetry for the periodicity of axions and/or the compact $U(1)$ gauge symmetries for gauge bosons, which are compatible with the 5D diffeomorphism and Lorentz symmetry. Specifically, one may start with a 5D theory invariant under
\bea
\label{5dsymmetry_basis}
&T_\Phi:&\Phi\rightarrow \Phi+2\pi f_5^{3/2},\nonumber
 \\ &U(1)_{5D}:& A_M\rightarrow A_M + \partial_M\Lambda \quad \mbox{for}\quad \Lambda(x,y) \equiv \Lambda(x,y)+2\pi. \eea
Then, in the above field basis defined in terms of $T_\Phi$ and $U(1)_{5D}$,  the physical origin and meaning of the general CCW  model parameters, including the couplings to instantons and charged particles at the boundaries, are unambiguously identified. In this prescription,  there are two model parameters $\mu_1,\mu_2$ describing the effects of background geometry and dilaton field profile, and one parameter $\mu$ (or $\tilde \mu$) describing the localization of zero mode axion (or gauge boson)  by fine-tuned bulk and boundary masses.

\item Only when $\mu=\tilde\mu=0$, the infinitesimal localized CW symmetries to protect zero modes can be embedded into a sensible 5D symmetry compatible with the 5D spacetime symmetries.

\item For $\mu\neq 0$ (or $\tilde\mu\neq 0$), the localized infinitesimal CW symmetry for zero mode axion (or gauge boson) corresponds to an accidental approximate symmetry not respected by the metric and dilaton fluctuations over the background vacuum values.
\end{enumerate}

\section{5D models for continuum clockwork} \label{sec:5Dmodel}

In this section, we provide a concrete framework to realize  general CCW discussed in the previous section. 
We first introduce the {\it generalized linear dilaton model} which is obtained by generalizing the dilaton couplings of the original linear dilaton model
\cite{Antoniadis:2011qw}. This model will provide the background geometry and dilaton profile necessary to implement
 general CCW.
 % described by three independent parameters $\mu_1, \mu_2$ and $\mu$.
We then analyze the mass spectrum and couplings of the KK graviton modes in general CCW.
It turns out that many of the results on KK gravitons apply also to the KK modes of CW axions and $U(1)$ gauge bosons,  reproducing many of the characteristic features of the DCW model. 
 In the subsequent subsections, we examine more detailed features of the CCW axions originating from a 5D angular field and/or a 5D gauge field with odd orbifold-parity, and also the CCW photon originating from a 5D gauge field with even orbifold-parity.

\subsection{Generalized linear dilaton model} \label{subsec:LDM}

We start from a five-dimensional (5D) model with the 5th dimension compactified on 
orbifold 
$S_1/{Z}_2$ with the fixed points at $y=\{0, \pi R\}$.
%boundary localized interactions and boundary localized fields can be introduced. 
All 5D fields in the model obey the boundary conditions 
%in a domain $y=[0, \pi R]$  is related with that in other regime as
\bea
\Phi_{\pm}(y+2\pi R)=\Phi_{\pm}(y), \quad \Phi_{\pm}(-y)=\pm \Phi_{\pm}(y),
\eea
where $\pm$ denotes the orbifold parity. 
%Bulk 5D fields in the model can be clasif also obey the  are periodic, $\Phi(x, y
%+2\pi R) = \Phi(x, y)$. 
%By orbifolding, the fields at $y$ are matched with those evaluated at $-y$. We have %two possibilities as
% $\Phi(y)= \pm \Phi(-y)$.
%In order to have zero mode solutions, the corresponding fields should be even under %the parity transformation, 
%$\Phi(x,y)=\Phi(x, |y|)$. If 
%$\Phi(x, y)$ is odd, there is no zero mode and only KK modes exist. 
%When we consider a geometry in 5D dimension, the light scalar fields can play the %important role. In our set-up,
The most general form of 5D metric invariant under the 4D Poincare symmetry is given by\footnote{The corresponding solution of the inflationary 4D spacetime for the linear dilaton model was studied in \cite{Kehagias:2016kzt, Im:2017eju}. This might be straightforwardly extended to our generalized model. }
\bea
\label{general_clockwork}
ds^2 = \langle G_{MN} \rangle dx^Mdx^N= \Omega_1(y)\eta_{\mu\nu}dx^\mu dx^\nu + \Omega_2(y) dy^2,
\eea
where $\Omega_1$ corresponds to the warp factor of the 4D metric component, while $\Omega_2=G_{55}$ defines the physical length of the 5th dimension: $L_5= \int_0^{\pi R} dy\sqrt{\Omega_2}$. Throughout this paper, we consider the case that underlying 5D gravity dynamics generates an exponential $y$-dependence in $\Omega_{1,2}$:
\bea
\Omega_1(y)=e^{2k_1y}, \quad \Omega_2(y)=e^{2k_2 y},\eea
where   $k_1$ and $k_2$ are generically independent mass parameters with the same sign.
As we will see, the above form of 5D metric can be obtained as a solution of linear dilaton model which is appropriately generalized from the well known model of Ref. \cite{Antoniadis:2011qw}.

Motivated by the UV origin of the linear dilaton model as a dual description of the little string theory \cite{Antoniadis:2011qw} (see also \cite{Kehagias:2017grx, Antoniadis:2017wyh}), 
we consider the following 5D model with a universal dilaton coupling, which might be interpreted as the 5D action in the string frame:
\dis{ \label{GLD_string}
S=  \int d^4 x \int_{-\pi R}^{\pi R} dy \sqrt{-G}\, e^{\xi S}  \Big(
 \frac{M_5^3}{2}\, R_5 \,+\,&  \frac{M_5^3}{2}  \partial_M S \partial^M S -  \Lambda_5 \\
 &- \frac{1}{\sqrt{G_{55}}}\left[\Lambda_0  \,\delta(y) 
+\Lambda_\pi  \,\delta(y-\pi R) \right] + \cdots  \Big),
}
where $M_5$ is the 5D cut-off scale,  $\Lambda_{5, 0, \pi}$ are the bulk and boundary cosmological constants, and the normalization of the dimensionless dilaton $S$ is fixed by the coefficient of the kinetic term.
Notice that we introduce a general dilaton coupling $\xi$, which was taken to be $1$ in the original model \cite{Antoniadis:2011qw}.
Here we assume that the underlying theory involves a small (string) coupling $g_{\rm st}^2 \propto e^{-\xi S}$, which would assure  that the above 5D action defined at the leading order in $g_{\rm st}^2$ provides a sufficiently good approximation. 
As we go to the Einstein frame with \bea
g_{MN} \rightarrow e^{-2\xi S/3} g_{MN},\eea and subsequently rescale the dilaton field to arrive at the canonically normalized kinetic term, the action reads
\dis{\label{GLD_einstein}
S=  \int d^5 x \sqrt{-G}\, \Big(
 \frac{M_5^3}{2} \, R_5 \,-\,& \frac{M_5^3}{2}\,   \partial_M S \partial^M S - e^{-2 c S} \Lambda_5 \\
 &- \frac{e^{-c S}}{\sqrt{G_{55}}}\left[\Lambda_0  \,\delta(y) 
+\Lambda_\pi  \,\delta(y-\pi R) \right] + \cdots  \Big),
}
where
\dis{
c \equiv \frac{\xi}{\sqrt{12 \,\xi^2-9}}\,.
}
We will see that the value of $c$ different from $1/\sqrt{3}\, (\xi=1)$ accounts for the background metric and dilaton profile necessary to realize general CCW.

Since the cosmological constants $\Lambda_{5, 0, \pi}$ break the dilatonic shift symmetry \bea
\label{dilaton_shift}
 S \rightarrow S+ \alpha,\eea one can take them to be soft symmetry breaking parameters whose scale is well below the 5D cut-off scale
 $M_5$. Let us parameterize them as
\bea
\label{GLD_parametrization}
\Lambda_5 = -2M_5^3 k^2, \quad
\Lambda_0 =  -4M_5^3 k_0, \quad
\Lambda_\pi =  4M_5^3 k_\pi,
\eea
with $k, k_{0, \pi}  < M_5$. Then the leading order action of the dilaton-graviton system is written as
\dis{ \label{GLD}
S_{\rm gravity}=  \int d^5 x \sqrt{-G}\, M_5^3 \left(
\, \frac{R_5}{2} \,-\, \, \frac{1}{2}  \partial_M S \partial^M S +2 e^{-2 c S} k^2 
 +4 \frac{e^{-c S}}{\sqrt{G_{55}}}\left[ k_0  \,\delta(y) 
- k_\pi  \,\delta(y-\pi R) \right]  \right).
}
Now we can solve the equations of motion  with the following ansatz\footnote{
By rescaling the soft parameters, $k\to e^{c S|_{y=0}}k$, we can set $S|_{y=0}=0$.
It is assumed that the additional boundary dynamics 
determines the radius, $R$,  of the 5th dimension.} 
\beq
\label{GLD_sol}
ds^2= e^{2 k_1|y|}(\eta_{\mu\nu} dx^\mu dx^\nu) 
+ e^{2 k_2 |y|} dy^2, \quad 
S = k_3|y|.
\eeq
We then find a solution with $k_i\, (i=1, 2 ,3)$  given by
\bea \label{k's}
&&k_1 = \frac{2 k}{\sqrt{12- 9\,c^2}}, \quad  k_2 = c k_3 = 3 \,c^2 k_1,
\eea
while the two boundary parameters $k_{0,\pi}$ are fixed by the boundary conditions as 
\bea
k_0=k_\pi = \frac{3}{2} k_1.
\eea
%So the solution is determined by two free parameters $k$, $c$. Alternatively %we can take $k_1$, $k_2$ as independent parameters instead of $k$, $c$.  
Note that $k$ is real ($\textrm{AdS}_5$ bulk space) for $c^2 < 4/3$, while it is imaginary ($\textrm{dS}_5$ bulk space) for $c^2 > 4/3$.

From the above solution, we can obtain the two well-known limits: 
\bea
&&c=0: \quad {\rm RS\ geometry \ with}\   \,\, k_2=0,\nonumber\\
&&c=\frac{1}{\sqrt{3}}:\quad {\rm Linear\ Dilaton\ geometry\ with}\ \,\, k_1=k_2.
\eea
The large extra dimension (LED) limit  can be obtained also by taking $c\gg 1$,  which yields $k_2\pi R\gg 1\gg  k_1\pi R$.  
%\blue{However, $k_2$ is bounded above by $4k_1$ for real $k$ ($\textrm{AdS}_5$ space) as mentioned above.
%Yet there is an enough room to get $e^{2k_2\pi R}/e^{2k_1\pi R}  = {\cal O}(100)$ for $e^{2k_1\pi R} = {\cal O}(1)$, 
%so that the geometry with $k_2 > k_1$ (or $ c > 1/\sqrt{3}$) can reproduce certain features of LED as we will show later.}

We are basically interested in the parameter space where
the background geometry provides a solution to the weak scale hierarchy problem. 
This is addressed with an appropriate size of the extra dimension depending on the values of $k_1, k_2$.  
The 4D Planck mass is evaluated as 
\dis{ \label{mpl0}
\Mpl^2 &= M_5^3 \int_{-\pi R}^{\pi R} dy \, e^{(2k_1+k_2)|y|} =
 \frac{M_5^3}{k_1+k_2/2} \left(e^{2(k_1+k_2/2) \pi R} -1 \right),
} 
while the physical size of the extra dimension is given by
\dis{ \label{L5}
L_5 = \int_{0}^{\pi R} dy \, e^{k_2 |y|} =\frac{1}{k_2} \left(e^{k_2 \pi R} -1\right).
}
From (\ref{k's}), one can see that $k_1$ and $k_2$ must have the same sign for real $c$, and their sign is determined 
by the sign of $k$. If we take them to be positive, the boundary at $y=0$ is identified as the IR boundary, while the boundary at $y=\pi R$ becomes the UV boundary. For negative $k_{1,2}$, the boundaries are flipped without changing physics.
In the following, 
we will take the convention of $k_{1, 2} \geq 0$.  
Then, combining (\ref{mpl0}) and (\ref{L5}), we can express the proper length $L_5$ of the 5-th dimension in terms of $\Mpl, M_5, k_1$ and $k_2$:
\dis{
L_5 &= \frac{1}{k_2} \left[\left( \frac{(k_1+\frac{k_2}{2}) \Mpl^2}{M_5^3} +1 \right)^{\frac{k_2}{2k_1+k_2}} -1 \right] \\
&\approx \frac{1}{k_2} \left[ \left(\frac{\Mpl}{M_5}\right)^{\frac{2(k_2/k_1)}{2+(k_2/k_1)}}-1\right] \approx \left\{ \begin{array}{l}
\frac{1}{k_1} \ln \frac{\Mpl}{M_5} ~~~{\rm as}~~~k_2/k_1\rightarrow 0 ~({\rm RS}) \\
\frac{1}{k_2}\frac{\Mpl^2}{M_5^2} ~~~{\rm as}~~~k_2/k_1 \rightarrow \infty ~({\rm LED})
\end{array}\right.,
}
where $0< k_2/k_1 < 4$ for real $k$. 
Imposing the experimental upper limit on $L_5 \lesssim 100 \,\mu$m \cite{Patrignani:2016xqp},
we find 
\dis{
k_2 \lesssim 2 k_1 \times \frac{1- b}{1+b},
}
where
\dis{
b \equiv \frac{\log_{10}(M_5/k_2)-2\log_{10}(M_5/{\rm TeV})}{15-\log_{10}(M_5/{\rm TeV})}.
}
Therefore, if one wishes to address the hierarchy problem with the lowest possible $M_5 \sim$ TeV,  $k_2$ cannot be larger than about $2 k_1$.

\subsection{KK spectrum and couplings}

Let us now discuss the spectrum and couplings of the bulk graviton KK modes in the generalized linear dilaton background. 
It turns out that much of the results apply also to the CW axions and $U(1)$ gauge bosons\footnote{The spectrum and couplings for KK modes of the dilaton and also radion  field
depend on the additional boundary dynamics that stabilize both fields. 
Such a model dependence is studied in \cite{Cox:2012ee} for the linear dilaton model.} .

To quadratic order of the 4D metric fluctuation   $ h_{\mu \nu} \, (\eta_{\mu \nu} \rightarrow \eta_{\mu \nu} + 2h_{\mu \nu})$ in the traceless transverse gauge $h_\mu^\mu = \partial^\mu h_{\mu\nu} =0$, the action (\ref{GLD}) gives rise to
\bea \label{graviton_cw1}
S_{\rm gravity} = - M_5^3 \int d^5 x\, 
e^{(2k_1 + k_2) |y|} \left[\frac{1}{2} 
(\partial_\rho h_{\mu\nu})^2  +\frac{1}{2} e^{2(k_1-k_2)|y|}(\partial_yh_{\mu\nu})^2   \right].
\eea
Notice that the (bulk or boundary) graviton mass term is absent due to the 4D diffeomorphism invariance even in the presence of non-trivial quadratic order graviton couplings to the dilaton background in $S_{\rm gravity}$. 
As we will show later, a generic quadratic action for  5D CW bosonic field $\Phi$ ($=$ graviton, axion or $U(1)$ gauge boson) takes the form:
\dis{ \label{CW_bosons}
S_\Phi = - \int d^5 x \, e^{ 2 k_{ \Phi} |y|} \left[ \frac{1}{2} (\partial_\mu \Phi)^2 
+ \frac{1}{2} e^{2(k_1- k_2)|y|} (\partial_y \Phi  - \mu_\Phi \,\epsilon(y) \Phi )^2\right],
}
where $k_{\Phi}$ is some combination of $k_1$ and $k_2$ describing the effects of background metric and dilaton profile, and $\mu_\Phi$ originates from the bulk and boundary masses. For graviton, 
$k_\Phi = k_1 + k_2/2$ and $\mu_{\Phi}=0$ as can be seen from (\ref{graviton_cw1}).
Notice that this corresponds to the defining lagrangian (\ref{gccw_axion}) of general CCW whose parameters are identified as \bea
\mu_1 = -k_\Phi,\quad \mu_2 = -(k_\Phi + k_1 - k_2),\quad  \mu = - \mu_\Phi.\eea 

\begin{figure}[t]
\begin{center}
\includegraphics[width=0.5\textwidth]{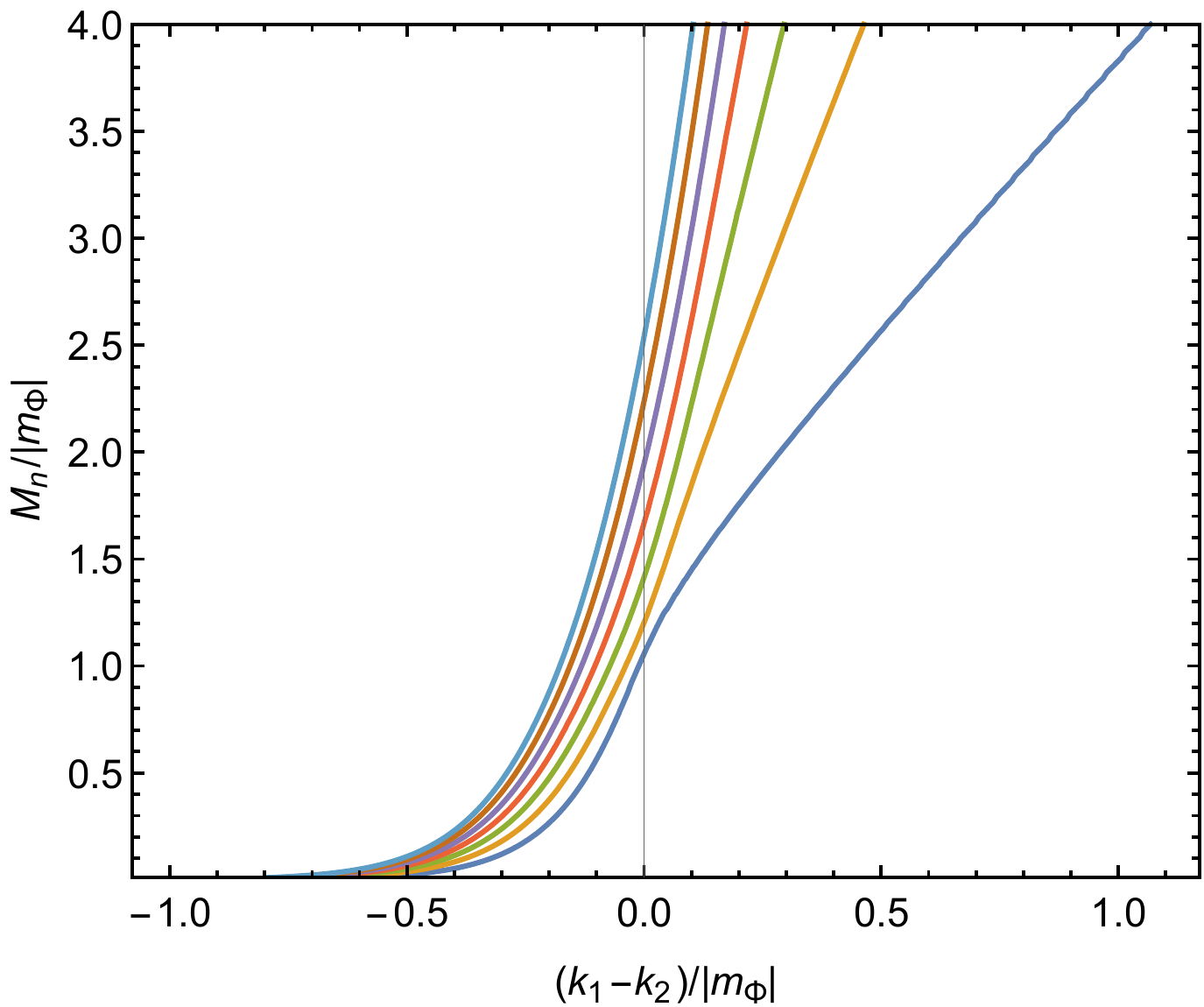}~
\includegraphics[width=0.5\textwidth]{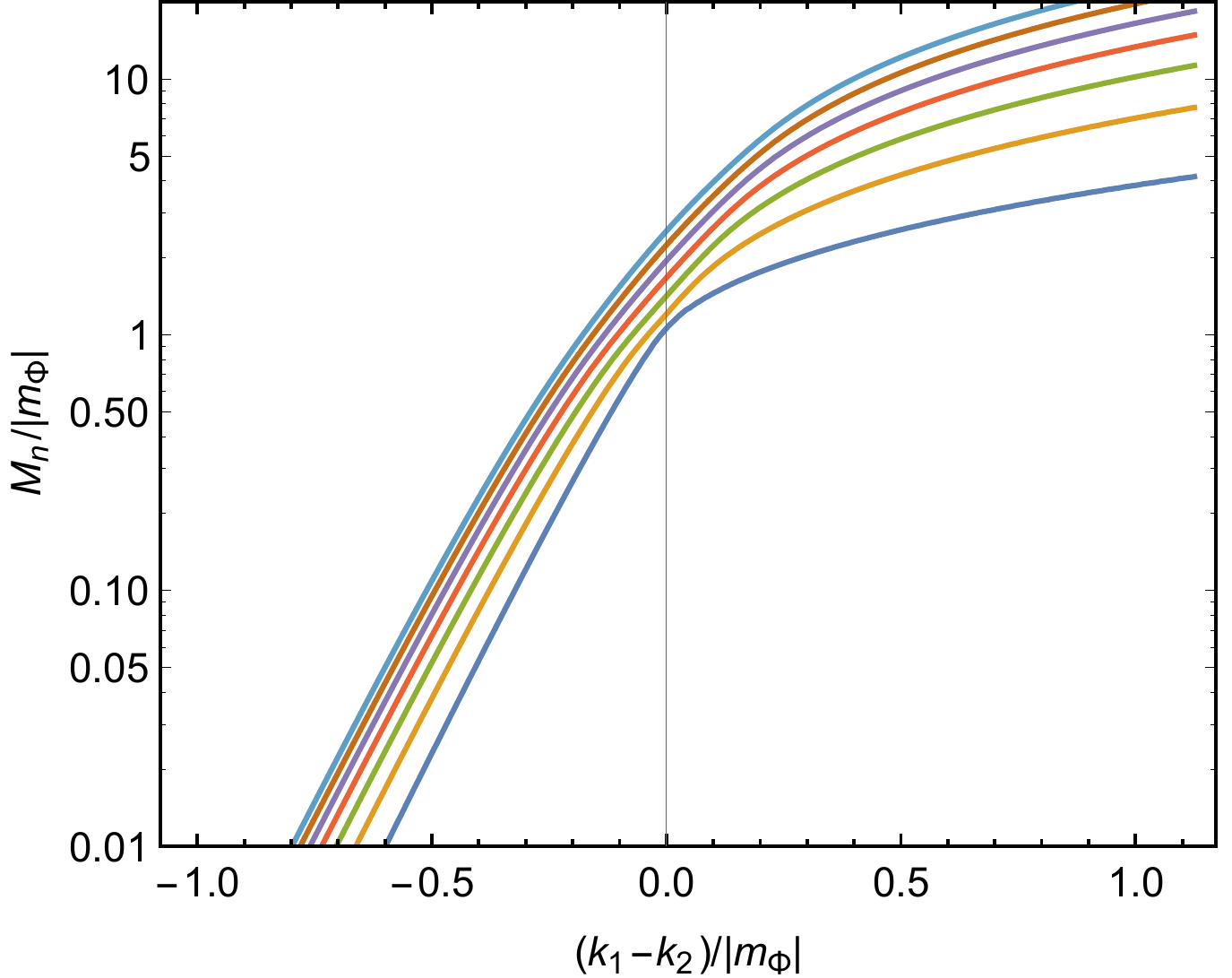}
\caption{ 
KK masses ($n=1,2,\cdots, 7$) as a function of $\kappa/|m_\Phi|\equiv (k_1-k_2)/|m_\Phi|$  for fixed values of $m_\Phi$ and $R$ (left: linear, right :log).
For graviton $(|m_\Phi| = k_1 + k_2/2)$, the RS limit corresponds to $(k_1 - k_2)/|m_\Phi|= 1$, while the LED limit is $(k_1- k_2)/|m_\Phi| = -2$. 
%In the plots, $m_\Phi$ and $R$ are set by the condition (\ref{mpl0}) with $k_1 + k_2/2 = m_\Phi$ while taking $m_\Phi R =3$ for clear representation of the spectrum. 
}
\label{KKmass}
\end{center}
\end{figure}

The field basis used in (\ref{CW_bosons}) corresponds to the one for which the 5D discrete shift or continuous $U(1)$
gauge symmetries are defined as (\ref{5dsymmetry_basis}), and therefore both the natural size of the corresponding couplings and the origin of  the mass parameters $ k_\Phi,\,  \mu_\Phi, \, k_1-k_2$  are unambiguously determined. 
To perform the KK analysis, however, it is convenient to use the other field basis with the field redefinition
\dis{ \label{phc}
\Phi_c = e^{k_\Phi |y|} \Phi,
}
for which the action (\ref{CW_bosons}) reads
\bea \label{CW_bosons_c}
S_\Phi = - \int d^5 x \,   \left[ \frac{1}{2} (\partial_\mu \Phi_c)^2 
+ \frac{1}{2} e^{2\kappa |y|} (\partial_y \Phi_c  - m_\Phi \, \epsilon(y) \Phi_c )^2\right], 
\eea
where
\bea
m_\Phi = k_{\Phi} + \mu_\Phi, \quad 
\kappa = k_1 - k_2.
\eea
%Then the KK mass spectrum is to be determined by $\kappa, \, m_\Phi$ and $R$ parametrizing the size of the extra dimension.
The equation of motion for $\Phi_c$ is then given by
\dis{
\left(\partial_y^2  + 2\kappa\, \epsilon(y) \partial_y  + e^{-2\kappa |y|}  \eta^{\mu \nu} \partial_\mu \partial_\nu - M_\Phi^2\right) \Phi_c(x, y) 
= 2 m_\Phi \left[\delta(y) - \delta(y-\pi R)\right] \Phi_c (x, y),
}
where
\dis{ \label{bulk_m}
M_\Phi^2 \equiv m_\Phi^2 + 2\kappa \,m_\Phi.
}
%Notice that we have the above relation between the bulk mass $M_\Phi$ and boundary mass $m_\Phi$ due to the structure in 
%(\ref{CW_bosons_c}).
Decomposing $\Phi_c$ as
\beq
\Phi_c(x, y) = \sum_{n=0}^\infty f_{n}(y)\, \phi_n(x),
\eeq  
the equation for the mode functions $f_n$ are given by
\dis{ \label{fn_eq}
\left(\partial_y^2  + 2\kappa\, \epsilon(y) \partial_y  + e^{-2\kappa |y|}  M_n^2 - M_\Phi^2\right) f_n(y)
= 2 m_\Phi \left[\delta(y) - \delta(y-\pi R)\right] f_n (y),
}
where $M_n$ denotes the 4D mass eigenvalues of the corresponding 4D field $\phi^{(n)}$:
\dis{
\eta^{\mu \nu} \partial_\mu \partial_\nu \phi_n(x) = M_n^2 \, \phi_n(x).
}
%The equation (\ref{fn_eq}) can be converted to more familiar form by writing $f_n(y)$ as
%\dis{ \label{fn_rescale}
%f_n(y) = e^{\kappa |y|} \chi_n(y),
%}
%which yields
%\dis{ \label{chin_eq}
%\left(\partial_y^2  + 4\kappa\, \epsilon(y) \partial_y  + e^{-2\kappa |y|}  M_n^2 - \tilde{M}_\Phi^2\right) \chi_n(y)
%= 2 \tilde{m}_\Phi \left[\delta(y) - \delta(y-\pi R)\right] \chi_n (y),
%}
%where
%\bea
%\tilde{M}_\Phi^2 &\equiv& M_\Phi^2 - 3\kappa^2 \\
%\tilde{m}_\Phi &\equiv& m_\Phi - \kappa
%\eea
%Then one can find that (\ref{chin_eq}) actually corresponds to the KK mode equation of the 5D scalar field with the bulk mass 
%$\tilde{M}_\Phi$ and boundary mass $\tilde{m}_\Phi$
%in the RS background of the warp factor $\kappa$, whose solution is given in \cite{Gherghetta:2000qt}. Therefore, our KK mass spectrum %$M_n$ must be same as the corresponding one, 
%and the wave function profile $f_n(y)$ is simply obtained by the rescaling relation (\ref{fn_rescale}).
The solution is then found to be\footnote{For $\kappa=0$, taking the limit $\kappa \rightarrow 0$ in (\ref{f_n}) is rather tricky, and
the corresponding explicit expression is given by
\dis{
f_n(y)|_{\kappa=0} = \frac{1}{N_n} \left[ \frac{n}{R} \cos \left(\frac{n}{R} |y| \right) + m_\Phi \sin \left( \frac{n}{R} |y| \right) \right].  \label{f_n_0}
}}
\bea
f_0(y) &=& \frac{1}{N_0} e^{m_\Phi |y|},\label{f_0} \nonumber \\
f_n(y) &=& \frac{1}{N_n}  e^{-\kappa |y|} \left[ J_{|\alpha|} \left(\frac{M_n}{|\kappa|} e^{-\kappa|y|}\right) + b_\alpha (M_n) Y_{|\alpha|}\left(\frac{M_n}{|\kappa|} e^{-\kappa|y|}\right)  \right] ~ (n=1,2,\dots), \label{f_n} \nonumber \\
\eea
where $N_n$ is the normalization factor, $J_{|\alpha|} (x)$ and $Y_{|\alpha|} (x)$ are the first and second kind Bessel functions, respectively, and 
\bea 
\alpha &=&  1+ \frac{m_\Phi}{\kappa}  \label{alpha}, \\
b_\alpha (M_n) &=& -\frac{J_{\textrm{sgn}(\alpha)( \alpha - 1)}\left(\frac{M_n}{|\kappa|}\right)}{Y_{\textrm{sgn}(\alpha)( \alpha - 1)}\left(\frac{M_n}{|\kappa|}\right)}. \label{ba}
\eea
The zero mode $\phi_0$ is massless as expected, and the mass eigenvalues $M_n$ of the massive modes $\phi_n \, (n=1,2,3, \dots)$ can be determined by the following boundary condition:
\dis{
b_\alpha(M_n e^{-\kappa \pi R})= b_\alpha(M_n),
}
which results in
\bea \label{KKm}
M_n \simeq \left\{
\begin{array}{ll} 
\left(n -\frac{1}{4} + \left| \frac{m_\Phi}{2\kappa} \right| \right) \pi \kappa, & \kappa \gtrsim |m_\Phi| \\
\sqrt{m_\Phi^2+ \frac{n^2}{R^2} },  &  \kappa\simeq0 \\
\left(n -\frac{1}{4} + \left| \frac{m_\Phi}{2\kappa} \right| \right) \pi |\kappa| e^{-|\kappa|\pi R}, &   \kappa \lesssim - |m_\Phi|
\end{array}\right.
\eea
%For $\kappa \neq 0$, the above expression is more precise for larger $n$. 
The above result shows that the spectrum is similar to the case of RS for $\kappa \gtrsim |m_\Phi|$, while the overall mass scale is suppressed for $\kappa \lesssim -|m_\Phi|$ as in the LED case. 
In Fig.~\ref{KKmass}, we depict how the KK masses behave as a function of $(k_1-k_2)/m_\Phi$ 
 for fixed values of $m_\Phi$ and $R$.

\begin{figure}[t]
\begin{center}
\includegraphics[width=0.5\textwidth]{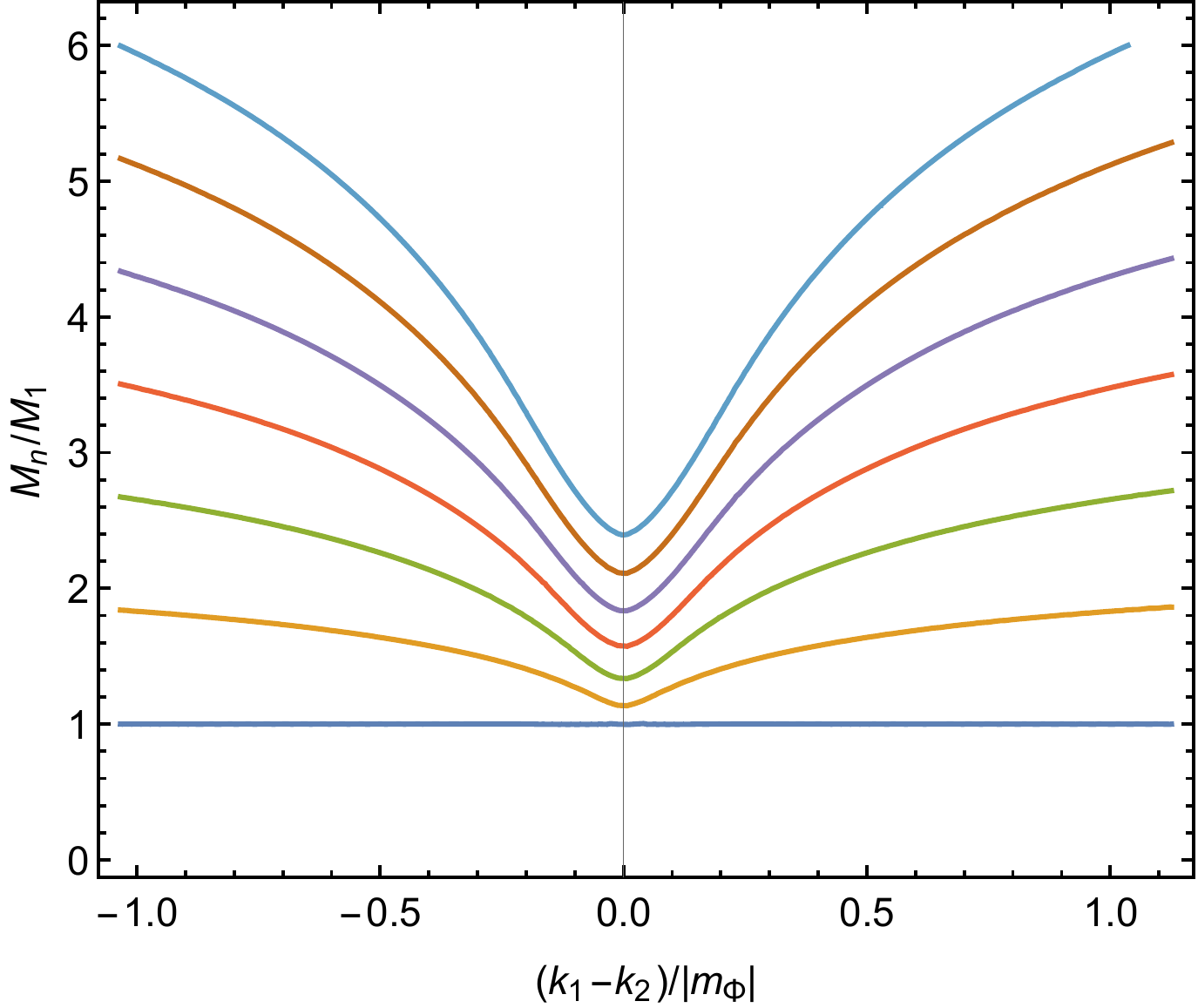}
\caption{KK masses ($n=1,2,\cdots, 7$) relative to the first KK  mass $M_{1}$ as a function of $\kappa/|m_\Phi|\equiv(k_1-k_2)/|m_\Phi|$ for fixed values of $m_\Phi$ and $R$.}\label{KKmass_ratio}
\end{center}
\end{figure}
In fact, the parameter point $\kappa\equiv k_1-k_2=0$  is special in regard to the KK mass pattern.  
In Fig.~\ref{KKmass_ratio}, we plot the mass ratios $M_n/M_1$ as a function of $\kappa/|m_\Phi|$, again  for fixed values of $m_\Phi$ and $R$.
It shows that the mass gap among the KK states is minimal in the unit of the first KK mass $M_1$ for $\kappa=0$.
This feature is related to the fact that the CW gear masses in the DCW are nearly degenerate at the CW threshold scale.
As we have shown in the previous section, the direct continuum limit of the DCW produces either CCW-I or CCW-II depending on which field basis is used to realize the 5D discrete shift or continuous $U(1)$ gauge symmetry as in (\ref{5dsymmetry_basis}).  
Notice that both 5D models correspond to the special case with $\kappa\equiv k_1-k_2=0$ in view of general CCW. Therefore, the relatively degenerate KK masses 
for a given KK threshold scale can be considered as a special feature of the CCW obtained as a direct continuum limit of the DCW, but not a generic feature of general CCW.
%On the other hand, a non-zero $\kappa \, (k_1 \neq k_2)$ provides the site-dependent CW mass scales $M_{\textrm{CW}, i}$ upon the %deconstruction of the continuum lagrangian (\ref{CW_bosons_c}), which was taken to be site-independent in the DCW for simplicity. %However, the CW masses in the DCW can be site-dependent 
%as long as they are comparable, and a non-zero $\kappa$ can parameterize such deviation from the simplest CW scenario.

\begin{figure}[t]
\begin{center}
\includegraphics[width=0.5\textwidth]{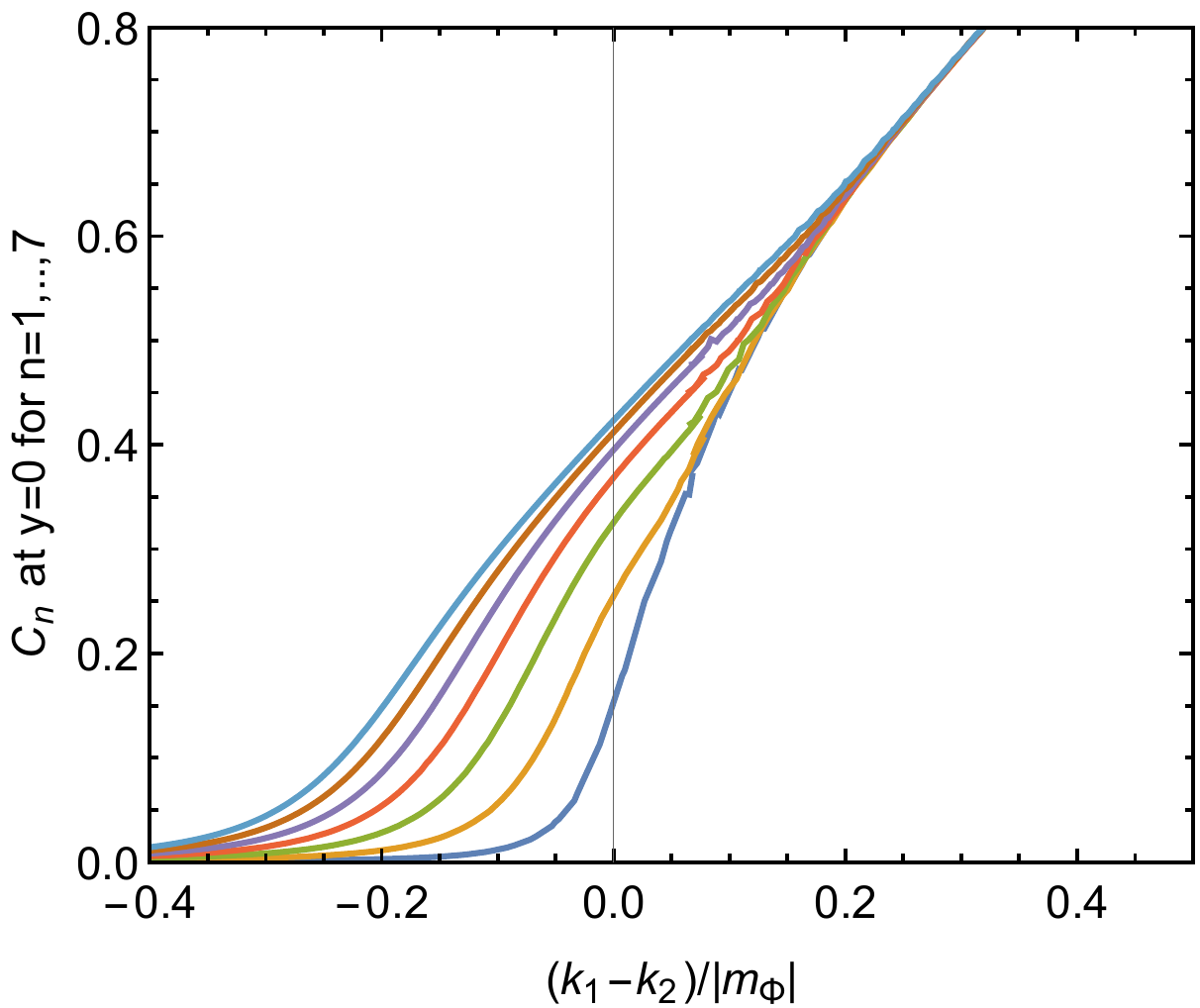}~
\includegraphics[width=0.5\textwidth]{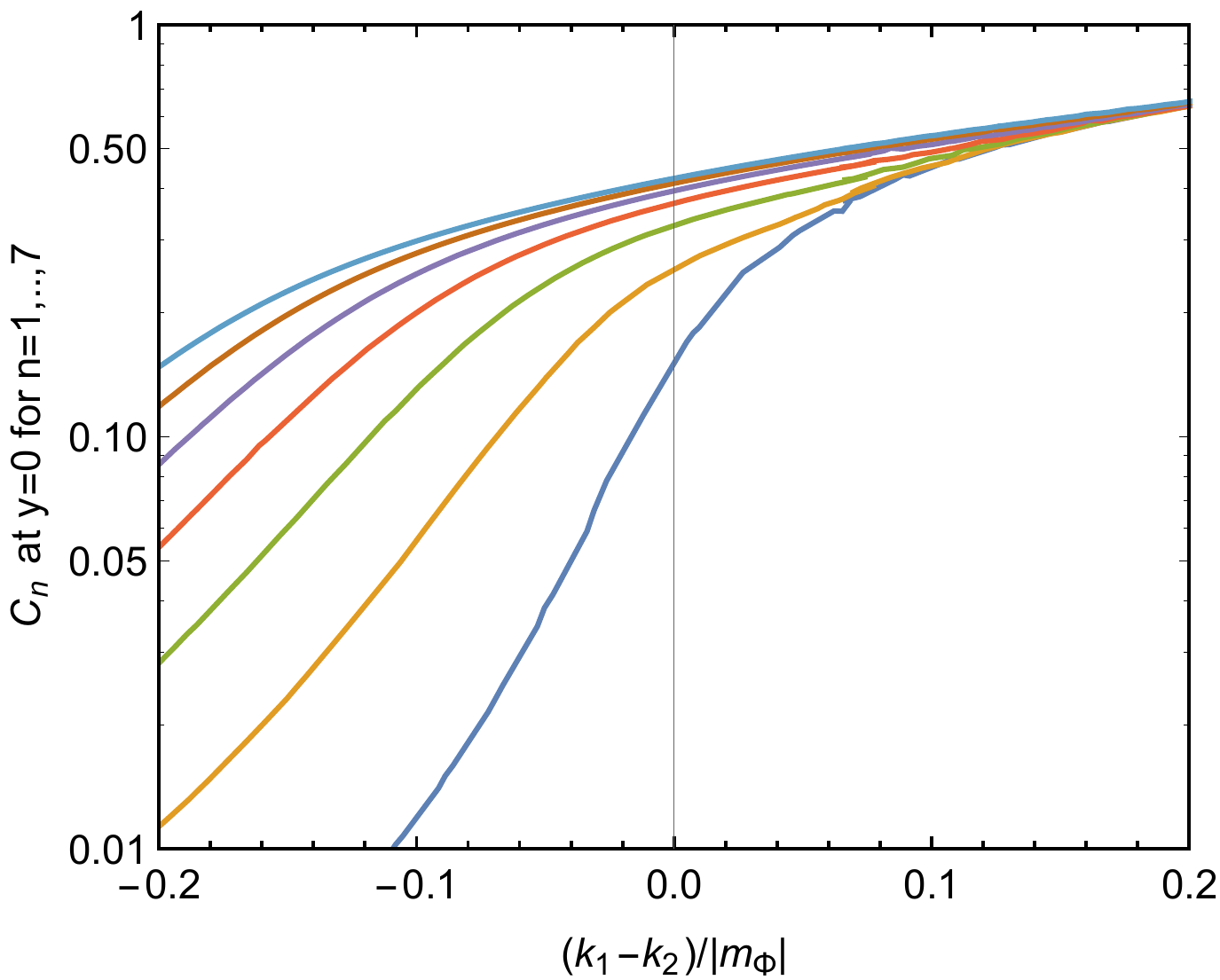}
\caption{ 
The couplings between the boundary localized fields (at $y=0$) and the KK modes ($n=1,2,\cdots, 7$) as a function of $\kappa/|m_\Phi|\equiv(k_1-k_2)/|m_\Phi|$ for fixed value of $m_\Phi$ and $R$.}\label{KKcoupling}
\end{center}
\end{figure}
Let us now turn to the couplings of the zero mode and KK modes to the boundary operators.
For this, we should go back to the original field basis $\Phi$ rather than using the canonical basis $\Phi_c$,
since the natural size of the relevant boundary couplings are determined in the original  field basis for which   
the 5D discrete shift and continuous $U(1)$ gauge symmetries are defined as (\ref{5dsymmetry_basis}).  
Then the mode couplings to the boundary operators can be read off by the component of the corresponding mode in $\Phi$.
Explicitly, if we consider the couplings on the boundary at $y=y_b$, the mode coupling $C_n (y_b)$ is extracted from
\beq
\Phi(x, y=y_b) = \sum_n \sqrt{|m_\Phi|}\,C_{n}(y_b)\, \phi_n(x),
\eeq
where $C_n (y_b)$ is determined to be
\beq \label{C_n}
C_{n}(y_b)=   \frac{f_n (y_b)}{\sqrt{|m_\Phi|}} e^{-k_\Phi |y_|}
\eeq
for $f_n(y)$ given by (\ref{f_n}).

Let us first discuss the case $m_\Phi > 0$, which applies to the graviton case with $m_\Phi = k_1 + k_2/2 > 0$ as
we take the convention $k_1, k_2 > 0$. 
The resulting couplings at $y=0$ are calculated to be
\bea
C_0(0) &=& \frac{1}{\sqrt{|e^{2m_\Phi \pi R} -1|}}, \label{C00}\\
C_{n}(0) &\simeq& \left\{
\begin{array}{ll}
(-1)^{n-1} \sqrt{\frac{\kappa}{m_\Phi}},&  \kappa \gtrsim m_\Phi \\
 \frac{1}{\sqrt{m_\Phi \pi R} }\frac{n}{\sqrt{ m_\Phi^2 R^2 + n^2}}, & \kappa\simeq0 \\
-\sqrt{\frac{|\kappa| \pi}{m_\Phi}} \frac{e^{-m_\Phi \pi R}}{\Gamma\left(m_\Phi/|\kappa|\right)}
\left[\frac{\pi}{2}(n-\frac{1}{4}+\left|\frac{m_\Phi}{2 \kappa}\right|) \right]^{\frac{m_\Phi}{|\kappa|}-\frac{1}{2}} ,
   &   \kappa \lesssim -m_\Phi 
\end{array}\right. \label{Cn0}
\eea
In Fig.~\ref{KKcoupling}, we depict the couplings of KK modes to boundary localized fields at $y=0$  as a function of $\kappa/m_\Phi\equiv(k_1-k_2)/m_\Phi$ for fixed value of $m_\Phi$ and $R$.\footnote{ In the figure, the absolute values $|C_n(0)|$ are depicted, since the sign of $C_n(0)$ is not physical so that it can be removed by redefining the mode field $\phi_n(x)$. On the other hand, note that the relative sign between the KK mode couplings at different boundaries has a physical meaning.} 
Notice that there is an exponential hierarchy  $\sim e^{-m_\Phi \pi R}$ between the zero mode couplings and the massive KK mode couplings at $y=0$ for $\kappa \gtrsim 0$, which is one of the main features of the DCW. 
This hierarchy is controlled by  $m_\Phi \equiv \mu_\Phi + k_\Phi$, i.e. the sum of the bulk and boundary mass parameter $\mu_\Phi$ and the warp factor parameter $k_\Phi$ in the  lagrangian (\ref{CW_bosons}).  
This means that both  CCW-I ($\mu_\Phi \neq 0, k_\Phi=0, \kappa=0$) and CCW-II ($\mu_\Phi=0, k_\Phi\neq0, \kappa=0$)
possess this property.  
On the other hand, this exponential hierarchy disappears for $\kappa \lesssim -|m_\Phi|$ which approaches to 
the LED limit. This can be easily understood by noting that deconstructing the 5D LED model gives a 4D lagrangian with the DCW parameter $q=1$ \cite{ArkaniHamed:2001ca, Hill:2000mu}, i.e. there is no clockwork. 

The zero mode and KK mode couplings at $y=\pi R$ for $m_\Phi >0$ are determined by
 \bea
C_0(\pi R) &=& C_0(0)\, e^{\mu_\Phi \pi R}, \label{C0pi}\\
C_{n}(\pi R)  &\simeq& (-1)^n \,C_n(0) \left\{
\begin{array}{ll}
 \frac{\sqrt{\pi}}{\Gamma\left(m_\Phi/\kappa\right)}\left[\frac{\pi}{2}(n-\frac{1}{4}+\left|\frac{m_\Phi}{2 \kappa})\right| \right]^{\frac{m_\Phi}{\kappa}-\frac{1}{2}} e^{-(m_\Phi+k_\Phi) \pi R},&  \kappa \gtrsim m_\Phi \\
e^{- k_\Phi \pi R}, & \kappa\simeq0 \\
  \Gamma\left(m_\Phi/|\kappa|\right)
\left[\frac{\pi}{2}(n-\frac{1}{4}+\left|\frac{m_\Phi}{2 \kappa}\right|) \right]^{-\frac{m_\Phi}{|\kappa|}+\frac{1}{2}} \,e^{(m_\Phi - k_\Phi) \pi R},
   &   \kappa \lesssim -m_\Phi
\end{array}\right. \label{Cnpi} 
\eea
At this boundary at $y=\pi R$, the exponential hierarchy between the zero mode couplings and the KK mode couplings vanishes
for $\kappa \equiv k_1-k_2\simeq 0$ as expected from the DCW, while the KK mode couplings for $\kappa \gtrsim m_\Phi$ 
are exponentially smaller than the zero mode couplings as in the RS case.  

For the case $m_\Phi < 0$, the couplings can be easily obtained using the  relation
\dis{
\left.f_n (y)\right|_{\kappa, - m_\Phi} \simeq \left.f_n ( \pi R -y)\right|_{-\kappa, m_\Phi}
} 
for $n \geq 1$.
The resultant couplings of the KK modes ($n \geq 1$) are qualitatively similar to 
(\ref{Cn0}) and (\ref{Cnpi}), but now $m_\Phi$ replaced by $|m_\Phi|$, while the expression for zero mode couplings are same as
 (\ref{C00}) and (\ref{C0pi}). 
In this case, the hierarchical structure of couplings turns out to be similar to flipping the boundaries to each other, while interchanging the RS limit and the LED limit in the case of $m_\Phi > 0$.
 Interestingly, this means that the KK couplings are similar to the LED limit while the KK spectrums
are similar to the RS limit for $\kappa \gtrsim |m_\Phi|$, and vice versa for $\kappa \lesssim -|m_\Phi|$, when $m_\Phi$ is negative. Thus the sign of $m_\Phi$ for a given $\kappa$ determines the pattern of the KK couplings, whereas the sign of $\kappa$ determines the pattern of the KK spectrum.

%Interesting difference between the CW and RS case is that 
%in CW case, the masses are nearly degenerated, but the couplings are different for few KK modes. 
%On the other hand, the KK masses are separated, but the couplings are degenerated in the RS case. 

Let us finally discuss the zero mode couplings at different boundaries. 
From (\ref{C_n}), the zero mode coupling at $y=y_b$ is determined as 
\dis{
 C_0(y_b) = \frac{e^{\mu_\Phi |y_b|}}{\sqrt{|e^{2m_\Phi \pi R} -1|}},
 }
 showing that a non-zero $\mu_\Phi$, which is obtained only by non-zero bulk and boundary masses, is necessary to obtain an exponential hierarchy between the zero mode couplings at different $y_b$.
% Therefore  CCW-II cannot reproduce this hierarchy, while the CCW-I does it. 
On the other hand, a non-zero $m_\Phi \equiv \mu_\Phi + k_\Phi$, which does not necessarily require non-zero $\mu_\Phi$,   yields  an exponential suppression of the zero mode coupling at  $y_b=0 \,(\pi R)$ for a positive (negative) $m_\Phi$, which is responsible for the hierarchy between the zero mode coupling and the massive KK mode coupling at $y_b=0\,(\pi R)$.

In the following three subsections,  we will examine  how the lagrangian (\ref{CW_bosons}) can be obtained from  a 5D diffeomorphism and Lorentz invariant theory for CCW axions and $U(1)$ gauge bosons,  and discuss further
issues such as the field range of CCW axion zero mode.

\subsection{CCW axion from 5D angular field}

A natural candidate of 5D field that would yield a periodic 4D axion  is a 5D angular field $\theta(x,y)$, with the periodicity: \bea
\theta \equiv \theta +2\pi.\eea
Including a periodic potential of $\theta$, as well as the couplings to gauge fields at the boundaries, the 5D action  can be written as 
\bea \label{eq:action_angle}
S &=& -\int d^5 x \sqrt{-G} \,\left[ \frac{f_5^3}{2}   G^{MN} \partial_M \theta \partial_N\theta   + 
 V(\theta)  \right.  \\ 
&&\hskip 2.8cm \left. + \frac{\theta}{32\pi^2}\left(\kappa_0\frac{\delta(y)}{\sqrt{G_{55}}} G_0\tilde G_0 
 + \kappa_\pi \frac{\delta(y-\pi R)}{\sqrt{G_{55}}} G_\pi \tilde G_\pi\right)\right],\label{eq:instanton_angle}  
\eea 
where $G_{0,\pi}$ are the gauge field strengths at the boundaries  $y=\{0,\pi R\}$,  $\tilde G_{0,\pi}$ are their duals,  
 $\kappa_{0,\pi}$ are restricted to be integers, and the dimensionful parameter  $f_5$ is 
 assumed to be well below the 5D  Planck scale $M_5$.

If $V(\theta)=0$,  there is an obvious massless 4D mode $\phi_0(x)$ protected by the shift symmetry $\theta\rightarrow \theta +{\rm constant}$,
and the corresponding mode function has a flat profile, i.e.  $f_0(y)={\rm constant}$ for $\theta(x,y) =\sum_{n=0} f_n(y) \phi_n(x)$.
In the presence of nonzero bulk and/or boundary potentials, generically there is no massless 4D mode. However, as we will show in the following, if the bulk and boundary potentials are appropriately related to each other, there can be a massless mode with exponentially localized profile over the 5th dimension, which reveals many features of the CW mechanism.

 To achieve a 5D axion potential yielding such a zero mode, let us assume that both the dilatonic shift symmetry
$S\rightarrow S+{\rm constant}$ and the axionic shift symmetry $\theta\rightarrow \theta+{\rm constant}$ are softly broken by 
a dimensionful  mass parameter $m$ through the combination \bea e^{-c S} e^{i\theta} m.\eea  As the exponential dilaton dependence originates from the dilaton-dependent string coupling  $g_{\rm st}^2 \propto e^{-\xi S}$,  at leading order in $g_{\rm st}^2$ the bulk dilaton potential in the Einstein frame should scale as $e^{-2cS}$, while the boundary dilaton potential scales as $e^{-cS}$. (See Eqs. (\ref{GLD_string}) and (\ref{GLD}).)  In addition to the soft breaking by $e^{-cS+i\theta} m$ of both the dilatonic and axionic shift symmetries, the generalized linear dilaton model involves additional soft breaking by   $ke^{-cS}$ of the dilatonic shift symmetry only, where
$k$ parametrize the bulk cosmological constant  as in (\ref{GLD_parametrization}).   
%\bea
%V(\theta) = f_5^3\, \tilde V( e^{-cS} k,  e^{-cS} e^{i\theta} m, e^{-cS}e^{-i\theta}m^*).
%\eea
Then the leading order bulk and boundary potentials of the 5D angular field $\theta$  is generically given by
\bea\label{eq:pot_angle}
V(\theta) &=& -f_5^3\left[ e^{-2 c S}  ( c_1 k\, m \cos\theta  + c_2 m^2 \cos2\theta) 
\right.\nonumber\\
&& \left.\hskip 1cm +\,   \frac{ e^{-cS}}{\sqrt{G_{55}}} m \cos\theta \left( c_0\delta(y) 
- c_\pi  \delta(y-\pi R)\right)\right], 
\eea
where we assume the CP invariance under $\theta\rightarrow -\theta$, and $\{c_i\}$ 
are dimensionless coefficients of order unity.
In order to have a localized massless mode under the metric and dilaton background given by
(\ref{GLD_sol}) and (\ref{k's}), one needs the following specific values of $\{c_i\}$:
\bea
\label{massless_a}
c_1= \frac{-4k_1 + k_2}{k}=-\sqrt{\frac{16}{3}-4c^2}, \quad c_2 =\frac{1}{4}, \quad c_0 = c_\pi = 2.\eea
Taking these particular parameter values, while replacing the spacetime metric and dilaton by their vacuum values, 
the 5D action  (\ref{eq:action_angle}) now takes the following form of general CCW:
\bea \label{eq:CCW_angle}
S =- \frac{f_5^3}{2}\int d^5 x  \,  e^{ (2k_1 + k_2)|y|} 
\left[ (\partial_\mu\theta)^2  + e^{ 2(k_1 -k_2)|y|} (\partial_y\theta - m \sin\theta )^2\right]
\eea 
which is manifestly invariant under the discrete shift $\theta\rightarrow \theta +2\pi$.

 To proceed, we can make a field redefinition 
\bea \label{redef_theta}\varphi(x,y)\equiv  \tan (\theta(x,y)/2),\eea
and then the 5D action (\ref{eq:CCW_angle}) is given by 
\beq \label{eq:CCW_nonlinear}
 - \frac{f_5^3}{2}\int d^5x \frac{ 4e^{(2k_1 + k_2)|y|}}{(1 + \varphi^2)^2}
\left[ (\partial_\mu\varphi)^2 + e^{2(k_1 - k_2)|y|} (\partial_y\varphi - m\varphi)^2\right]. 
\eeq   It is clear that the above action admits a localized massless 4D mode given by
\bea
\label{zero_varphi}
\varphi(x,y) =    e^{ my}u(a(x))\eea  
 where $a(x)$ is the canonically normalized massless 4D axion, and $u(a)$ is a function of $a$ introduced to make  the 4D axion $a$ have  the standard canonical kinetic energy {\it over its entire field range}.
 For flat metric and trivial  dilaton backgrounds, i.e. $k_1=k_2=0$, which we will be focusing on, the kinetic term of $a$ is given by
 \bea
 - \frac{ f_5^3  (e^{2\pi m R} - 1)(\partial_\mu u)^2}{m (u^2 + 1)(e^{2\pi m R}  u^2 + 1)} \equiv 
-\frac{1}{2} (\partial_\mu a)^2.
\eea
We then find
\bea
\label{zero_a}
u(a)=  i e^{- m \pi R} {\rm sn} \left( \left. \frac{a}{2i f} \right| e^{-2 m\pi R}\right),\eea
where 
$f= \sqrt{ {f_5^3(1 - e^{-2m\pi R})}/{2m}}$,
and
${\rm sn}(x|z) $ is the Jacobi elliptic function which has two periods as 
${\rm sn}(x|z) = (-1)^{n_1}{\rm sn}(x + 2 n_1 K(z) + 2 i n_2  K (1-z) | z) $ for integer-valued $n_1, n_2$ and
 $K(z)$ being the complete elliptic integral of the first kind.

%For $\theta\ll 1$, we can expand Eq.~(\ref{eq:pot_angle}) about the origin, which results in
%the following mass terms:
%\bea\label{eq:pot_quad}
%V_{\rm quad}(\theta) =  \frac{f_5^3}{2} \left[ e^{- 2c S}( c_1  k\, m + 4 c_2m^2) \theta^2  
%+ \frac{e^{-c S}}{\sqrt{G_{55}}} m\theta^2\left( c_0\delta(y)   - c_\pi\delta(y-\pi R) \right)   \right].
%\eea
%Keeping the quadratic order of the infinitesimal fluctuation around $\theta=0$,  
%it is easily identified that the quadratic action  of Eq.~(\ref{eq:action_angle}) can be organized as 
%\bea\label{eq:CCW_quad}
%S_{\rm quad} = -\frac{f_5^3}{2}\int d^5 x   e^{ (2k_1 + k_2)|y|} 
%\left[ (\partial_\mu\theta)^2  + e^{ 2(k_1 -k_2)|y|} (\partial_y\theta - m \theta)^2\right],
%\eea

With the above model for CCW axion, we can address the following three issues: 
\begin{itemize} 
\item  Wavefunction profile and the KK mass spectrum, $\{f_n(y),\, M_n\}$ for a given  background value of $\langle\theta\rangle$.
\item  Field range of the canonically normalized 4D axion $a$.
\item  Effective potentials of the  4D axion $a$ generated by the model parameters $\{c_i\}$  slightly deviated from the 
values of (\ref{massless_a}).
\end{itemize} 
It is in principle straightforward to examine these issues for generic metric and dilaton background. However, for simplicity in the following we will focus on the simplest background with flat metric and constant dilaton, i.e. $k_1=k_2=0$.
% while giving a short remark on the general case in the last stage.  

\begin{figure}[t]
\begin{center}
\includegraphics[width=0.325\textwidth]{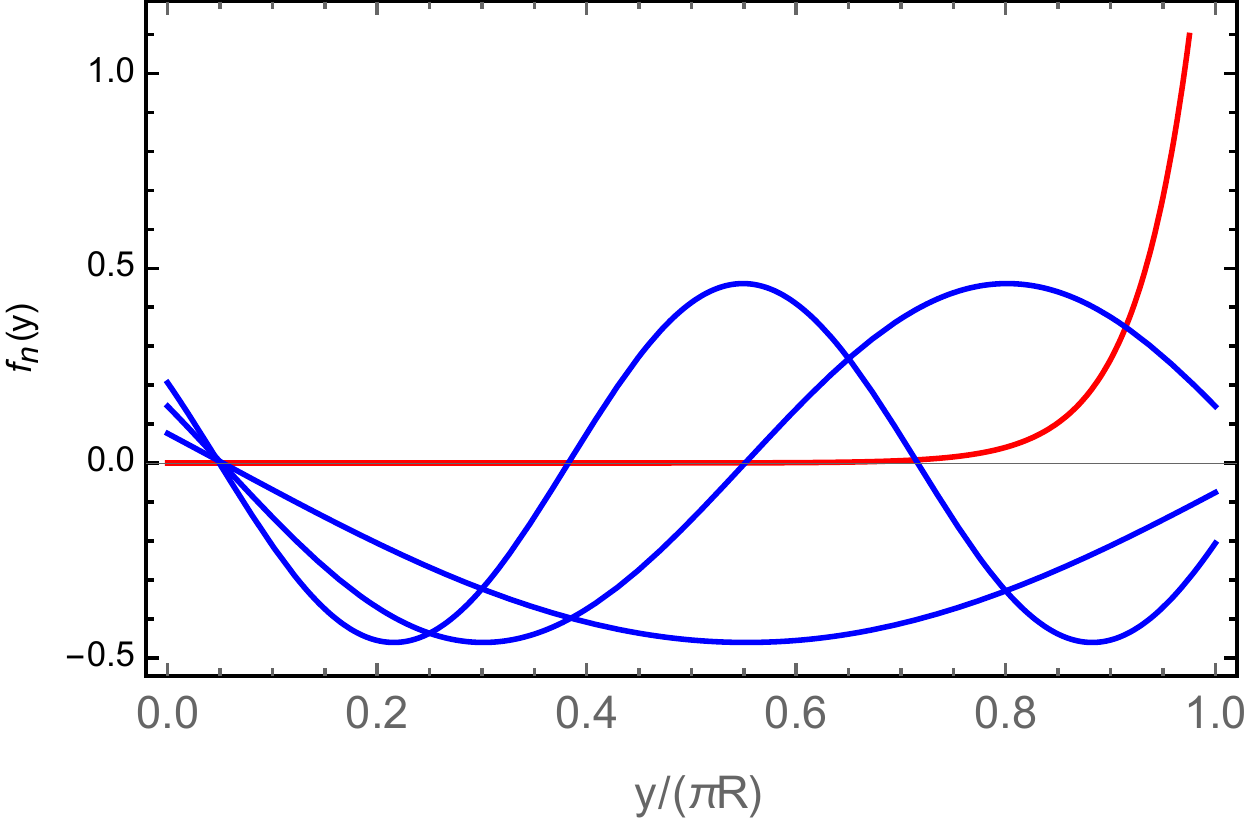} 
\includegraphics[width=0.325\textwidth]{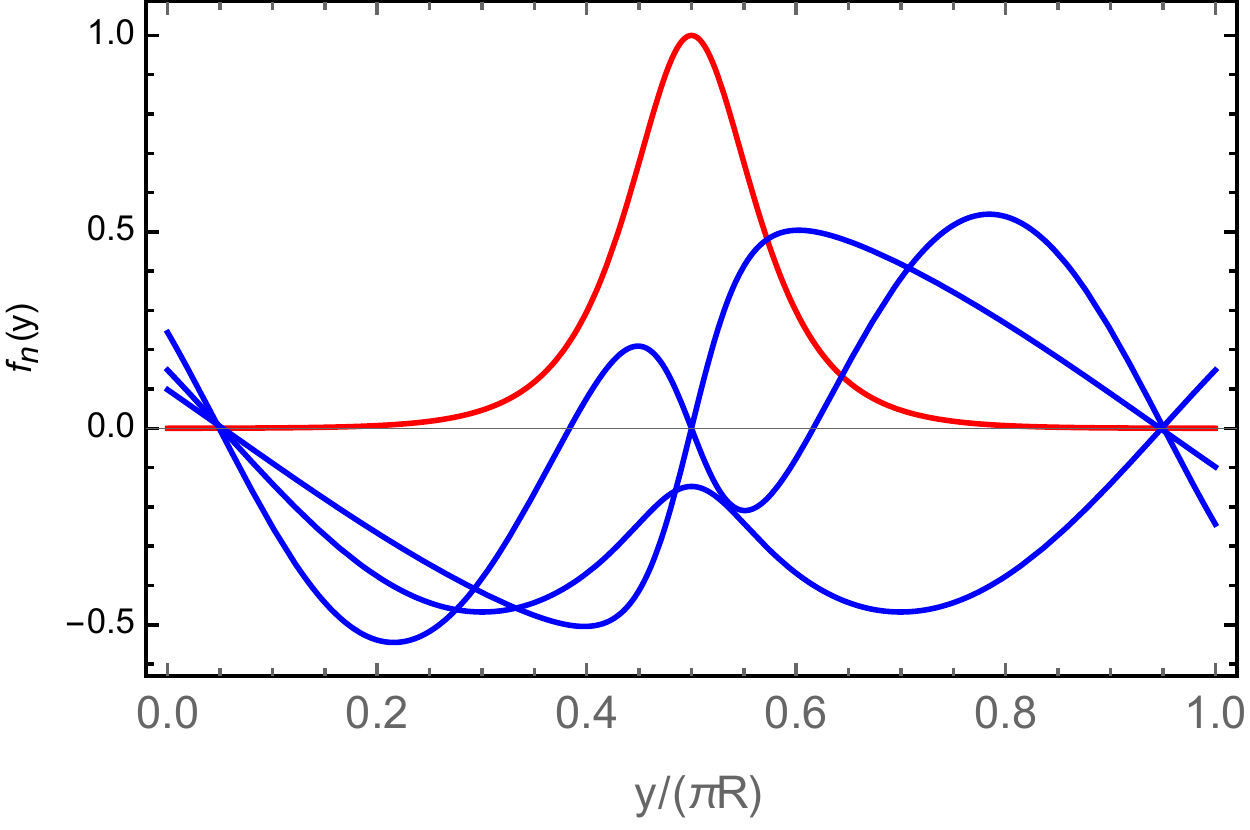} 
\includegraphics[width=0.325\textwidth]{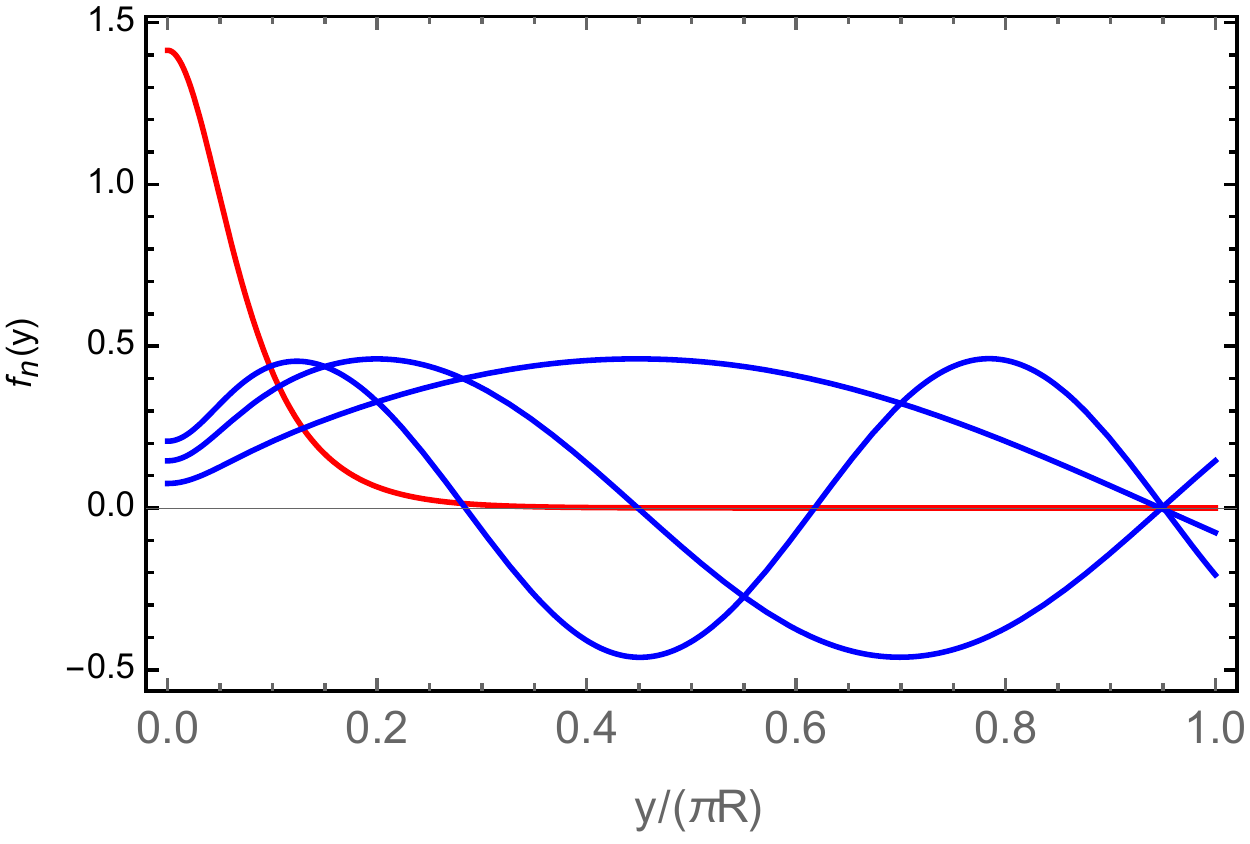} 
\caption{ Wavefunction profile (\ref{eq:wavefunction}) for the zero mode fluctuation ($n=0$: red) and massive mode fluctuations 
(n=$1,2,3$: blue) for  $y_0 = 2\pi R$ (left), $y_0=\pi R/2$ (middle) and $y_0=0$ (right).  
For illustration, we take   $m R = 6$.}\label{CCW_axion_wavefunctions}
\end{center}
\end{figure}

For the KK analysis,  let us divide $\theta(x,y)$ into the background vacuum value $\langle \theta\rangle$ and a small fluctuation $\delta\theta$:
\bea
\theta(x,y) = \langle \theta\rangle + \delta\theta,\eea
where  the vacuum solution of $\theta$ is given by 
\bea
\langle\theta\rangle =  2\tan^{-1}[e^{my}\langle u(a)\rangle]= 2\tan^{-1}[e^{m(y-y_0)}].\eea
Here the parameter $y_0$ is introduced to describe the vacuum value of the massless 4D axion $a$. As we will see later, $y_0$  describes also the position where the zero mode wavefunction is localized. 
Then the action (\ref{eq:CCW_angle}) can be written as
\bea
-\frac{f_5^3}{2} \int d^5 x  \Big[ (\partial_\mu\delta\theta)^2 + 
(\partial_y\delta\theta + m\tanh[m(y-y_0)] \delta\theta)^2  + {\cal O}(\delta\theta^3) \Big].
\eea  At the quadratic order in $\delta\theta$, the above action reveals an infinitesimal shift symmetry
\beq
\delta\theta(x,y) \to \delta\theta(x,y) +  c_0\, {\rm sech}[m(y-y_0)] \quad (|c_0|\ll 1), 
\eeq
which insures that there exists a massless mode for  any value of $y_0$.  
Taking the mode expansion 
\bea
\delta\theta(x,y) = \sum_{n=0}^\infty f_n(y) \delta\phi_n(x),
\eea
where $\delta\phi_n$ correspond to the canonically normalized 4D field fluctuations in mass eigenstates,
we find the wavefunction profiles given by   
 \bea \label{eq:wavefunction}
 f_0(y) &=& \frac{1}{N_0}\,{\rm sech}[m (y-y_0)],\nonumber\\ 
 f_n(y) &=& \frac{1}{N_n}\, \left[ n \cos\Big(\frac{n y}{R}\Big) + m R \tanh[m (y-y_0)] \sin\Big(\frac{n y}{R}\Big)   \right]\quad (n = 1,\,2,\,..)
\eea and the mass eigenvalues: 
\bea
M_0=0,\quad M_n = \sqrt{ m^2 + \frac{n^2}{R^2}}.
\eea
 In Fig.~\ref{CCW_axion_wavefunctions}, we depict the wavefunction profiles of the zero and few KK modes for
 certain values of $y_0$. 
 Note that $y_0$ can have any value in the range $-\infty < y_0 < \infty$, while the orbifold coordinate $y$ is defined only over the fundamental domain: $0\leq y\leq \pi R$.  
%For $y_0\rightarrow \pm \infty$,  we recover the zero mode wavefunction localized at the boundaries, which is the result of   

 From  the above wavefunction profile, we can obtain also the  zero mode coupling to boundary operators.  For this, let us introduce a boundary coupling of the form 
\bea \int dy\delta(y-y_b)\kappa_b \theta(x,y) {\cal O}(x),\eea
where ${\cal O}(x)$ is a local operator constructed by generic boundary fields at $y=y_b$. The resulting coupling of the
zero mode fluctuation $\delta\phi_0\equiv \delta a$, which corresponds to the small fluctuation of the canonically normalized 4D axion field $a$ which was defined in Eqs.~\ref{redef_theta}, \ref{zero_varphi}, and \ref{zero_a},
is given by 
 \bea
 \label{zero_coupling}
 \kappa_b\sqrt{\frac{m}{f_5^3}} \frac{   {\rm sech}[m (y_b-y_0)]}{\sqrt{ \tanh[m(\pi R - y_0)] + \tanh(m y_0)}} 
\delta a(x){\cal O}(x). 
 \eea
This then yields 
 \beq
 \label{zero_coupling_1}
 \frac{ \kappa_be^{-m (\pi R - y_b)} }{f} \delta a(x) {\cal O}(x) \eeq
in the limit $y_0\to -\infty$ where the zero mode axion is exponentially localized at $y=\pi R$.
 With the above form of boundary couplings of the zero mode axion fluctuation $\delta a$, one can easily generate an exponential hierarchy between the zero mode couplings at different boundaries, i.e. 
$y_b=0$ and $y_b=\pi R$.
 Although the analytic expression of the massive KK mode couplings to the boundary operator ${\cal O}(x)$ is  rather complicated,  one   can generate also a similar hierarchy between the zero mode coupling and the KK mode coupling to the boundary fields at a given $y_b$ with an appropriately chosen value of $y_0$.

 Let us now discuss the field range of the canonically normalized 4D axion $a$, which is embedded in the 5D angular field $\theta$ as
  \bea \theta = 2 \tan^{-1} [e^{m y} u(a)]\eea
  where
\bea
u=  i e^{- m \pi R} {\rm sn} \left( \left. \frac{a}{2i f} \right| e^{-2 m\pi R}\right) \quad {\rm for}\,\,\,  
f= \sqrt{ \frac{f_5^3(1 - e^{-2m\pi R})}{2m}}.
\eea
From the periodicity of the Jacobi elliptic function:
\bea
{\rm sn}(x|z) = (-1)^{n_1}{\rm sn}(x + 2n_1  K(z) + 2 i n_2  K (1-z) | z)  \quad ( n_1, n_2={\rm integers}),\eea
where $K(y)$ is the complete elliptic integral of the first kind, we find the field range of $a$, which is matched to the $2\pi$ range of $\theta$, i.e. $\theta\equiv \theta +2\pi$, is given by  
\beq
\label{a_range}
a\equiv a+2\pi f_{\rm eff} \quad {\rm for}\quad  2\pi f_{\rm eff} =  4 f  K(1-e^{-2m\pi R}).
\eeq
In the limit of small $m$ ($m\pi R \ll 1$), this results in \bea
f_{\rm eff} \simeq\sqrt{ f_5^3\pi R},\eea while for a large value of $m$ ($m\pi R \gg1$),  the field range becomes
\bea
f_{\rm eff} \simeq  \sqrt{ \frac{2 m R}{\pi }}\sqrt{ f_5^3 \pi R} 
= \sqrt{\frac{2m R f_5^3}{\pi M_5^3 }} M_{\rm Pl}  .
\eea
Note that there is {\it no} exponential enhancement of the field range of $a$ relative to the 5D axion scale $f_5$. Yet, an exponentially localized  profile of the zero mode fluctuation $\delta a$ could be successfully realized,
yielding  an exponential hierarchy among the couplings of $\delta a$ at different boundaries as dictated by (\ref{zero_coupling}) or (\ref{zero_coupling_1}), as well as a similar exponential hierarchy between the zero mode coupling and the KK mode couplings at the same boundary.

\begin{figure}[t]
\begin{center}
\includegraphics[width=0.5\textwidth]{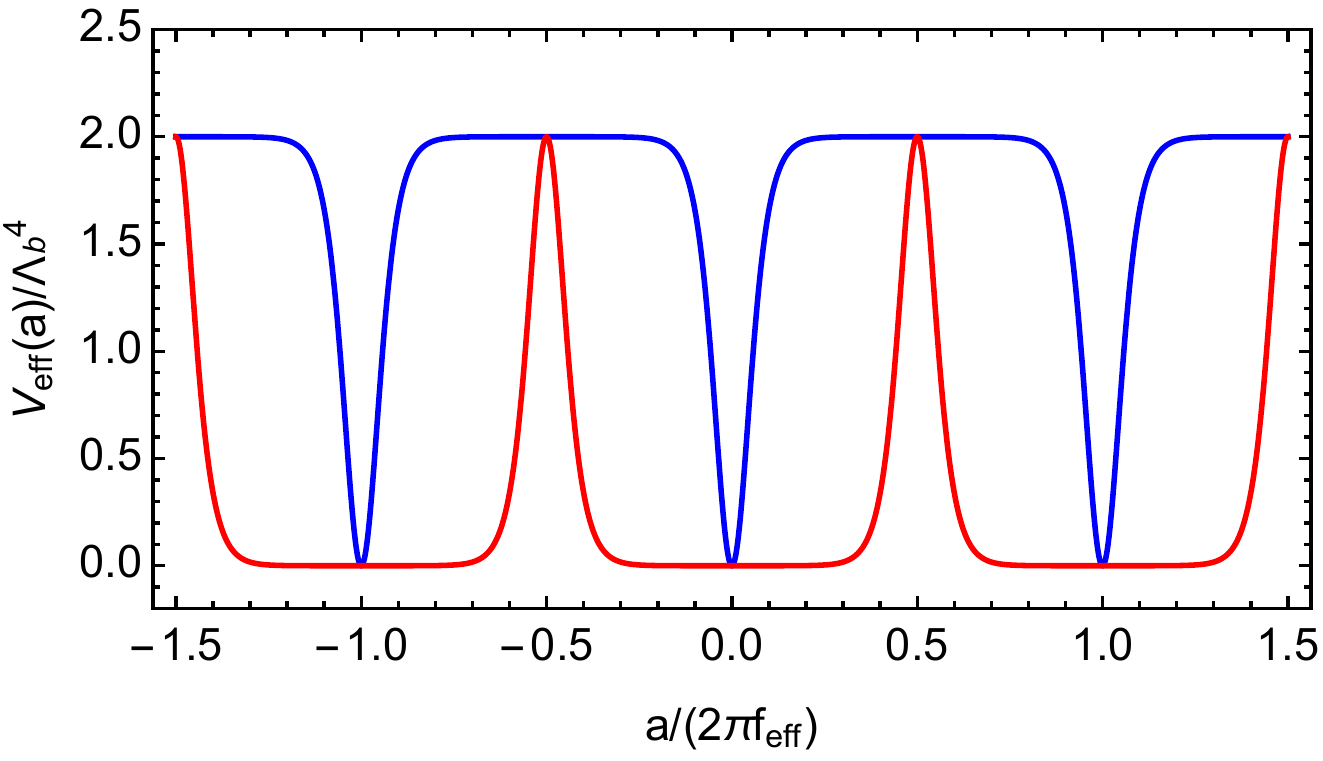} 
\caption{ The axion potentials induced by small parameter deviation at $y_b=0$ (red) and $y_b=\pi R$ (blue). 
For illustrative purposes, we take a moderated value of $m$ giving  $m R = 2$.}\label{CCW_axion}
\end{center}
\end{figure}

Exponential hierarchy among the couplings of $\delta a$ {\it without} an exponentially enhanced range of $a$ yields a rather unusual form of the effective potential of $a$, which might be generated by a deviation of model parameters from  
(\ref{massless_a}) or by an axion coupling to the confining gauge sector at the boundaries. To be specific, here we consider the case that the model parameter $c_0$ and $c_\pi$ are slightly deviated from the values of (\ref{massless_a}), which would result in the following effective potential 
\bea \label{eq:effpot}
V_{\rm eff}(a)&=&  \sum_b -\Lambda_b^4\cos \theta (x, y_b)+... \nonumber\\
&=& - 2\sum_b \Lambda_b^4 \left[\frac{    e^{ - 2m (\pi R- y_b)}  {\rm sn}^2(  \frac{a}{2i f}|  e^{-2m\pi R}) }{ 1- e^{ - 2m (\pi R- y_b)}  {\rm sn}^2(\frac{a}{2if} |  e^{-2m\pi R})   } \right]+.., 
\eea where $y_b=\{0,\pi R\}$ and we ignored irrelevant constant parts.
% is  the position of instanton\footnote{We simplify the instanton induced potential just as $-\Lambda_b^4\cos\theta$. There could be %sizable contributions such as  $\cos( n\theta)$ for $n=1,2,\cdots$. This contribution could be suppressed if there is the light quarks %with $m_q \ll \Lambda_b$. 
%The effective potential Eq.~(\ref{eq:effpot}) can be considered as the result of small mismatch between the boundary and bulk mass %terms.}.    
We stress that the above axion potential is valid over the full range of $a$ given by (\ref{a_range}).
 In Fig.~\ref{CCW_axion}, we depict this effective potential originating from $y_b=0$ (red) and $y_b=\pi R$ (bluee). 
The unusual shape of this axion potential can be understood from that the zero mode fluctuation $\delta a \equiv a-\langle a\rangle$  is exponentially localized at different position for different value of $\langle a\rangle$.  For $\langle a\rangle=0$, $\delta a$ is localized at $y_b=\pi R$. However for $\langle a\rangle = \pi f_{\rm eff}$, $\delta a$ is localized at $y=0$. For $0 < \langle a\rangle < \pi f_{\rm eff}$,  $\delta a$ is localized at a certain position between $0$ and $\pi R$. 
Since $\delta a$ is localized at different position in the 5th dimension for different $\langle a\rangle$, 
its overlap with $\theta(x, y_b)$ also changes, which results in the change of the curvature of the scalar potential along different value of $\langle a\rangle$.

%So far, we have been focusing on the case with $k_1=k_2=0$, which has flat geometry and constant dilaton profile.
%For the case of  nonzero $k_1$ and $k_2$, there is 
%procedure. However they could be quite complicated, so we skip this part in this paper. One of the differences between the flat and %warped cases is that the masses of the  KK modes depend on the background value of the axion. For example, when $k_1=k_2$, 
%the KK masses for $\langle\theta\rangle=0$ are $\sqrt{(m + 3k_1/2)^2 + n^2/R^2}$, while for the background value of $\theta$ is $\pi$, %they become $\sqrt{(m-3k_1/2)^2 + n^2/R^2}$. This could generate the additional bulk potential for the zero modes at one-loop level. 

\subsection{CCW axions from 5D gauge  and St\"uckelberg  fields}\label{sec:CCW_axion_2}

Another candidate 5D field that could give a periodic 4D axion  is a 5D gauge field $C_M$ with odd orbifold parity  \cite{Choi:2003wr, Flacke:2006ad,Ahmed:2016viu}.
%Sincethe 5th component of $C_M$ is a 4D scalar and the gauge parameter of $U(1)$ satisfies $\Lambda \equiv \Lambda+2\pi$, we could %expect to find some interesting consequences related with CCW axion. 
%The 5D gauge symmetry is of course not fully preserved in the 5D action with an orbifold. 
%It is explicitly broken by the  orbifold boundary conditions for the gauge parameter, $\Lambda(x,y) 
%=\Lambda(x, y+2\pi R)=\pm \Lambda(x,-y)$, and by  those for the 5D fields. 
%Under the 5D gauge transformation consistent with the boundary conditions, 
%the 5D effective action is  taken to be invariant.  Eventually,  this gauge symmetry is broken spontaneously by the vacuum value of the %scalar component of $C_M$ and those of the other charged scalars.
%After KK decompositions, the massive 4D vector modes have the correct longitudinal degrees of freedom consistent with the gauge %symmetry.
%If there is no charged matter with a nonzero vacuum value, $\langle C_5(x,y)\rangle $ is the only source of  spontaneous breaking of %$U(1)$. 
To realize continuum CW axions in more general context, in this section we consider the case that the 5D gauge field $C_M$ has a St\"uckelberg mass involving an angular field $\theta$ which might originate from the phase degree of freedom
of $U(1)$-charged complex scalar field.
Under the $U(1)$ gauge symmetry,  $C_M$ and  $\theta$ transform as
\bea
U(1)_C:\  C_M \to C_M  + \partial_M\Lambda,\ \  \theta \to \theta + q\epsilon(y)
 \Lambda, 
\eea
where $\theta$ and $\Lambda$ are periodic as \bea
\theta\equiv \theta +2\pi, \quad  \Lambda\equiv \Lambda + 2\pi,\eea 
$q$ is an integer-valued charge,
and the fields and transformation function obey the following orbifold boundary conditions:
\bea
&& C_\mu(x, y) = C_\mu(x, y+2\pi R)=- C_\mu(x,-y) ,\quad
C_5(x,y) = C_5(x, y+2\pi R) =  C_5(x,- y), \nonumber \\
&& \theta(x,y)= \theta(x, y+2\pi R) = \theta(x, -y),\quad
 \Lambda(x,y)=\Lambda(x, y+2\pi R) =-\Lambda(x, -y). \quad
\eea
 Note that the gauge  transformation of $\theta$ involves a parity odd step function $\epsilon(y)=\epsilon(y+2\pi R)=-\epsilon(-y)$ with $\epsilon(y)=1$ for $0<y<\pi R$, which is necessary to be compatible with the above orbifold boundary conditions. 

As in Sec.~\ref{subsec:LDM}, one can start with a 5D action of $C_M$ and $\theta$ at leading order in $g_{\rm st}^2 \propto e^{-\xi S}$ in the context of generalized linear dilaton model in the string or Jordan frame, for which the 5D action including the boundary terms involves the overall dilaton factor $e^{\xi S}$. Moving to the Einstein frame, one finds the relevant part of
the  5D action is given by      
\bea
\label{5D_C_theta}
&&S_C = -\int d^5 x \sqrt{-G}\left[ \frac{e^{ 2cS}}{4 g_C^2}C^{MN} C_{MN}  
+ \frac{f_5^3}{2}   (\partial^M\theta - q\epsilon(y) C^M)(\partial_M\theta - q\epsilon(y) C_M)\right.    \nonumber \\ 
&&\hskip 1cm  +  \frac{\delta(y)}{\sqrt{G_{55}}} 
\left(\frac{\gamma_{0}}{M_5} \frac{C_{\mu 5}}{\sqrt{G_{55}}} \tilde J_0^\mu+  \frac{\kappa_{0}}{32\pi^2}\theta G_0\tilde G_0 
+ c_0\partial_\mu\theta J_0^\mu   \right)  \nonumber \\ 
&&\hskip 1cm \left. +  \frac{\delta(y-\pi R)}{\sqrt{G_{55}}} 
\left(\frac{\gamma_{\pi}}{M_5} \frac{C_{\mu 5}}{\sqrt{G_{55}}} \tilde J_\pi^{\mu}+  \frac{\kappa_{\pi}}{32\pi^2}\theta G_\pi\tilde G_\pi 
+ c_\pi \partial_\mu\theta J_\pi^\mu   \right) \right], \label{eq:ypi_C5angle}
\eea   where 
$\tilde J_{0,\pi}^\mu$, $J^\mu_{0,\pi}$ are model-dependent boundary-localized current operators.  As for the natural size of model parameters, the 5D gauge coupling $g_C$ might be estimated as $1/g_C^2 \sim M_5$ for the 5D Planck mass $M_5$, the bulk St\"uckelberg mass of $C_M$, i.e. 
$ q g_Cf_5^{3/2} $, is supposed to be well below $M_5$, and the dimensionless boundary couplings $\gamma_{0,\pi}$, $\kappa_{0,\pi}$ and $c_{0,\pi}$ can be chosen to be of order unity. 
%The factor $e^{2cS}$ in the gauge kinetic term is required to make the action invarinat under the dilatonic-scale transformation. 
We note that the $U(1)_C$ invariance of the bulk kinetic term of $\theta$ enforces that the gauge transformation function $\Lambda$ should be continuous at the boundaries, so 
\bea
\Lambda(x,0)=\Lambda(x,\pi R)=0.\eea  As a consequence,  $\delta_{U(1)_C}\theta=0$ at the boundaries, and therefore  the $U(1)_C$ gauge symmetry admits an arbitrary boundary potential of  $\theta$ at $y=0, \pi R$. One can  forbid such arbitrary boundary potential
by assuming a global $U(1)$ symmetry at each boundary, 
\bea\label{global_sy_boundary}
&U(1)_0:& \,\, \delta(y)\theta(x, y) \to \delta(y)(\theta(x,y) + c_0), \nonumber\\
&U(1)_\pi:&\,\, \delta(y-\pi R)\theta(x,y)\to \delta(y-\pi R)(\theta(x,y) + c_\pi) ,\eea
which is explicitly  broken by the YM instantons at the boundaries.
We already assumed such global $U(1)$ symmetries in the 5D action (\ref{5D_C_theta}).
%\beq\label{eq:Sym_C5theta}
%[U(1)_{\Lambda_0=\Lambda_\pi =0}]_{\rm local} \times [U(1)_0]_{\rm global}\times [U(1)_\pi]_{\rm global}, 
%\eeq
%where
%\beq\label{eq:global_C5theta}
% [U(1)_0]_{\rm global}\times [U(1)_\pi]_{\rm global}: \ 
%C_5\to C_5 + \frac{c_\pi - c_0}{\pi R} ,\ \ \theta \to \theta +  \left(1-\frac{|y|}{\pi R}\right)c_0 
%+ \frac{|y|}{\pi R} c_\pi.
%\eeq
%We identify this transformation as $U(1)$s, because the range of $c_{0,\pi}$ is defined from the  transformation at boundaries, 
%$\{\theta(x,0)\to \theta(x,0)+c_0$, $\theta(x, \pi R) \to \theta(x, \pi R) + c_\pi\}$. From the periodicity of $\theta$, the parameters %are constrained as  $c_0\equiv c_0+2\pi$, and $c_\pi\equiv c_\pi + 2\pi$.  
%Terms allowed by gauge and global symmetries are written in Eq.~(\ref{eq:y0_C5angle}) and (\ref{eq:ypi_C5angle}). For the source of %breaking global symmetry defined in Eq.~(\ref{eq:global_C5theta}), we add  anomalous couplings whose coefficients, 
Note that the axion-instanton couplings $\kappa_{0,\pi}$ should be integer-valued, while  the derivative axion couplings  $\gamma_{0,\pi}$ and $c_{0,\pi}$ can have any real value.

%In this model, we can give the relation between $[U(1)]_{\rm global}^{N+1}$  of the DCW and the corresponding transformation in the 5D %theory, and find the $[U(1)_{\rm CW}]_{\rm global}$ in the 5D action. In order to answer this question and  see the related KK %decomposition for the 5D field fluctuation, 

To proceed, we choose a specific gauge in the 5D action,  introducing $R_\xi$ gauge fixing term to remove the mixing between $C_\mu$ and $\{C_5,\, \theta\}$, and take the Feynman-'t Hooft gauge ($\xi=1$). This gauge choice makes the 5D action to take CCW form, from which the KK mass spectrum and wavefunction profiles  can be easily extracted.  
Specifically, we add
\bea \label{eq:gauge_fixing}
S_{\rm g.f} = -\int d^5 x \sqrt{-G} \left[ \frac{e^{2cS}}{g_C^2}\Big(
g^{\mu\nu}\partial_\mu C_\nu + G^{55}\chi^{-1}\partial_y(\chi C_5)- e^{-2cS}  g_C^2qf_5^3 \theta\Big)^2\right] 
\eea to the action (\ref{eq:ypi_C5angle}), where
 $\chi = e^{(2k_1-k_2)|y|}$, and
 replace the 5D metric $G_{MN}$ and the dilaton field $S$ with their vacuum values (\ref{GLD_sol}). 
 Although this removes the mixing between $C_\mu$ and $\{C_5,\theta\}$, the mixing between $C_5$ and $\theta$ is not removed yet.
We then make the field redefinition: 
\bea
\theta(x,y) &=& e^{-(k_1+ \frac{k_2}{2})|y|} f_5^{-3/2}(\cos\beta\, \varphi_+(x,y) -\sin\beta\, \varphi_-(x,y)),  \nonumber\\
C_5(x,y)  &=&  e^{-(k_1+\frac{k_2}{2})|y|} g_C ( \sin\beta\, \varphi_+(x,y) +\cos\beta\, \varphi_-(x,y) ),
\eea
which results in the following CCW form of 5D action:
\bea
S_C+S_{\rm g.f}&=&-\frac{1}{2} \int d^5 x   \Big[(\partial_\mu\varphi_+)^2  
  + e^{2(k_1-k_2)|y|}(\partial_y\varphi_+ - m_C\varphi_+)^2  \nonumber\\
&& \hskip 1.6cm  
+(\partial_\mu\varphi_-)^2  + e^{2(k_1-k_2)|y|}(\partial_y\varphi_-  + m_C\varphi_-)^2 \Big],
\eea 
where
\beq
m_C = \sqrt{ q^2g_C^2 f_5^3+ \big(k_1 + \frac{k_2}{2}\big)^2},\quad  \tan2\beta = \frac{ 2qg_C  f_5^{3/2}}{2k_1-k_2}.\eeq 
Obviously the above CCW lagrangian enjoys  the localized shift symmetries 
\beq
\label{local_shift_sy}
\varphi_\pm (x,y) \to \varphi_\pm(x,y) + c_\pm e^{\pm m_C y}, 
\eeq which originate essentially from the global shift symmetry (\ref{global_sy_boundary}) at the boundaries, which 
  insure the existence of two massless 4D axions.

One may do the standard KK analysis in this prescription with 
\bea
\label{KK_varphi}
\varphi_{\pm} = \sum_{n=0}^\infty f_{\pm n}(y) \delta\phi_{\pm n}(x),\eea and then the resulting mass eigenvalues and wavefunction profiles  can be immediately read off from  
Eq.~(\ref{f_n}) and (\ref{KKm}).  Here, the two zero modes $\delta\phi_{\pm 0}$  clearly correspond to massless 4D axions.
The zero mode axion $\delta\phi_{+0}(x)$  has a wavefunction profile exponentially localized at $y=\pi R$, while   $\delta\phi_{-0}(x)$  does at $y=0$.
With the localized shift symmetry (\ref{local_shift_sy}) and the associated zero mode axions $\delta \phi_{\pm 0}$, one can achieve hierarchical axion couplings at different boundaries.
 We stress that  the localized symmetry (\ref{local_shift_sy}) for CCW axions is obtained as a simple consequence of the underlying symmetries of the model, i.e.
$U(1)_C\times U(1)_0\times U(1)_\pi$, {\it without} any fine tuning of model parameters.

 As for the massive modes, one combination of $\delta\phi_{\pm n}$ $(n\geq 1)$ corresponds to the longitudinal  mode of massive 4D vector field, while the other is a massive 4D scalar.
 Since we are working at a particular gauge, at this level we can  not identify  which combination corresponds to the longitudinal  mode of massive vector field. This can be done by solving the 5D equations of motion as in Sec.~\ref{sec:CCW_photon}, however 
we will not perform it here.  Yet we can  easily show that  each massive vector and scalar modes have the same mass. 
For instance,
for $k_1=k_2$, the massive vector and scalar modes  have a common mass $M_n=\sqrt{m_C^2 + n^2/R^2}$ for $n\geq 1$,

Although convenient for the KK analysis, the above prescription is not ideal for a discussion of the field range of the canonically normalized zero mode axions. 
%In this representation with gauge fixing, it is not clear to discuss the full period of the canonically normalized 4D axion $a(x)$
% and  possible form of the boundary interactions for $a(x)$.  
A straightforward way to address this issue, as well as the low energy effective couplings of zero mode axions, is to integrate out massive KK modes in a gauge invariant manner and construct the effective action of zero mode axions.
The equations of motion of $C_\mu$ and $\theta$ have a gauge-covariant solution obeying 
\bea
&&C_{\mu }(x,y) = \partial_\mu \Gamma(x,y), \quad \partial_y \theta (x,y)= q\epsilon(y)C_{5}(x,y), 
\eea
where $\Gamma(x,y)$ is a scalar field with the boundary condition $\Gamma(x,0)= \Gamma(x,\pi R)=0$,
which transforms as $\Gamma\to \Gamma + \Lambda$ under $U(1)_C$. We then find that a gauge invariant solution 
given by
\bea 
\label{gauge_inv_sol}\theta(x,y)-  q\epsilon(y)\Gamma(x,y) &= & e^{-(k_1+ \frac{k_2}{2}) y}\Big[\,{\rm csch}(m_C\pi R)\sinh[m_C(\pi R- y)]\,\theta_1(x)  \nonumber \\
&&  \hskip 1.7cm +\,{\rm csch}(m_C\pi R) \sinh(m_C y)\,\theta_2(x)\,\Big], 
\eea
which is described by the two gauge invariant 4D scalar degrees of freedom $\theta_i(x)$ ($i=1,2$). 
Note that the matching condition \bea
\theta_1(x)=\theta(x, y=0), \quad \theta_2(x)=e^{(k_1+\frac{k_2}{2})\pi R}\theta(x, y=\pi R)\eea determines 
the periodicity of 4D fields $\theta_i(x)$ as\bea
\theta_1\equiv \theta_1 +2\pi, \quad \theta_2\equiv \theta_2 + 2\pi e^{(k_1+\frac{k_2}{2})\pi R}.\eea
The 4D effective Lagrangian of $\theta_i$  can be obtained by inserting the solution (\ref{gauge_inv_sol}) into the original 5D action (\ref{5D_C_theta}).  
The more conventional form can be obtained in terms of the canonically normalized axion fields $a_{1,2}$ (up to  kinetic mixing):\bea
{\cal L}_{\rm eff} &=& -\frac{1}{2}\Big( \partial_\mu a_1 \partial^\mu a_1 +\partial_\mu a_2 \partial^\mu a_2+\hat\epsilon\partial_\mu a_1 \partial^\mu a_2\Big) +\frac{\kappa_0}{32\pi^2}\frac{a_1}{f_1}G_0\tilde G_0 + \frac{\kappa_\pi}{32\pi^2}\frac{a_2}{f_2}G_\pi\tilde G_\pi \nonumber \\
&& +\, \frac{\partial_\mu a_1}{f_1}\left( c_{11} \tilde J^\mu_0 + c_{12} \tilde J^\mu_\pi -c_0 J_0^\mu \right) - \frac{\partial_\mu a_2}{f_2}\left( c_{21} \tilde J^\mu_0 + c_{22} \tilde J^\mu_\pi +c_\pi J_\pi^\mu \right),
\eea
where the axion decay constants $f_i$  define the axion periodicity as 
\bea
a_i\equiv a_i+2\pi f_i  \,\,\,\, (i=1,2).\eea
Using the parameters defined as 
\bea \label{eq:C5angle_wavefunctions}
Z_{11} &=&   m_C \coth(m_C\pi R) + (k_1+ k_2/2),\nonumber\\
Z_{12} &=& Z_{21}= - m_C\, {\rm csch}(m_C\pi R), \nonumber\\
Z_{22}&=&  m_C \coth(m_C\pi R) - (k_1+ k_2/2), \eea
we obtain explicit expressions for the axion decay constants, 
\bea   f_1=\frac{\sqrt{Z_{11}}}{qg_C},
 \quad f_2 =e^{(k_1+\frac{k_2}{2})\pi R}\frac{\sqrt{Z_{22}}}{qg_C},\eea
and  the nontrivial kinetic mixing coefficient $\hat\epsilon$,  
 \bea
\hat\epsilon = \frac{2Z_{12}}{\sqrt{Z_{11}Z_{22}}},\eea
and the dimensionless couplings $c_{ij}$  between the axions and the currents,
\bea
&& c_{11}=\frac{\gamma_0 Z_{11}}{M_5}, \quad c_{12} =e^{-(k_1+\frac{k_2}{2})\pi R}\frac{\gamma_\pi Z_{12}}{M_5}, \nonumber \\
&&c_{22}=e^{-k_2\pi R} \frac{\gamma_\pi Z_{22}}{M_5},\quad  c_{21}= e^{(k_1+ \frac{k_2}{2})\pi R}\frac{\gamma_0 Z_{21}}{M_5}.\eea

To examine the possible hierarchical structure of axion decay constants and the couplings, let us take the limit $m_C\pi R \gg 1$, and assume the following natural size of the underlying model parameters:
\bea
k_i \lesssim m_C < M_5,  \quad \gamma_{0,\pi}\sim c_{0,\pi} \sim \kappa_{0,\pi}={\cal O}(1). \eea
The resulting low energy axion  decay constants are estimated as
\bea
&& \hskip -0.5cm f_1\sim \sqrt{m_CM_5}, \quad \frac{f_2}{f_1} \sim e^{(k_1+\frac{k_2}{2})\pi R}, \quad \frac{f_2}{M_{\rm Pl}} \sim \frac{\sqrt{m_C k_i }}{M_5}\,\,\,\, {\rm or}\,\,\,\, \sqrt{\frac{m_C}{M_5^2\pi R}}. \eea
The kinetic mixing is exponentially suppressed, 
$\hat\epsilon \sim e^{- m_C \pi R}$, and 
the axion couplings have the following hierarchical structure: 
\bea
&& \hskip 1cm c_{11}\sim \frac{m_C}{M_5}, \quad c_{12} \sim e^{-(m_C+k_1+\frac{3}{2}k_2)\pi R} \frac{m_C}{M_5},\nonumber \\
&& \hskip 1cm c_{22}\sim e^{-(m_C+k_2)\pi R}\frac{m_C}{M_5},\quad c_{21}\sim e^{-(m_C-k_1-\frac{k_2}{2})\pi R}\frac{m_C}{M_5}.\eea
The above results show that the 5D model (\ref{5D_C_theta}) for CCW axions can generate a variety of exponential hierarchies among the axion derivative couplings, as well as an exponentially small kinetic mixing. The model can also generate an exponential hierarchy between the two axion scales $f_i$ ($i=1,2$) which define the field range of the canonically normalized 4D axions $a_i$. However, as long as the underlying 5D mass parameters are sub-Planckian, e.g. $m_C, k_i < M_5$,  the model can {\it not} generate a super-Planckian axion field range in the effective 4D theory, i.e. $f_i < M_{\rm Pl}$. Note that a super-Planckian effective axion scale could be easily   achieved in the original DCW axion.  In regard to this point, the key difference between DCW and CCW is that the extra spacial dimension for CCW should be a part of gravitational dynamics, while the linear quiver for DCW is completely decoupled from gravity. It is thus likely  that generically CCW can not provide a super-Planckian axion field range.   Our specific 5D model (\ref{5D_C_theta}) does not generate also an exponential hierarchy among the axion-instanton couplings.

%However, there is no such a mixing for the instanton couplings. 
%The boundary induced scalar potential for $\theta_i$ is just $V(\theta_1,\, \theta_2)= V_1(\theta_1) + V_2(\theta_2)$. If we assume %that $\theta_1$ becomes heavy with a mass $m_{\theta_1}$ due to the other boundary instanton potential,
%we can integrate out $\theta_1$ by the equation of motion:
%\bea
%[\theta_1]_{\rm heavy} = -\frac{Z_{21}}{Z_{11}}\frac{\partial^2\theta_2}{m_{\theta_1}^2}
%\sim e^{- m\pi R} \frac{\partial^2\theta_2}{m_{\theta_1}^2}.
%\eea
%The  role of the kinetic mixing term is to generate the additional derivative interactions for the lighter axion suppressed by 
%$e^{-m\pi R}$. Therefore the effective potential of the remaining lighter axion, $\theta_2$ is just given by the instanton coupling 
%at $y=\pi R$, and there is no freedom to give such a potential at $y=0$. 

Finally, we provide the relation between the canonically normalized zero modes $\delta\phi_{\pm 0}(x)$ defined in (\ref{KK_varphi})
and the axion fluctuation $\delta a_1\equiv f_1\delta\theta_1$ and $\delta a_2 \equiv f_2 e^{-(k_1+\frac{k_2}{2})\pi R}\delta\theta_2$ for $\theta_i$ defined as the massless fluctuation of the gauge-invariant combination (\ref{gauge_inv_sol}):
\bea
\delta\phi_{+0} &=& \sqrt{\frac{f_5}{(1- e^{-2m_C\pi R})(2m_C+ 2k_1+ k_2 )}}\left(
\frac{f_5e^{(k_1+\frac{k_2}{2})\pi R}}{f_2} \delta a_2 -\frac{f_5e^{-m_C\pi R}}{f_1} \delta a_1 \right) , \nonumber\\
\delta\phi_{-0} &=& -\sqrt{\frac{f_5}{(1- e^{-2m_C\pi R})(2m_C-2k_1- k_2)}}\left(
\frac{f_5}{f_1}\delta a_1 -\frac{f_5 e^{-(m_C-k_1-\frac{k_2}{2})\pi R}}{f_2} \delta a_2 \right).\eea

\subsection{CCW photon from 5D gauge field with even orbifold-parity}\label{sec:CCW_photon}

In this section, we study a 5D model for CCW $U(1)$ gauge boson. The model involves 
a $U(1)$ gauge symmetry and the associated gauge field $A_M= \{ A_\mu, A_5\} $ which transforms as
\bea
U(1)_A:\, A_M\to A_M + \partial_M\Lambda.
\eea
To have a light 4D vector boson with localized profile over the extra dimension, we need to introduce the bulk and boundary masses of $A_M$, which may take the St\"uckelberg  form for simplicity. Then, again at leading order in $g_{\rm st}^2\propto e^{-\xi S}$ in the generalized linear dilaton model, the 5D action is given by
\bea\label{eq:CCWphoton}
S_A&=& - \int d^5 x\sqrt{-G}\left[ \frac{e^{2cS }}{4 g_A^2}  
F^{MN} F_{MN}  + \frac{m_B^2}{2g_A^2}   (\partial^M\theta - A^M)(\partial_M \theta -   A_M) \right. \nonumber \\
&&\left.\hskip 1.5cm +\,
  \frac{e^{cS} m_b}{g_A^2}\, (\partial^\mu\theta -   A^\mu)(\partial_\mu\theta -   A_\mu)
\left(\frac{\delta(y)}{\sqrt{G_{55}}} - \frac{\delta(y-\pi R)}{\sqrt{G_{55}}}\right) \right],
\eea
where $\theta$ is a periodic Goldstone field for gauge invariant St\"uckelberg mass, which transforms under $U(1)_A$ as
\bea
\theta \rightarrow \theta +\Lambda, \eea
and $m_B$ and $m_b$ are the St\"uckelberg bulk and boundary mass terms, respectively. 
The field  variables and the gauge transformation
function obey the following orbifold boundary condition and the periodicity condition: 
\bea
&& A_\mu (x, y) = A_\mu(x, y+ 2\pi R) = A_\mu(x, - y),\nonumber \\
&& A_5(x,y) = A_5(x, y+2\pi R) =- A_5(x, -y),\nonumber \\   
&& \theta(x,y) = \theta(x, y+ 2\pi R) =\theta(x, -y)=\theta(x, y) + 2\pi, \nonumber \\
&& \Lambda(x,y)=\Lambda(x, y+2\pi R) =\Lambda(x, -y) =\Lambda(x, y) + 2\pi.
\eea

% If there is no other source to break the 5D $U(1)$ gauge symmetry spontaneously except the the vacuum value of the $A_5$, the unbroken %4D gauge symmetry is restricted by the 5D gauge parameter to make $A_5$ invariant, $\partial_y\Lambda =0 $. The solution becomes 
%$\Lambda_{\rm eff}(x,y)= Z_A \alpha(x)$, and the corresponding massless 4D  gauge field, $A_{0\mu}$ has the flat profile as 
%$A_{\mu{\rm eff}}(x,y) = Z_A A_{0\mu}(x)$ in the 5D field basis. 
%The exponential localization of the 4D vector field could be realized if there is the boundary localized mass terms like $  A_\mu A^\mu %(m_0\delta(y) + m_\pi\delta(y-\pi R))$. Including the bulk mass term $M_V^2 A_M A^M$  in the 5D action,  we could tune the mass of the %lightest vector field to be vanishing, which might yield an accidental 4D gauge symmetry, 
%$A_{0\mu}\to A_{0\mu} + \partial_\mu\alpha(x)$ with exponentially different charges for the boundary localized charge matters. 

%In order to get a 5D action which shows the CCW structure manifestly, one may fix the gauge properly, e.g. $A_5=0,\partial_y\theta = %A_5$, e.t.c, to decouple the gauge neutral degrees. Here we take a different approach in order to keep all degrees of freedom in the %action.  
For a gauge covariant KK analysis, we decompose the  relevant 5D fields as
\bea
A_{\mu}(x,y) &=&  \sum_{n=0}^\infty f_n(y) A_{n\mu}(x), \nonumber\\
A_5(x,y) &=& \sum_{n=0}^\infty f'_n(y) \pi_n(x) +   \sum_{n=1}^\infty (h_n(y) + g'_n(y)) \phi_n(x),\nonumber\\
\theta(x,y) &=& \sum_{n=0}^\infty f_n (y) \pi_n(x) + \sum_{n=1}^\infty g_n (y) \phi_n(x), 
\eea
and solve the equations of motion.  One can similarly decompose the gauge transformation function as
\beq
\Lambda(x,y) = \sum_{n=0}^\infty f_n(y) \alpha_n(x), \eeq 
and then the 4D fields $A_{n\mu}, \pi_n, \phi_n$ transform under $U(1)_A$ as
\bea
A_{n\mu}\to A_{n\mu} + \partial_\mu \alpha_n,\quad \pi_n\to \pi_n+\alpha_n, \quad \phi_n\to \phi_n.\eea
This shows that  $\pi_n$ can be identified as the longitudinal component  of the massive 4D vector fields $A_{n\mu}$, while the gauge invariant $\phi_n$ denote physical massive 4D scalar fields.  
After integrating  over $y$,  the lagrangian of all 4D fields  take the form
\bea
{\cal L}  =- \sum_{n=0}^{\infty} \left[\frac{1}{4 } (F_{n\mu\nu})^2  +\frac{M_n^2}{2} (\partial_\mu\pi_n - A_{n\mu})^2 \right] - \sum_{n=1}^{\infty}\left[
\frac{1}{2} (\partial_\mu\phi_n)^2  +\frac{M_n^2}{2} \phi_n^2\right],
\eea
where the mass eigenvalues $M_n$ will be determined later.
Here we assume that $M_n$ (including $n=0$) is nonzero in order to take into account all physical degrees of freedom correctly. We then find 
from the 5D equations of motion that the mode functions  $f_n, h_n$ and $g_n$ satisfy
\bea
\left[ \partial_y^2- m_B^2 - \big(k_1 + \frac{k_2}{2}\big)^2 +  e^{-2(k_1 - k_2)y} M_n^2\right] \big(e^{(k_1+\frac{k_2}{2})y} f_n(y)\big) &=&0,\nonumber\\
 \left[ \partial_y^2-  m_B^2 - \big(k_1 + \frac{k_2}{2}\big)^2  +  e^{-2(k_1 - k_2)y} M_n^2\right] \big( e^{3(k_1-\frac{k_2}{2})y} h_n(y) \big)&=&0,  \nonumber\\
 \Big[\partial_y + k_1 + \frac{k_2}{2}\Big] \big(e^{3(k_1 -\frac{k_2}{2})y}h_n(y)\big) - M_n^2 \big( e^{(k_1 +\frac{k_2}{2})y} g_n(y) \big)
\ &=& 0,
\eea
together with the boundary conditions 
\bea
\Big[\partial_y - m_V- k_1 - \frac{k_2}{2} \Big]\big(e^{(k_1+ \frac{k_2}{2})y}f_n(y)\big)|_{y=0, \pi R} &=&0, \nonumber\\
\Big[\partial_y - \frac{m_B^2}{m_b} +  k_1 + \frac{k_2}{2} \Big]\big( e^{3(k_1 -\frac{k_2}{2})y} h_n(y)\big)|_{y=0,\pi R} &=&0.
\eea

Although the boundary conditions for $f_n(y)$ and $h_n(y)$ are different, 
it could be verified that they share the same mass spectrum except the mass $M_0$ of $A_{0\mu}$. For generic values of $\{k_1,\, k_2,\, m_B,\, m_b\}$,  $M_0$ can be even larger than $M_n$ ($n\geq 1$). However there exists a wide range of model parameters yielding
$M_0 \ll M_n$ ($n\geq 1$). In particular,  for $m_B^2 = m_b(m_b + 2k_1 + k_2)$, we have a massless 4D vector ($M_0=0$) with a localized wave function $f_0(y)\propto e^{ m_by}$.

For simplicity, let us consider the case  $k_1=k_2$ for which it is straightforward  to obtain the mode functions and mass eigenvalues. 
We then find \bea
f_0(y) &=& e^{m_b y} \sqrt{\frac{m_b+ \frac{3}{2}k_1 }{e^{(2m_b +  3k_1 )\pi R} - 1}} 
\nonumber  \\
f_n (y) &=&  \frac{e^{-\frac{3k_1}{2} y}}{\sqrt{\pi R \big((m_b+\frac{3}{2}k_1)^2R^2 + n^2 \big) }}
\left[n\cos\left(\frac{n y}{R}\right) 
+  \Big(m_b+\frac{3}{2}k_1\Big)R  \sin\left(\frac{n y}{R}\right) \right] , 
\nonumber\\
h_n(y) &=& \frac{e^{-\frac{3}{2}k_1 y}}{\sqrt{\pi R  \big((\frac{m_B^2}{m_b}-\frac{3}{2}k_1 )^2R^2 + n^2 \big)}}
\left[n\cos\left(\frac{n y}{R}\right) 
+  \Big(\frac{m_B^2}{m_b}-\frac{3}{2}k_1\Big)R  \sin\left(\frac{n y}{R}\right) \right], \nonumber\\ \eea 
for $n\geq 1$, and the corresponding mass eigenvalues are given by
\bea
M_0^2 &=& m_B^2 - m_b(m_b + 3k_1), \nonumber\\
M_n^2 &=& m_B^2 + \Big(\frac{3}{2}k_1\Big)^2 +\frac{n^2}{R^2} \quad (n\geq 1).
\eea

Focusing on the exponential $y$-dependence in the limit $m\pi R\gg 1$, we have  
\bea
f_0(y) \sim  e^{-\frac{3}{2} k_1\pi R} e^{ m_b(y-\pi R)}, 
\quad  f_n(y) \sim  h_n(y) \sim e^{- \frac{3}{2} k_1 y}.
\eea
This behavior of mode functions  shows clearly the role of each parameter.  
 $A_{0\mu}$ has an exponentially localized  profile only when the boundary St\"uckelberg mass $m_b$ is non-vanishing. 
On the other hand, the mode function ratio $f_0(y)/f_n(y) \sim e^{(m_b+\frac{3}{2}k_1)(y-\pi R)}$ ($n\geq 1$) indicates that
a possible exponential hierarchy between the zero mode ($n=0$) coupling and the non-zero mode $(n\geq 1)$  coupling at a boundary with $y=0$ depends on both the boundary mass $m_b$ and the background geometry described by $k_1$. Actually, all KK modes are exponentially localized at $y=0$ in our field basis due to the background geometry. 
We also note that the hierarchical structure of the boundary couplings of mass eigenstate modes is independent of the existence of massless or particularly light mode, i.e. 
it is valid even when $M_0 \sim M_1$. 

Related to the CCW, the most interesting parameter region of the model is the one yielding\footnote{One may even take the limit $M_0\rightarrow 0$, while keeping $M_n$ ($n\geq 1$) well above the characteristic energy scale. It is in fact not clear if the Goldstone zero mode $\pi_0$ has  a sensible behavior when $M_0\rightarrow 0$
as its kinetic term vanishes, while still there could be non-vanishing interactions. Such potentially singular behavior might be a consequence of another massless degree of freedom which is already integrated out at the starting point. In this paper, we do not pursue this issue further as our major concern is the CCW features of $A_{0\mu}$. }  
\bea
M_0 =\sqrt{m_B^2-m_b(m_b+3k_1)}\ll 1/R \ll m_B.\eea
In such case, the 5D model (\ref{eq:CCWphoton}) successfully provides a light $A_{0\mu}$ with localized profile whose boundary couplings at different boundaries reveal an exponential hierarchy\footnote{Such exponential hierarchy among the couplings of $A_{0\mu}$ suggests that there is no globally well defined compact 4D $U(1)$ gauge symmetry associated with 
 $A_{0\mu}$.} as \bea
 \frac{g_0(y=0)}{g_0(y=\pi R)}\propto e^{-m_b\pi R},\eea together with approximately degenerate massive KK modes whose boundary couplings at $y=0$
 reveal another exponential hierarchy \bea
 \frac{g_0(y=0)}{g_n(y=0)} \propto e^{-(m_b+\frac{3}{2}k_1)\pi R} \quad (n\geq 1),\eea thereby realizing the key features of the CW $U(1)$ gauge bosons.   
Note that the mass eigenvalues of non-zero modes, $M_n$ for $n\geq 1$, are independent of the boundary  St\"uckelberg  mass $m_b$, and the shape of mode functions is rather insensitive to the bulk St\"uckelberg mass $m_B$.

\section{Conclusions} \label{sec:conclusion}

In this paper, we have studied the continuum limit of the discrete clockwork (DCW) mechanism, dubbed the continuum clockwork (CCW). 
To be specific, we focused on the possible continuum realizations of CW axions and $U(1)$ gauge bosons.
There are two different prescriptions for the corresponding CCW lagrangian 
as previously discussed in \cite{Giudice:2016yja,Craig:2017cda}: CCW-I associated with localized profile of zero modes in a field basis for which the 5D $U(1)$ and discrete gauge symmetries take the standard form  as (\ref{5d_gauge_symmetry}), which is induced by appropriately tuned bulk and boundary masses, and CCW-II associated with the background geometry and dilaton profile of the linear dilaton model \cite{Antoniadis:2011qw}. 
We discussed in detail that each prescription has its own limitation. For instance, assuming a discrete shift symmetry which assures that the involved axion is a periodic scalar, and also taking into account the  $U(1)$ gauge symmetry associated with the involved 5D vector field, in CCW either the localized CW symmetries to protect zero modes are not respected by gravitational interactions (CCW-I), or an exponential hierarchy between the zero mode couplings at different boundaries can not be generated (CCW-II). 
As other limitations relative to the DCW, CCW could yield  neither an exponentially enhanced trans-Planckian field range of the zero mode axion, nor an exponential hierarchy among the 4D gauge charges, while keeping the quantized nature of gauge charges of an unbroken 4D $U(1)$ gauge symmetry. 

Motivated by those two  prescriptions,  we  examined a more general CCW which can incorporate  both CCW-I and CCW-II within a common framework.
For this, we proposed a generalized 5D linear dilaton model providing a concrete realisation of the setup, in which the three independent parameters describing the general CCW  are identified as geometric warp factor,
large volume, and bulk/boundary mass responsible for localized zero mode profile. 
We then  analyzed the KK spectra and the couplings of zero and massive KK modes to boundary-localized operators. It is found that many of the characteristic features of the DCW can be reproduced in a certain parameter region which might be identified as the CCW limit, and our setup allows a continuous deformation from the CCW limit to other limits such as the well known Randall-Sundrum and large extra dimension limits.  

We subsequently discussed concrete 5D models for CCW axions and $U(1)$ gauge bosons in generalized linear dilaton background. 
The 4D zero mode axion originating from a 5D angular field has quite different properties from the DCW axion since the corresponding CW symmetry to protect zero mode has a different form due to the periodicity of the 5D angular field. As a result, the couplings of the  zero mode axion fluctuation to different boundary operators can exhibit an exponential hierarchy, while there is no exponentially enlarged field range of the corresponding  zero mode axion.
We also examined the CCW axions originating from a 5D gauge field having  nonzero St\"uckelberg mass and
odd orbifold parity. It turns out that the resulting zero mode axions can have an exponential hierarchy among the 
{\it derivative} couplings at different boundaries, while there is no such hierarchy for the axion-instanton couplings due to the restriction imposed by the underlying 5D gauge symmetry. 
One can obtain also a light 4D vector field and its KK excitations, which originate from a 5D gauge field with even orbifold parity and nonzero bulk and boundary  St\"uckelberg masses, and reproduce many of  the characteristic features of the DCW $U(1)$ gauge bosons.
 
%%%%%%%%%%%%%%%%%%%%%%%%%%
\section{Acknowledgement}
%%%%%%%%%%%%%%%%%%%%%%%%%%
The work of [KC] and [CSS] was supported by IBS under the project code, IBS-R018-D1. [CSS] acknowledges the support from the Korea Ministry of Education, Science and Technology, Gyeongsangbuk-Do and Pohang City for Independent Junior Research Groups at the Asia Pacific Center for Theoretical Physics. [CSS] was also supported by the Basic Science Research Program through the NRF Grant (No. 2017R1D1A1B04032316). [SHI] was supported by the German Science Foundation (DFG) within the SFB-Transregio TR33 ``The Dark Universe".


\begin{thebibliography}{99}
 
 %\cite{Choi:2014rja}
\bibitem{Choi:2014rja} 
  K.~Choi, H.~Kim and S.~Yun,
  {\it Natural inflation with multiple sub-Planckian axions},
  Phys.\ Rev.\ D {\bf 90}, 023545 (2014)
  %%%doi:10.1103/PhysRevD.90.023545
  [arXiv:1404.6209 [hep-th]].
  %%CITATION = %%doi:10.1103/PhysRevD.90.023545;%%
  %101 citations counted in INSPIRE as of 30 Oct 2017
 
 %\cite{Choi:2015fiu}
\bibitem{Choi:2015fiu} 
  K.~Choi and S.~H.~Im,
  {\it Realizing the relaxion from multiple axions and its UV completion with high scale supersymmetry},
  JHEP {\bf 1601}, 149 (2016)
  %%%doi:10.1007/JHEP01(2016)149
  [arXiv:1511.00132 [hep-ph]].
  %%CITATION = %%doi:10.1007/JHEP01(2016)149;%%
  %69 citations counted in INSPIRE as of 30 Oct 2017
 
 %\cite{Kaplan:2015fuy}
\bibitem{Kaplan:2015fuy} 
  D.~E.~Kaplan and R.~Rattazzi,
  {\it Large field excursions and approximate discrete symmetries from a clockwork axion},
  Phys.\ Rev.\ D {\bf 93}, no. 8, 085007 (2016)
  %%%doi:10.1103/PhysRevD.93.085007
  [arXiv:1511.01827 [hep-ph]].
  %%CITATION = %%doi:10.1103/PhysRevD.93.085007;%%
  %77 citations counted in INSPIRE as of 30 Oct 2017
 
 
    \bibitem{Saraswat:2016eaz} 
  P.~Saraswat,
  {\it Weak gravity conjecture and effective field theory},
  Phys.\ Rev.\ D {\bf 95}, no. 2, 025013 (2017)
  %%doi:10.1103/PhysRevD.95.025013
  [arXiv:1608.06951 [hep-th]].
 

     %\cite{Giudice:2016yja} 
\bibitem{Giudice:2016yja} 
  G.~F.~Giudice and M.~McCullough,
  {\it A Clockwork Theory},
  JHEP {\bf 1702}, 036 (2017)
  %%%doi:10.1007/JHEP02(2017)036
  [arXiv:1610.07962 [hep-ph]].
  %%CITATION = %%doi:10.1007/JHEP02(2017)036;%%
  %33 citations counted in INSPIRE as of 30 Oct 2017
 

  %%%%%%%%Applications%%%%%%%%%%%%%%%%%%

%\cite{Higaki:2015jag}
\bibitem{Higaki:2015jag} 
  T.~Higaki, K.~S.~Jeong, N.~Kitajima and F.~Takahashi,
  {\it The QCD Axion from Aligned Axions and Diphoton Excess},
  Phys.\ Lett.\ B {\bf 755}, 13 (2016)
  %%%doi:10.1016/j.physletb.2016.01.055
  [arXiv:1512.05295 [hep-ph]].
  %%CITATION = %%doi:10.1016/j.physletb.2016.01.055;%%
  %147 citations counted in INSPIRE as of 13 Nov 2017
  
    %\cite{Fonseca:2016eoo}
\bibitem{Fonseca:2016eoo} 
  N.~Fonseca, L.~de Lima, C.~S.~Machado and R.~D.~Matheus,
  {\it Large field excursions from a few site relaxion model},
  Phys.\ Rev.\ D {\bf 94}, no. 1, 015010 (2016)
  %doi:10.1103/PhysRevD.94.015010
  [arXiv:1601.07183 [hep-ph]].
  %%CITATION = %doi:10.1103/PhysRevD.94.015010;%%
  %18 citations counted in INSPIRE as of 13 Nov 2017

%\cite{Kehagias:2016kzt} 
\bibitem{Kehagias:2016kzt} 
  A.~Kehagias and A.~Riotto,
  {\it Clockwork Inflation},
  Phys.\ Lett.\ B {\bf 767}, 73 (2017)
  %%doi:10.1016/j.physletb.2017.01.042
  [arXiv:1611.03316 [hep-ph]].
  %%CITATION = %%doi:10.1016/j.physletb.2017.01.042;%%
  %17 citations counted in INSPIRE as of 13 Nov 2017
  

%\cite{Farina:2016tgd}
\bibitem{Farina:2016tgd} 
  M.~Farina, D.~Pappadopulo, F.~Rompineve and A.~Tesi,
  {\it The photo-philic QCD axion},
  JHEP {\bf 1701}, 095 (2017)
  %%doi:10.1007/JHEP01(2017)095
  [arXiv:1611.09855 [hep-ph]].
  %%CITATION = %%doi:10.1007/JHEP01(2017)095;%%
  %26 citations counted in INSPIRE as of 13 Nov 2017
  

%\cite{Ahmed:2016viu}
\bibitem{Ahmed:2016viu} 
  A.~Ahmed and B.~M.~Dillon,
  {\it Clockwork Goldstone Bosons},
  Phys.\ Rev.\ D {\bf 96}, no. 11, 115031 (2017)
  %%doi:10.1103/PhysRevD.96.115031
  [arXiv:1612.04011 [hep-ph]].
  %%CITATION = doi:10.1103/PhysRevD.96.115031;%%
  %26 citations counted in INSPIRE as of 14 Aug 2018


%\cite{Hambye:2016qkf}
\bibitem{Hambye:2016qkf} 
  T.~Hambye, D.~Teresi and M.~H.~G.~Tytgat,
  {\it A Clockwork WIMP},
  JHEP {\bf 1707}, 047 (2017)
  %%doi:10.1007/JHEP07(2017)047
  [arXiv:1612.06411 [hep-ph]].
  %%CITATION = %%doi:10.1007/JHEP07(2017)047;%%
  %16 citations counted in INSPIRE as of 13 Nov 2017
  
  %\cite{vonGersdorff:2017iym}
\bibitem{vonGersdorff:2017iym} 
  G.~von Gersdorff,
  {\it Natural Fermion Hierarchies from Random Yukawa Couplings},
  JHEP {\bf 1709}, 094 (2017)
  % doi:10.1007/JHEP09(2017)094
  [arXiv:1705.05430 [hep-ph]].
  %%CITATION = doi:10.1007/JHEP09(2017)094;%%
  %1 citations counted in INSPIRE as of 30 Nov 2017

%\cite{Teresi:2017yrp}
\bibitem{Teresi:2017yrp} 
  D.~Teresi,
  {\it Clockwork Dark Matter},
  arXiv:1705.09698 [hep-ph].
  %%CITATION = ARXIV:1705.09698;%%
  %2 citations counted in INSPIRE as of 13 Nov 2017


%\cite{Coy:2017yex}
\bibitem{Coy:2017yex} 
  R.~Coy, M.~Frigerio and M.~Ibe,
  {\it Dynamical Clockwork Axions},
  JHEP {\bf 1710}, 002 (2017)
  %%doi:10.1007/JHEP10(2017)002
  [arXiv:1706.04529 [hep-ph]].
  %%CITATION = %%doi:10.1007/JHEP10(2017)002;%%
  %9 citations counted in INSPIRE as of 13 Nov 2017
  
    %\cite{Ben-Dayan:2017rvr}
\bibitem{Ben-Dayan:2017rvr} 
  I.~Ben-Dayan,
  {\it Generalized Clockwork Theory},
  arXiv:1706.05308 [hep-ph].
  %%CITATION = ARXIV:1706.05308;%%
  %4 citations counted in INSPIRE as of 13 Nov 2017

%\cite{Hong:2017tel}
\bibitem{Hong:2017tel} 
  D.~K.~Hong, D.~H.~Kim and C.~S.~Shin,
  {\it Clockwork graviton contributions to muon $g-2$},
  Phys.\ Rev.\ D {\bf 97}, no. 3, 035014 (2018)
  %%doi:10.1103/PhysRevD.97.035014
  [arXiv:1706.09376 [hep-ph]].
  %%CITATION = doi:10.1103/PhysRevD.97.035014;%%
  %12 citations counted in INSPIRE as of 14 Aug 2018
  
%\cite{Park:2017yrn}
\bibitem{Park:2017yrn} 
  S.~C.~Park and C.~S.~Shin,
  {\it Clockwork seesaw mechanisms},
  Phys.\ Lett.\ B {\bf 776}, 222 (2018)
  %%doi:10.1016/j.physletb.2017.11.057
  [arXiv:1707.07364 [hep-ph]].
  %%CITATION = doi:10.1016/j.physletb.2017.11.057;%%
  %12 citations counted in INSPIRE as of 14 Aug 2018


%\cite{Lee:2017fin}
\bibitem{Lee:2017fin} 
  H.~M.~Lee,
  {\it Gauged $U(1)$ clockwork theory},
  Phys.\ Lett.\ B {\bf 778}, 79 (2018)
  %%doi:10.1016/j.physletb.2018.01.010
  [arXiv:1708.03564 [hep-ph]].
  %%CITATION = doi:10.1016/j.physletb.2018.01.010;%%
  %17 citations counted in INSPIRE as of 14 Aug 2018
  
%\cite{Agrawal:2017eqm}
\bibitem{Agrawal:2017eqm} 
  P.~Agrawal, G.~Marques-Tavares and W.~Xue,
  {\it Opening up the QCD axion window},
  JHEP {\bf 1803}, 049 (2018)
  %%doi:10.1007/JHEP03(2018)049
  [arXiv:1708.05008 [hep-ph]].
  %%CITATION = doi:10.1007/JHEP03(2018)049;%%
  %13 citations counted in INSPIRE as of 14 Aug 2018
  
%\cite{Kim:2017mtc}
\bibitem{Kim:2017mtc} 
  J.~Kim and J.~McDonald,
  {\it Clockwork Higgs portal model for freeze-in dark matter},
  Phys.\ Rev.\ D {\bf 98}, no. 2, 023533 (2018)
  %%doi:10.1103/PhysRevD.98.023533
  [arXiv:1709.04105 [hep-ph]].
  %%CITATION = doi:10.1103/PhysRevD.98.023533;%%
  %13 citations counted in INSPIRE as of 14 Aug 2018
  
  
%\cite{Agrawal:2017cmd}
\bibitem{Agrawal:2017cmd} 
  P.~Agrawal, J.~Fan, M.~Reece and L.~T.~Wang,
  {\it Experimental Targets for Photon Couplings of the QCD Axion},
  JHEP {\bf 1802}, 006 (2018)
  %%doi:10.1007/JHEP02(2018)006
  [arXiv:1709.06085 [hep-ph]].
  %%CITATION = doi:10.1007/JHEP02(2018)006;%%
  %15 citations counted in INSPIRE as of 14 Aug 2018
  
%\cite{Ibarra:2017tju}
\bibitem{Ibarra:2017tju} 
  A.~Ibarra, A.~Kushwaha and S.~K.~Vempati,
  {\it Clockwork for Neutrino Masses and Lepton Flavor Violation},
  Phys.\ Lett.\ B {\bf 780}, 86 (2018)
  %%doi:10.1016/j.physletb.2018.02.047
  [arXiv:1711.02070 [hep-ph]].
  %%CITATION = doi:10.1016/j.physletb.2018.02.047;%%
  %8 citations counted in INSPIRE as of 14 Aug 2018
  
%\cite{Patel:2017pct}
\bibitem{Patel:2017pct} 
  K.~M.~Patel,
  {\it Clockwork mechanism for flavor hierarchies},
  Phys.\ Rev.\ D {\bf 96}, no. 11, 115013 (2017)
  %%doi:10.1103/PhysRevD.96.115013
  [arXiv:1711.05393 [hep-ph]].
  %%CITATION = doi:10.1103/PhysRevD.96.115013;%%
  %6 citations counted in INSPIRE as of 14 Aug 2018

  
  
  
  %%%%%%%%%%%%%%%%%%%%%%%%%%%%%%%
  
  %\cite{ArkaniHamed:2006dz}
\bibitem{ArkaniHamed:2006dz} 
  N.~Arkani-Hamed, L.~Motl, A.~Nicolis and C.~Vafa,
  {\it The String landscape, black holes and gravity as the weakest force},
  JHEP {\bf 0706}, 060 (2007)
  %doi:10.1088/1126-6708/2007/06/060
  [hep-th/0601001].
  %%CITATION = %doi:10.1088/1126-6708/2007/06/060;%%
  %325 citations counted in INSPIRE as of 13 Nov 2017
  
%\cite{Ibanez:2017vfl}
\bibitem{Ibanez:2017vfl} 
  L.~E.~Ibanez and M.~Montero,
  {\it A Note on the WGC, Effective Field Theory and Clockwork within String Theory},
  JHEP {\bf 1802}, 057 (2018)
  %%doi:10.1007/JHEP02(2018)057
  [arXiv:1709.02392 [hep-th]].
  %%CITATION = doi:10.1007/JHEP02(2018)057;%%
  %15 citations counted in INSPIRE as of 14 Aug 2018
 
 
%\cite{Craig:2017cda}
\bibitem{Craig:2017cda} 
  N.~Craig, I.~Garcia Garcia and D.~Sutherland,
  {\it Disassembling the Clockwork Mechanism},
  JHEP {\bf 1710}, 018 (2017)
  %%%doi:10.1007/JHEP10(2017)018
  [arXiv:1704.07831 [hep-ph]].
  %%CITATION = %%doi:10.1007/JHEP10(2017)018;%%
  %14 citations counted in INSPIRE as of 30 Oct 2017
  
%\cite{Giudice:2017fmj}
\bibitem{Giudice:2017fmj} 
  G.~F.~Giudice, Y.~Kats, M.~McCullough, R.~Torre and A.~Urbano,
  {\it Clockwork/linear dilaton: structure and phenomenology},
  JHEP {\bf 1806}, 009 (2018)
  %doi:10.1007/JHEP06(2018)009
  [arXiv:1711.08437 [hep-ph]].
  %%CITATION = doi:10.1007/JHEP06(2018)009;%%
  %14 citations counted in INSPIRE as of 14 Aug 2018
  
    %\cite{Antoniadis:2011qw}
\bibitem{Antoniadis:2011qw} 
  I.~Antoniadis, A.~Arvanitaki, S.~Dimopoulos and A.~Giveon,
  {\it Phenomenology of TeV Little String Theory from Holography},
  Phys.\ Rev.\ Lett.\  {\bf 108}, 081602 (2012)
  %%%doi:10.1103/PhysRevLett.108.081602
  [arXiv:1102.4043 [hep-ph]].
  %%CITATION = %%doi:10.1103/PhysRevLett.108.081602;%%
  %37 citations counted in INSPIRE as of 30 Oct 2017
  
     %\cite{Giudice:2017suc}
\bibitem{Giudice:2017suc} 
  G.~F.~Giudice and M.~McCullough,
  {\it Comment on "Disassembling the Clockwork Mechanism"},
  arXiv:1705.10162 [hep-ph].
  %%CITATION = ARXIV:1705.10162;%%
  %8 citations counted in INSPIRE as of 30 Oct 2017
  
  
    %\cite{Randall:1999ee}
\bibitem{Randall:1999ee} 
  L.~Randall and R.~Sundrum,
  {\it A Large mass hierarchy from a small extra dimension},
  Phys.\ Rev.\ Lett.\  {\bf 83}, 3370 (1999)
  %doi:10.1103/PhysRevLett.83.3370
  [hep-ph/9905221].
  %%CITATION = %doi:10.1103/PhysRevLett.83.3370;%%
  %7729 citations counted in INSPIRE as of 13 Nov 2017
  
  %\cite{ArkaniHamed:1998rs}
\bibitem{ArkaniHamed:1998rs} 
  N.~Arkani-Hamed, S.~Dimopoulos and G.~R.~Dvali,
  {\it The Hierarchy problem and new dimensions at a millimeter},
  Phys.\ Lett.\ B {\bf 429}, 263 (1998)
  %doi:10.1016/S0370-2693(98)00466-3
  [hep-ph/9803315].
  %%CITATION = %doi:10.1016/S0370-2693(98)00466-3;%%
  %6132 citations counted in INSPIRE as of 13 Nov 2017
  
    %\cite{Gherghetta:2000qt}
\bibitem{Gherghetta:2000qt} 
  T.~Gherghetta and A.~Pomarol,
  {\it Bulk fields and supersymmetry in a slice of AdS},
  Nucl.\ Phys.\ B {\bf 586}, 141 (2000)
  %%%doi:10.1016/S0550-3213(00)00392-8
  [hep-ph/0003129].
  %%CITATION = %%doi:10.1016/S0550-3213(00)00392-8;%%
  %1061 citations counted in INSPIRE as of 30 Oct 2017
  
  %\cite{Choi:2003wr} 
\bibitem{Choi:2003wr} 
  K.~Choi,
  {\it A QCD axion from higher dimensional gauge field},
  Phys.\ Rev.\ Lett.\  {\bf 92}, 101602 (2004)
  %%%doi:10.1103/PhysRevLett.92.101602
  [hep-ph/0308024].
  %%CITATION = %%doi:10.1103/PhysRevLett.92.101602;%%
  %27 citations counted in INSPIRE as of 30 Oct 2017
  
  %\cite{Flacke:2006ad}
\bibitem{Flacke:2006ad} 
  T.~Flacke, B.~Gripaios, J.~March-Russell and D.~Maybury,
  {\it Warped axions},
  JHEP {\bf 0701}, 061 (2007)
  %%%doi:10.1088/1126-6708/2007/01/061
  [hep-ph/0611278].
  %%CITATION = %%doi:10.1088/1126-6708/2007/01/061;%%
  %22 citations counted in INSPIRE as of 30 Oct 2017
  
%\cite{Im:2017eju}
\bibitem{Im:2017eju} 
  S.~H.~Im, H.~P.~Nilles and A.~Trautner,
  {\it Exploring extra dimensions through inflationary tensor modes},
  JHEP {\bf 1803}, 004 (2018)
  %%doi:10.1007/JHEP03(2018)004
  [arXiv:1707.03830 [hep-ph]].
  %%CITATION = doi:10.1007/JHEP03(2018)004;%%
  %12 citations counted in INSPIRE as of 14 Aug 2018
  
  
%\cite{Kehagias:2017grx}
\bibitem{Kehagias:2017grx} 
  A.~Kehagias and A.~Riotto,
  {\it The Clockwork Supergravity},
  JHEP {\bf 1802}, 160 (2018)
  %%doi:10.1007/JHEP02(2018)160
  [arXiv:1710.04175 [hep-th]].
  %%CITATION = doi:10.1007/JHEP02(2018)160;%%
  %7 citations counted in INSPIRE as of 14 Aug 2018
  
%\cite{Antoniadis:2017wyh}
\bibitem{Antoniadis:2017wyh} 
  I.~Antoniadis, A.~Delgado, C.~Markou and S.~Pokorski,
  {\it The effective supergravity of Little String Theory},
  Eur.\ Phys.\ J.\ C {\bf 78}, no. 2, 146 (2018)
  %%doi:10.1140/epjc/s10052-018-5632-4
  [arXiv:1710.05568 [hep-th]].
  %%CITATION = doi:10.1140/epjc/s10052-018-5632-4;%%
  %5 citations counted in INSPIRE as of 14 Aug 2018
  
  %\cite{Patrignani:2016xqp}
\bibitem{Patrignani:2016xqp} 
  C.~Patrignani {\it et al.} [Particle Data Group],
  {\it Review of Particle Physics},
  Chin.\ Phys.\ C {\bf 40}, no. 10, 100001 (2016).
  %doi:10.1088/1674-1137/40/10/100001
  %%CITATION = doi:10.1088/1674-1137/40/10/100001;%%
  %1905 citations counted in INSPIRE as of 13 Nov 2017
  
       %\cite{Cox:2012ee}
\bibitem{Cox:2012ee} 
  P.~Cox and T.~Gherghetta,
  {\it Radion Dynamics and Phenomenology in the Linear Dilaton Model},
  JHEP {\bf 1205}, 149 (2012)
  %doi:10.1007/JHEP05(2012)149
  [arXiv:1203.5870 [hep-ph]].
  %%CITATION = doi:10.1007/JHEP05(2012)149;%%
  %11 citations counted in INSPIRE as of 29 Nov 2017 
  
  %\cite{ArkaniHamed:2001ca}
\bibitem{ArkaniHamed:2001ca} 
  N.~Arkani-Hamed, A.~G.~Cohen and H.~Georgi,
  {\it (De)constructing dimensions},
  Phys.\ Rev.\ Lett.\  {\bf 86}, 4757 (2001)
  %doi:10.1103/PhysRevLett.86.4757
  [hep-th/0104005].
  %%CITATION = %doi:10.1103/PhysRevLett.86.4757;%%
  %687 citations counted in INSPIRE as of 13 Nov 2017
  
  %\cite{Hill:2000mu}
\bibitem{Hill:2000mu} 
  C.~T.~Hill, S.~Pokorski and J.~Wang,
  {\it Gauge invariant effective Lagrangian for Kaluza-Klein modes},
  Phys.\ Rev.\ D {\bf 64}, 105005 (2001)
  %doi:10.1103/PhysRevD.64.105005
  [hep-th/0104035].
  %%CITATION = %doi:10.1103/PhysRevD.64.105005;%%
  %386 citations counted in INSPIRE as of 13 Nov 2017
  

  

  
\end{thebibliography}
\end{document}